\documentclass[a4wide,11pt,oneside]{book} % 'master' to combine all chapters
\input{add/settings.tex}

\usepackage{color,xcolor}
\usepackage[export]{adjustbox}

\definecolor{amethyst}{rgb}{0.6, 0.4, 0.8}
\definecolor{green}{rgb}{0.55, 0.71, 0.0}
\definecolor{apricot}{rgb}{0.98, 0.81, 0.69}
\definecolor{auburn}{rgb}{0.43, 0.21, 0.1}
\definecolor{babyblueeyes}{rgb}{0.63, 0.79, 0.95}
\definecolor{bittersweet}{rgb}{1.0, 0.44, 0.37}
\definecolor{officegreen}{rgb}{0.0, 0.5, 0.0}
\definecolor{darkcandyapplered}{rgb}{0.64, 0.0, 0.0}
\definecolor{blue(munsell)}{rgb}{0.0, 0.5, 0.69}

\newcommand{\strike}{\bgroup\markoverwith{\textcolor{red}{\rule[0.5ex]{2pt}{0.4pt}}}\ULon}
\usepackage{soul} % for strikethrough
\begin{document}
%\title{}\\

%\bigskip
\noindent
\textbf{{\Large Dark Matter and Fundamental Physics Searches with IACTs}}

\bigskip\noindent
Chapter 8 of \textit{`Advances in Very High Energy Astrophysics -- The Science Program of the Third Generation IACTs for Exploring Cosmic Gamma Rays`}, Mukherjee \& Zanin, World Scientific (2024) \url{https://www.worldscientific.com/worldscibooks/10.1142/11141} and \url{https://doi.org/10.1142/11141}

\bigskip\bigskip
\noindent
Michele Doro$^1$, Miguel Angel S\'anchez-Conde$^2$, Moritz~H\"utten$^3$

\medskip
\noindent
$^1$ University of Padova, Department of Physics and Astronomy, I-35131 Padova (Italy) \url{michele.doro@unipd.it}\\
$^2$ Instituto de Física Teórica, IFT UAM-CSIC, Departamento de Física Teórica, Universidad Autónoma de Madrid, ES-28049 Madrid (Spain)
 \url{miguel.sanchezconde@uam.es} \\
$^3$ Institute for Cosmic Ray Research, The University of Tokyo, Kashiwa, Chiba 277-8583 (Japan) \url{huetten@icrr.u-tokyo.ac.jp }

%\maketitle
\tableofcontents

%\chapter*{Dark Matter and Fundamental Physics\label{ch7}}
%\newpage
\setcounter{chapter}{8}

\section*{Introduction\label{sec:ch7_intro}}

Gamma rays make up the highest energy end of the electromagnetic
spectrum and are generated when interactions involve highly energetic
leptons or hadrons. Those accelerated particles reaching us from outer space are collectively called cosmic rays, and are subject of intense studies for almost a century now. Gamma radiation is created in diverse non-thermal environments in the Universe. Today, cosmic-ray studies are performed with several tens of different instruments of different classes, including \glspl{iact}.  The astrophysical reach of gamma-ray astronomy is huge, as largely proven in the past decade of scientific production and comprehensively resumed in this book.  Besides for non-thermal astrophysical environments, gamma rays are an excellent probe to search  for new physics.  Some of these exotic and exciting scenarios
 are the subject of this contribution.

 All current \glspl{iact} have invested a great deal of time and resources in
searching for these signatures over the past two decades, investing a large fraction of their observation time, so precious for the other astrophysical target observations described in the book. In this chapter, we focus on a selection of topics in fundamental physics such as the indirect search for   weakly interacting massive particles (WIMPs; Sec.~\ref{sec:dm}) and the search for
axion-like particles (ALPs; Sec.~\ref{sec:alp}). We follow with less
debated, yet interesting studies on the
search for primordial black holes (PBHs; Sec.~\ref{sec:pbh}), tau-neutrinos
(Sec.~\ref{sec:tau}), and magnetic
monopoles (Sec.~\ref{sec:monopoles}). Final remarks will close the chapter. Other topics in Fundamental Physics such as Lorentz Invariance tests are reported elsewhere in this book (see Chap.~XX).

We structured the contribution trying to provide a self-contained (yet minimal) theoretical framework, as well as a complete report of all \glspl{iact} published contribution to the topics under consideration. Our aim is to provide a reference review as well as take a photograph of the effort of the current generation of IACTs and the challenges taken, in preparation of the next-generation instrument to come, the Cherenkov Telescope Array.

 \begin{comment}

% \bigskip
%\mhc{This comes a bit out of context to me. In case, should be come earlier in intro, where wording "fundamental physics" is first mentioned} The term new physics is decisively vague, however, all new physics searches share some characteristics. First, one rarely possesses a complete and accurate theoretical mapping of
 what to expect. This often translate into an additional complexity in 
 developing optimal observational strategies or develop ad-hoc technological
 solutions.  Second, until a detection is achieved, it is hard to
 convince the community to be anywhere close to discovery. Third,
 often experimental constraints rely on some particular model to
 explain new physics, and therefore the validity of any derived
 constraint is limited and restricted to that peculiar theoretical
 framework. This situation is somewhat exacerbated for \glspl{iact}, that are
 multi-purpose instruments with limited available observation time and
 strong competition for time-allocation.  
 
 In this document, we tried to report a complete picture of the history of such fundamental physics searches with IACTs, trying to surface arguments that allow the reader to learn the complexity of carrying programs for such searches. We hope that this review could serve as platform to carry on future searches with even more strength and determination.
 \end{comment}

\section{Massive Particle Dark Matter}
\label{sec:dm}
\glsreset{wimp}

%\paragraph{A brief history of dark matter.}
For several decades now, the term `dark matter' is not only known from all physicists worldwide, but also probably by all high-school students and many citizens, and has entered to pop culture\footnote{\textit{“I don’t want to be human. I want to see gamma rays, I want to hear X-rays, and I want to smell dark matter\ldots"} Cylon Model Number One in the Fiction TV Serie "Battlestar Galactica".}. This is due to the innumerable studies written about the evidence for \gls{dm}, the so far inconclusive effort to identify its nature (but knowing today a lot about what \gls{dm} \textit{is not}), and the theoretical mapping of potential candidates.
%the study in search for it \mhc{ "the study" not clear to me, also repetition}, and the theoretical mapping of our knowledge (or ignorance) about. 
There are also countless reviews of this knowledge. It is therefore a challenging task to summarize even a fraction of what we learned so far about the \gls{dm}. For the purpose of this book, we will review the  search for massive \gls{dm} candidates, so-called \glspl{wimp}, performed with \glspl{iact}. Within this scope, we will limit ourselves to the basic information that will be needed for such review. In particular, we will limit ourselves to \textit{particle} \gls{dm} as opposed to theories of modified gravity~\citep[as, e.g., by][]{Milgron:1983} that would have no evident signatures for \glspl{iact}. Beyond that, we refer the interested reader to the literature: we recommend \citet{Bertone:2016nfn} for a profound, detailed account on the history of \gls{dm} in which also some myths are  dismounted, such as that affirming that the term itself was coined by F.~Zwicky in 1933. Other comprehensive reports are~\citet{Bertone:2005a,Feng:2008ya}, and many more. Fine lecture notes taken from schools are e.g.~\citet{Profumo2013,Hooper2017,Lisanti2017a}. Dedicated textbooks are, e.g., \citet{2006dcsu.book.....H,2014dmp..book.....S,2017ipdm.book.....P}. 

\subsection*{Basics of gamma-ray searches for WIMP DM}
\label{sec:dm_foundations}

To date, we still experience \gls{dm} only through its gravitational manifestation. We do not know its microscopic character, its composition and coupling to ordinary matter, nor if it displays at all interaction with the non-dark sector. However, we know that \gls{dm} does \textit{not} interact electromagnetically, and any scattering cross-section must be very weak. Therefore, having excluded neutrinos to make up the \gls{dm}, the  nature of this particle -- or several particles -- must go beyond the ensemble of particles of the \gls{sm}. Such strange particles would have
permeated the whole Universe from its beginning, and shaped the
Universe's evolution and structure formation due to their dominance. In fact, we have today a remarkably accurate knowledge about these structures and the distribution of \gls{dm} throughout the Cosmos. We measure that \gls{dm} is
clustered in {\it halos} at very diverse cosmic scales, hosting in turn astrophysical systems from the scale of dwarf galaxies
to that of galaxy clusters. We know that, despite being elusive, weakly interacting and very much stable, there is room for many theoretical models predicting interactions that could be observable with different classes of instruments, from X-ray to gamma-ray telescopes.

\paragraph{Tracking the particle \gls{dm} evolution.}
One of the most compelling evidence of \gls{dm} is its imprint  on the spatial anisotropy spectrum of the \gls{cmb}. This
diffuse radiation encodes the acoustic oscillations of the Universe matter and radiation fields at the epoch of matter recombination, when the Universe was about
380,000~years old. The power of a specific scale of anistropy sensibly depends on the composition of matter species. For example, a particle such as \gls{dm} exhibits a different anisotropy spectrum that a particle such as a quark that additionally couples to photons. Today's reference for the measurement of the  \gls{cmb} anisotropy is obtained with the \emph{Planck} satellite detector. By decoding the matter composition at the recombination epoch and extrapolating to current times, this measurement  yields values of ~\citep{Aghanim:2018eyx}:
\begin{equation} 
    \Omega_{\text{DM}}\,h^2=\frac{\rho_{\text{DM}}}{\rho_{\text{crit}}}\,h^2=0.1198\pm0.0012 \quad\rm{and}\quad\Omega_{SM}\,h^2=0.02233\pm0.00015,
    \label{eq:planck2018}
\end{equation}
where
$\rho_{\text{crit}}=3\,H^2_0/8\,\pi\,G=10^{-6}\;\rm{GeV\;cm}^{-3}$, $h=10^{-2}H_0$~km s$^{-1}$Mpc$^{-1}=0.6737\pm 0.0054$ the reduced Hubble parameter at redshift zero, and $G$ is the Gravitational Constant. That is,
roughly 26\% of today's Universe's energy density is in the form of \gls{dm} and
\gls{dm} makes up 84\% of the total mass of the Universe. 

Being
over-abundant with respect to ordinary matter and having left thermal
equilibrium before baryons, \gls{dm} would then drive and shape
galaxy formation: after their decoupling, baryons would fall into
already existing and growing \gls{dm} overdensities. These
overdensities would gather and merge, slowly giving rise to
the large-scale filamentary structure that is supported by current
observations. Being the observed matter distribution very clumpy, this puts strong requirements on any \gls{dm} candidate made of elementary particles: generically, particle \gls{dm} candidates are classified into \emph{hot, warm and cold} \gls{dm} (HDM, WDM, CDM), according to whether they were highly relativistic (hot) or non-relativistic (cold) at the time they decoupled in the early Universe by thermal freeze-out or any other production mechanism.\footnote{HDM must consist of very light, $\lesssim$~eV particles, while CDM can be at any mass scale, depending on the production mechanism \citep{Primack:2000iq}.}
%particle mass was in the eV, keV or GeV scale when it was thermally frozen out or produced by some other mechanism. This condition has in
%turn important consequences in structures formation: a hot \gls{dm}
%candidate would have been almost relativistic at freeze-out or production, therefore
Due to their high velocity, hot \gls{dm} particles inherit a large free-streaming length which would have brought to
extremely diluted \gls{dm} clustering, which is in contradiction with observations. This was clarified only
thanks to N-body cosmological simulations~\citep{Gershtein:1966,White:1984yj}, which for example ruled out cosmogenic neutrinos as \gls{dm} candidates. 
%Even thermally produced keV \gls{dm} candidates would have been a
%too large free-streaming length, failing at the Mpc scale, compared to
%bservation. 
In turn, CDM provides at large a better agreement with
observations. 

%However, there is still a tension between the WDM and CDM case, because in many situations the former one provides a better explanation that the latter \emph{at small (galactic) scales}. \mhc{Isn't the Missing Satellites dispute to some extent settled, see e.g. \citet{Read:2018gpi}?} 
  %any
%\gls{dm} candidate should have (or have had) a not too large free
%streaming length, and a mass at least of the order of the GeV. This
%scenario is the standard \gls{cdm}, in which \gls{dm} is not
%relativistic. 
%with an average similar to those in galaxies
%like our own, $\sim200$~km~s$^{-1}$. 
%The average density of \gls{dm} particles in the solar neighborhood is about $0.4$ GeV cm$^{-3}$ and the expected velocity distribution is close to a Maxwellian distribution.~\citep{Dodelson:2003ft}.

It is important to track the evolution of \gls{dm} in the early Universe in order to assess the observability now. The evolution of a comoving particle field density
$n_\chi$ in a Universe with Hubble parameter $H$ is governed by the law~\citep{Dodelson:2003ft}:
\begin{equation}
  \frac{dn_\chi}{dt}+3\,H\,n_\chi=-\langle\sigma v\rangle\left(n_\chi^2-n_{\chi,\text{eq}}^2\right)\,,
  \end{equation}
where $\langle\sigma v\rangle$ is the self-annihilation cross-section
between two  particles in the field averaged over their relative velocity, and the
density at equilibrium has reached $n_{\chi,\text{eq}}$ when particles
are in thermal and chemical equilibrium with the environment. The Hubble
parameter, expressing the expansion rate of the Universe, evolves with the effective fields composition as well as the
equilibrium temperature $T$. As a consequence, the number density of a given
particle species evolves with the effective Universe composition and
temperature as well. Specifically, when the temperature decreases to
values below a certain particle mass, $T\lesssim m_\chi$, the comoving density of
that particle starts decreasing exponentially as:\footnote{In \autoref{eq:density-decrease}, $g_{\chi}$ denotes the degrees of freedom of the particle species $\chi$. In the following, we presume  Majorana fermions (Spin 1/2 and $\bar{\chi}=\chi$), such that $g_{\chi}=2$.}
\begin{equation}
\label{eq:density-decrease}
n_{\chi,\text{eq}} = g_{\chi}\left(\frac{m_\chi\,T}{2\pi}\right)^{3/2}\;e^{-m_\chi/T}\,.
\end{equation}
This trend stops when the particle self-annihilation becomes
inefficient, that is, when the Hubble expansion rate gets larger than
the particle annihilation rate. This event is called \emph{freeze-out}
of the particle density which subsequently stays constant in a comoving
volume. For particle masses at the GeV scale, the freeze-out happens at a temperature $T_{\text{F}}$ roughly
20 to 30 times smaller than the particle mass,
$$
T_{\text{F}}\sim\frac{m_\chi}{25}\,,
$$
which means that these particles are at that moment already non-relativistic, i.e., cold. Considering such frozen-out particle species $\chi$ as \gls{dm} candidate, a curious fact emerges. If one
evolved the \gls{dm} density from freeze-out to today, considering plausible
values for the total number of degrees of freedom $g^*$ according to the known particles in the Universe, one obtains for $m_\chi\sim$ GeV:
\begin{equation}\label{eq:density-evolution}
\rho_\chi\,h^2\simeq
0.12\,\rho_{\text{crit}}\left(\frac{80}{g^*}\right)^{1/2}\left(\frac{m_\chi}{25\,T_{\text{F}}}\right)\left(\frac{2.2\times10^{-26}\rm{cm}^3\rm{s}^{-1}}{\langle\sigma
v \rangle}\right)\,,
\end{equation}
which tells the interesting fact that if that particle $\chi$ made up all the \gls{dm} content of the Universe found according to \autoref{eq:planck2018},
  $$
\Omega_{\text{DM}} = \Omega_\chi=\frac{\rho_\chi}{\rho_{\text{crit}}}=0.12\,,
  $$
then, a velocity-averaged annihilation cross-section at freeze-out of 
\begin{equation}
\langle\sigma v \rangle \approx 2.2\times10^{-26}\rm{cm}^3\rm{s}^{-1}
\label{freezeout-xsection}
\end{equation}
is required. Such value is remarkably close to the leading order scattering cross-section at the weak scale, $\langle\sigma v \rangle\sim \hbar^2/c\times\alpha_W^2/m_W^2$.  This means that in a \gls{dm} scenario where \gls{dm} is made up of a yet unknown particle species of mass at the weak scale,  a weak-scale freeze-out annihilation cross-section would be naturally required.  This fact is commonly dubbed as
\emph{\acrshort{wimp} miracle}. 

This picture is only one possible realization. There are theories of valid particle \gls{dm} alternatives that were not in thermal equilibrium in the Universe, or interacted in more complicated ways with the ambient fields exist. Yet, in addition to being one of the preferred and most studied DM particle model, the \gls{wimp} scenario provides a very generic framework, and several of the search methods developed for this scenario can be easily reformulated for other cases. As such, we limit ourselves to this case hereafter.

\paragraph{Dark Matter Particle Candidates.}

The appeal of such generic \gls{wimp} as \gls{dm} candidate was even increased with the advent of the theory of \gls{susy}. Originally proposed to solve open questions unrelated to \gls{dm} like the electroweak hierarchy problem or the unification of
interactions in particle physics, it was soon recognized that supersymmetric particles could also account for \gls{wimp} \gls{dm} \citep[e.g.,][]{Pagels:1981ke}.\footnote{See \citet{Bertone:2016nfn} for a full historic account.}
%
%Some factors may alter the simple approach used
%above and allow for different scenarios, 
%
 \gls{susy} depicts a
symmetry of SM particles to a new sector populated with
particles of similar quantum numbers but spin changed by $\pm1/2$, thus
swapping fermions with bosons, and by this expanding the
number of elementary particles. 
%The theory was appealing because it
%could solve with a simple quantum rule most issues within the Standard
%Model such as the electroweak hierarchy problem or the unification of
%interactions. 
However, why supersymmetric particles have never been
observed? Clearly a quantum rule must protect the
couplings between the SM and \gls{susy} sectors: the R-parity, which,
combining particle spin, baryon and lepton number, provides SM particle with parity number $R=1$ and \gls{susy} particles with
$R=-1$. The conservation of this symmetry not only shields the
SM from the \gls{susy} sector, but also automatically makes the lightest  \gls{susy}
particle stable. Here is the most studied \gls{wimp} candidate, the \gls{lsp}. There is not one single \gls{lsp} candidate. Yet the
one that emerged the most as the best candidate for \gls{wimp} \gls{dm} is the lightest
neutralino, mass eigenstate of the superpartners of the photon, the W,Z
particle and the two Higgs fields. In principle, if the R-parity was slightly broken, a \gls{lsp} could have a finite lifetime, as long as it is much larger than the age of Universe or in agreement with lower limits for specific models \citep[e.g.][for gravitino \gls{dm}]{Buchmuller:2007ui}. This also holds to any other mechanism to provide stability to a heavy particle. Therefore, it is worth to also watch out for \gls{dm} decays, as long instruments are able to probe them for such sufficiently large lifetimes. As we will see later, we indeed can probe lifetimes larger than these lower bounds.

\paragraph{Gravitational anomalies and the \gls{dm} distribution in astrophysical targets.}
The term \gls{dm} was probably first used by
H. Poincar\'e in 1906 albeit with the more
  classical meaning of `non emitting light' \citep[\emph{mati\`ere obscure} in][]{Poincare:1906}.\footnote{See \citet{Bertone:2016nfn}.} In fact,
at the turn of last century, astronomers started 
to investigate the distribution of visible objects in the sky and
their motion, trying to understand from first principles why matter was clustered in
specific form. 
Early studies by Lord Kelvin, using thermodynamics arguments, and
Poincar\'e, using the virial equilibrium hypothesis, applied to the \gls{mw}, started to show that
a significant fraction of mass was not attributable to stars but
needed to be explained by different  \emph{dark bodies} such as
meteors, or gas clouds. Evidence of missing mass started also to pile
up at a much larger scale, that of galaxy clusters. First
\citet{Hubble:1931} and then \citet{Zwicky:1933} in the first decades
of the last century found a large scatter in the velocities of galaxies
within galaxy clusters, up to 2,000~km/s, difficult to explain with only the visible matter in such clusters. Indeed, Zwicky used the virial
theorem on a sample of a thousand galaxies in the Coma cluster and
arrived at the conclusion that 500 times more mass was present in the cluster than the one inferred from the visible counterpart\footnote{This value was based on the estimate of the Hubble parameter $H_0$ at the time, that was overestimated by a factor of
  $\sim$10. With the present estimation the imbalance noted by Zwicky would
  reduce down to a factor 8.}. This is the so-called \textit{mass-to-light ratio}, that is, the ratio of the total estimated mass with respect to the mass expected from the luminous matter alone. Thus, this mass-to-light ratio is a measure of how much an astrophysical object is \gls{dm} dominated. Similar gravitational evidence
continued accumulating by the middle of the 20th century, but the ``missing mass'' problem
started to become more intriguing as soon as standard
explanations started to be ruled out: X-ray observation limited the
amount of hot gas in galaxies to be below 2\% \citep{Meekins:1971} and
stellar nucleosynthesis could not generate enough helium to
accommodate the observed abundance~\citep{Burbidge:1957}. The advent
of precise spectrographs in the 1970s  allowed to measure the velocity
of gas and stars within galaxies, which in turn allowed determining the mass of individual galaxies. In particular, in spiral galaxies, stars and gas move approximately on circular orbits around the galaxy's centre. Newtonian dynamics infers that an object placed at distance $r$ from a barycenter should display a circular velocity

\begin{equation}
    v_{\text{c}}(r)=\sqrt{\frac{G\,M(<r)}{r}}\,.
\end{equation}

That is, objects well outside the size $R$ of a mass distribution, where the total mass $M(<R)$ no longer grows, should have circular velocities following simply $v_{\text{c}}(r)\sim\sqrt{1/r}$. However, in the 1970s, Bosma, Rubin and colleagues
\citep{Bosma:1978,Rubin:1978,Rubin:1980zd} used samples of ten to dozens of spiral galaxies and found unambiguous evidence of  flat circular ($v_{\text{c}}(r)\sim k$) velocities of gas regions well
outside the size $R$ determined from the optical extension of the galaxies. These observations
allowed to robustly infer for the first time that the distribution of the missing
mass extends by several factors beyond the visible mass in individual galaxies, confirming previous findings by  \citet{Einasto:1974} and early findings for the M31 galaxy \citep{Babcock1939}. Spectrography is still the
best technique nowadays to infer the \gls{dm} content in galaxies, and
is being widely used with instruments such as the Dark Energy Spectroscopic Instrument \citep[DESI,][]{2016arXiv161100036D}.

Driven by insights from N-body cosmological simulations, the \gls{dm} density profiles of cosmic \gls{dm} overdensities can be commonly parametrized with the
Zhao-Hernquist expression~\citep{Hernquist:1990be,Zhao:1995cp}: 

\begin{equation}
\rho(r) = \frac{\rho_0}{\left(\frac{r}{r_{\text{s}}}\right)^\gamma\; \left[ 1+\left(\frac{r}{r_{\text{s}}}\right)^\alpha\right]^{(\beta-\gamma)/\alpha}},
\label{zhao}
\end{equation}

where $r$ is the distance from the \gls{dm} halo centre, $r_{s}$ and
$\rho_{0}$ are the so-called scale radius and characteristic \gls{dm}
density, respectively, and $\alpha,\beta,\gamma$ are free
parameters~\citep{Kuhlen:2012ft}. A specific case of this profile,
widely used in the literature, is the \gls{nfw}
parametrization~\citep{Navarro:1995iw} with fixed $\alpha=1$, $\beta=3$ and
$\gamma=1$. A profile very similar to the \gls{nfw}
parametrization, but describing the innermost halo slopes in simulations slightly better with one additional parameter $\alpha$ \citep{2004MNRAS.349.1039N,Springel:2008cc,Wang:2019ftp} is the so-called Einasto-profile,

\begin{equation}
\rho(r) = \rho_0 \exp\left\{\cfrac{-2}{\alpha} \left[\left(\frac{r}{r_{\text{s}}}\right)^\alpha -1 \right] \right\}.
\label{Einasto}    
\end{equation}

A different analytic description was found for some dwarf spiral galaxies to better fit the inferred \gls{dm} profile, namely, the so-called Burkert profile \citep{Burkert:1995yz},
\begin{equation}
\rho(r) =  \cfrac{\rho_0}{\left(1+\cfrac{r}{r_{\text{s}}}\right)\left(1+\cfrac{r^2}{r_{\text{s}}^2}\right)}\,,
\label{burkert}
\end{equation}
which shows a density plateau in the inner halo centre.

\paragraph{Gamma-ray signatures of \gls{dm}.}
Besides trying to produce \gls{dm} in collider events and to observe their annihilation or decay into more known particles, \gls{dm} can be observed through nuclear or atomic recoils in noble gas or semiconductors. This method is usually called ``direct detection'' technique, and there is ample literature about it, e.g., \citet{Schumann:2019eaa}. 
A third avenue is to explore \gls{dm} signatures \textit{``indirectly''} by observing \gls{vhe} gamma rays generated in relic annihilations or decays of the \gls{dm} throughout the Cosmos. 

\begin{figure}[h!t]
\centering
\includegraphics[width=0.5\linewidth]{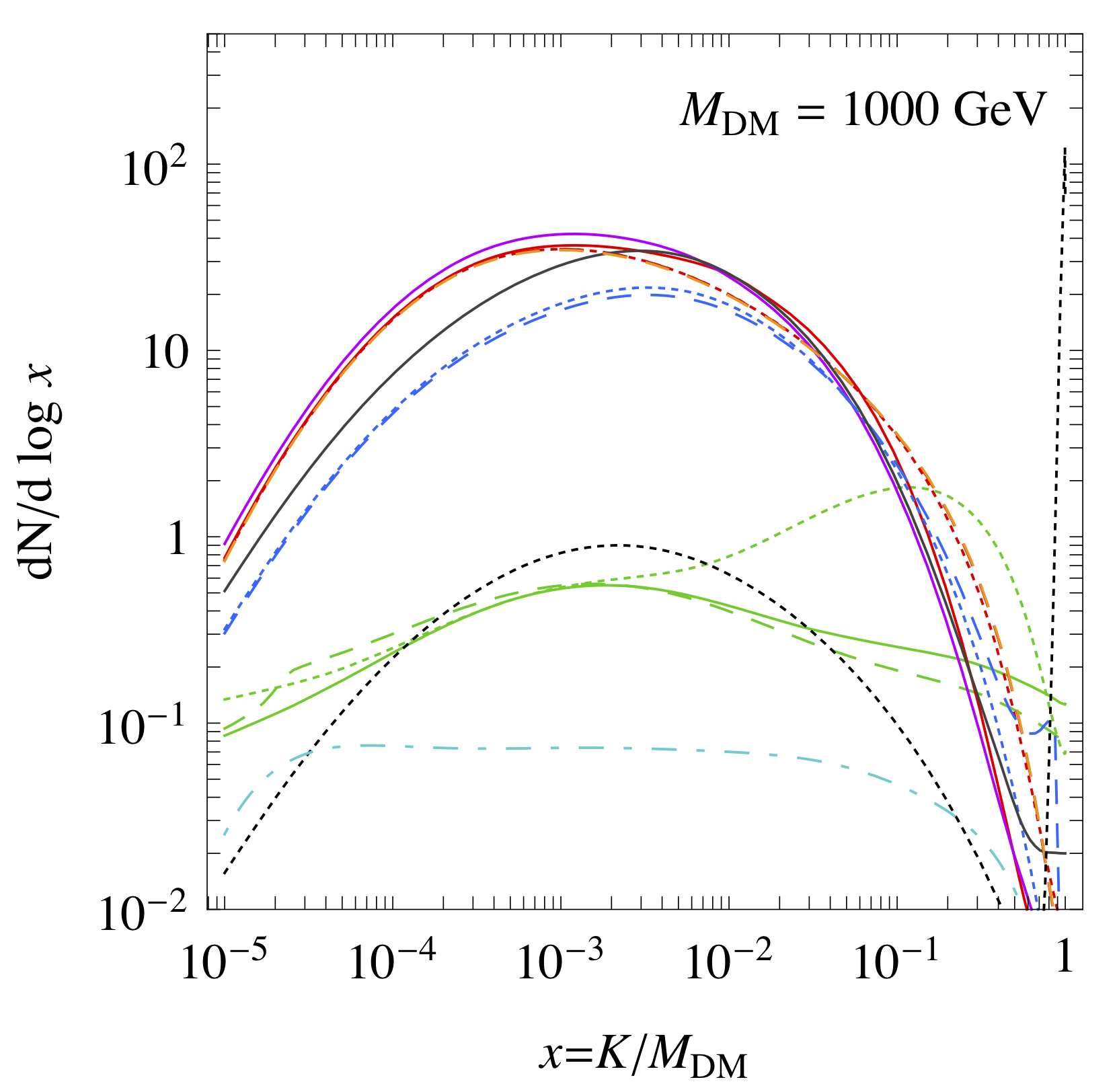}
\includegraphics[width=0.285\linewidth]{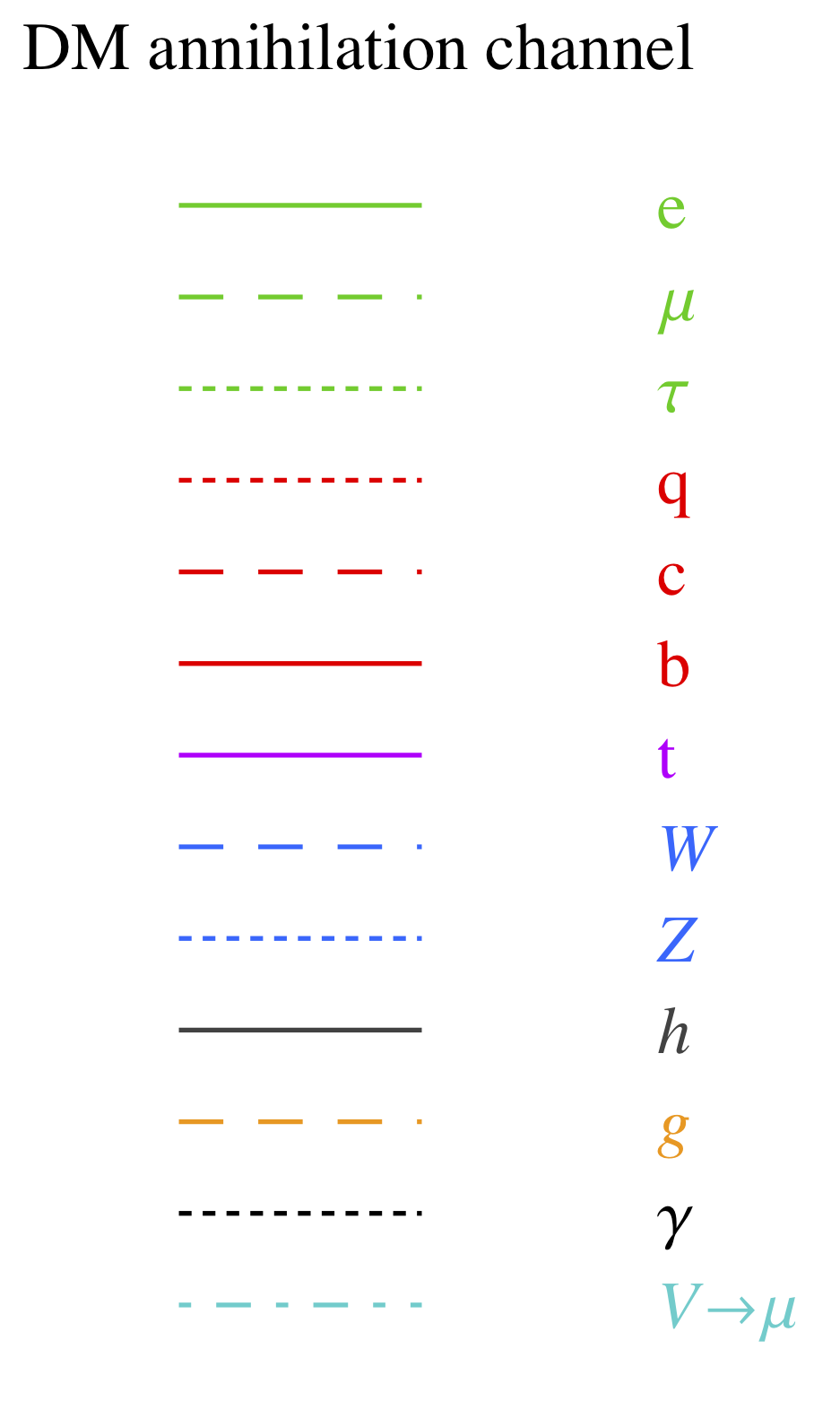}
\caption{\label{fig:spectra} Primary \gls{dm} gamma-ray spectra for various annihilation models. For \gls{dm} decays, the same spectra are expected, yet with the energy scale shifted to half these values, as the energy is equally distributed in two decay particles. Extracted from Fig. 3 of \citet{Cirelli:2010xx} with the permission of the authors.}
\end{figure}

High-energy photons
are found primarily as a result of neutral pion decay, where the pions are created after fragmentation of quarks expected to be produced in \gls{dm} annihilations or decays. Such origin of gamma radiation results in a continuous spectrum
(Fig.~\ref{fig:spectra}). As such, the latter may be indistinguishable
from other astrophysical processes in which pion decays are found.  However,
if the \gls{dm} is cold, i.e. non-relativistic, an abrupt spectral cutoff is
expected at the \gls{dm} mass for annihilating \gls{dm}, or at its half for
decaying \gls{dm}. In more sophisticated scenarios, enhanced or even line-like emission may occur at the termination energy set by the particle mass \citep{Bergstrom:1994mg,Bringmann:2007nk}. Because of the peculiarity of searches for line-like signals with \glspl{iact}, these will be discussed separately in the following. In any case, all these cut-off features make the resulting spectra from \gls{dm} annihilation or decay very distinct from those of other astrophysical phenomena. Another consequence arises from the energy scale of the annihilation products set by the mass of the cold \gls{dm}
particle: should the latter be of the order of the GeV to TeV,
gamma rays would be generated at these same energies, and
ground-based \gls{vhe} gamma-ray detectors like \glspl{iact} would be perfectly suited for such
indirect searches for \gls{dm}. Indeed, at \gls{wimp} masses above several TeV, \glspl{iact} currently are the most sensitive instruments to search for such particles. 
%Indeed, theories predict plausible particle candidates in this regime of very heavy \gls{wimp} masses [CITE]. 
Finally, there are several other avenues for indirect \gls{dm} searches, by looking for other of \gls{dm} annihilation or decay products, namely charged cosmic rays or neutrinos. While these searches are not in the scope of this review, on page~\pageref{pageref:antiparticles} we will mention the possibility to use \glspl{iact} to study the cosmic-ray electron and positron flux to search for \gls{dm} signatures.

Although \gls{dm} possesses a nearly unique or
very characteristic gamma-ray spectrum, the amount of
annihilation or decay expected events -- and therefore the
detectability at Earth -- ultimately depends on the integrated density, squared in the annihilating case, of \gls{dm} along the line of
sight. Because of that, overdense \gls{dm} regions are clearly preferred,
such as those at the centre of galaxies or clusters of
galaxies. The computation of the double differential gamma-ray \gls{dm} flux
at Earth can therefore be written as:\footnote{The annihilation formula is valid for self-conjugate \gls{dm} particles ($\bar{\chi}=\chi$), otherwise there is an additional factor $1/2$.} 

\begin{eqnarray}\label{eq:dm_flux}
  \nonumber
  {\rm Annihilation:} && \frac{d\Phi}{dE \, d\Omega}= \frac{1}{4\pi}\, \frac{\sv}{ 2\mdm^2}\; \frac{dN_\gamma}{dE} \frac{dJ_{\text{ann}}}{d\Omega},\;\;{\rm with}\\
&&\frac{dJ_{\text{ann}}}{d\Omega} =   \int_\mathrm{l.o.s.}  \,dl\,
  \rho^2(l,\Omega).\\
%\end{eqnarray}%
%
%\begin{eqnarray}\label{eq:diffflux}
\nonumber
{\rm Decay:} && \frac{d\Phi}{dE \, d\Omega}= \frac{1}{4\pi}\, \frac{1}{\taudm\; \mdm}\; \frac{dN_\gamma}{dE} \frac{dJ_{\text{dec}}}{d\Omega},\;\;{\rm with}\\
\nonumber
&&\frac{dJ_{\text{dec}}}{d\Omega} =   \int_\mathrm{l.o.s.}  \,dl\,
  \rho(l,\Omega).
\end{eqnarray}

Here, $\mdm$ is the \gls{dm} mass, $\taudm$ the \gls{dm} lifetime, $\sv$ is the
velocity-averaged annihilation cross-section; $dN_{\gamma}/dE$ is the
average number of photons per unity energy per reaction, 
$J_{\text{dec}}$ is called the \textit{astrophysical
  factor}~\citep{Bergstrom:1998a} for decaying \gls{dm}, and $J_{\text{ann}}$ is
the astrophysical factor for the annihilating case. Note that, in
order to derive the expected \gls{dm}-induced flux from a given target, one needs to
integrate along the line of sight ($l.o.s.$) and the aperture
$\Delta\Omega$. Note also that Eq.~\ref{eq:dm_flux} is 
factorized so that one factor entails information about the
microscopic \gls{dm} properties (mass, cross-section or lifetime, average
photon emission) while the astrophysical factor encloses separately all the
information on the \gls{dm} spatial distribution.\footnote{Strictly speaking, the factorization only holds for integration over a small redshift span, $\Delta z / z \ll 1$.} In most cases, for 
collisionless \gls{dm}, the latter is almost independent on the specific
microscopic nature of \gls{dm} and thus the astrophysical factor is very
general.\footnote{However, the mass of the
  smallest predicted C\gls{dm} halos is ultimately set by the properties of
  the \gls{dm} particle itself; this fact may have important implications in
  computations of the \gls{dm}-induced gamma-ray flux involving halo
  substructure, as those that will be commented later on in this 
  section.} Yet, other cases are also possible, such as
self-interacting \gls{dm}, in which the exact \gls{dm} density profile will depend
on the \gls{dm} particle's characteristics \citep{Kahlhoefer:2019oyt}. In all cases, the expected gamma-ray spectra should look the same
across different astrophysical targets. Thus, observing two
independent targets with identical spectra would constitute a strong
case in favour of the \gls{dm} interpretation.

The density integral in Eq.~\ref{eq:dm_flux} is very sensitive to the density scaling in the halo centre, in particular for integrating over the square-density in case of annihilation.
Therefore,
if interested in maximizing the \gls{dm}-induced gamma-ray signals,
the telescope should be pointed to
those astrophysical environments exhibiting the highest \gls{dm}
concentrations at their centres. Also, since fluxes decrease as the inverse
of the distance squared, the closer the targets, the better.\footnote{The inverse square-distance scaling is contained, but hidden in \autoref{eq:dm_flux} by the factor $1/l^2$ cancelling out the $l^2$ in the volume element.}  Fortunately, a plethora of  objects are such viable candidates for \gls{dm}
searches with \glspl{iact}: the closest and probably brightest in terms of
\gls{dm}-induced flux is the centre of the \gls{mw}. Further out, dwarf spheroidal galaxies (dSphs) are
optimal targets as well, especially those nearby and with large
mass-to-light ratios. Much further out, galaxy clusters host a huge \gls{dm}
content, which could provide a significant emission at Earth as
well. Less favored or more debated objects may also be interesting,
such as Galactic globular clusters: their origin is still a matter of
debate and there is the possibility that they could retain significant
\gls{dm} density spikes at their very centres \citep{Brown:2018pwq}. \gls{iact} \gls{dm}
searches in all these objects will be presented in the following.

%%%%%%%%%%%%%%%%%%%%
%
%
%
%
%
%
%
\subsection*{Review of WIMP DM searches with IACTs}
The search for \gls{wimp} \gls{dm} with \glspl{iact} started along with the birth of this class of instruments, and strongly evolved in expectations, methods and dedications over the years. In the following, we  gather all these observations as a summary of what it has been done so far and in order to guide our experience in the future.

\bigskip

Classical targets were first considered: the
central region of the \gls{mw}, the \gls{mw}  
\glspl{dsph} and galaxy clusters. The interest expanded to less
established targets such as \glspl{imbh} and dark satellites. In
general, searches have been conducted looking for \gls{dm}-induced continuum
emission and line-like emission, and both for decaying and annihilating
\gls{dm}. Not only gamma-ray signatures, but also antiparticle signatures which may originate from \gls{dm} annihilation or decay have been
searched for, demonstrating the capability of \glspl{iact} as
antiparticle detectors from the ground. It is interesting to see how
early efforts invested relatively low observation times of the
order of few hours, while later observation campaigns show more than a hundred
hours devoted to single targets. Considering the total available
observation time in a year of a single \gls{iact} experiment is about $1,000$~h, it is
obvious that a large time investment has been made, with more than $1,500$~h of
observation time devoted exclusively to \glspl{dsph} as of today. 

In Table~\ref{tab:targets} we have collected, to the best of our knowledge, all the \gls{dm} search campaigns carried out by \glspl{iact}. The table, an update of the one in \citet{Doro:2014pga}, now including results up to 2021, gathers the
numerous targets scrutinized as of today. The content is
organized by target class and year of observation, and also the
duration of observation is reported. In the following, we decide to discuss in detail all these observations according to target class.

\begin{center}
\begin{longtable}{llclcp{4.2cm}}
\captionsetup{width=.95\linewidth}
\caption{\label{tab:targets} Summary of the campaigns carried out by \glspl{iact} to indirectly search for \gls{dm}
  with gamma rays. We report observations from the Whipple, \glsentryshort{hess}, \glsentryshort{magic}, and
  \glsentryshort{veritas} experiments, sorted by astrophysical target class. First column is the name of the target. Second and third columns refer
  to the year and duration (in hours) of observations, respectively. The fifth column informs about the type of limit, either annihilation or decay. Unconfirmed \acrshort{dsph} candidates are marked with an asterisk. \glsentryshort{hess} data including the \glsentryshort{hess} Phase II telescope are marked with a ${\dagger}$, \glsentryshort{magic} data from monoscopic observations with \glsentryshort{magic}-I are marked with a $\ddag$. Times in parentheses denote data which are also a subset of other analyses. We also include yet unpublished data  presented at conferences.}  \\

\hline
\rowcolor[gray]{.8} 
{\bf Target} & {\bf Year} & {\bf Time$\,$[h]} & {\bf \glsentryshort{iact}} & {\bf Limit} &{\bf Ref.} \\
\hline 
\endfirsthead

\multicolumn{6}{l}{\tablename\ \thetable{} -- continued from previous page} \\
\hline
\rowcolor[gray]{.8} 
{\bf Target} & {\bf Year} & {\bf Time$\,$[h]} & {\bf \glsentryshort{iact}} & {\bf Limit} &{\bf Ref.} \\
\hline 
\endhead

\hline
\multicolumn{6}{r}{{\tablename\ \thetable{} -- Continued on next page}} \\
\endfoot

\hline \hline
\endlastfoot

\multicolumn{6}{>{\columncolor[gray]{.92}}c}{{\bf The Milky Way
    central region \& halo}} \\
\glsentryshort{mw} Centre      & 2004        & (48.7) & \glsentryshort{hess} & Ann. & \citet{Aharonian:2006:gc} \\
\glsentryshort{mw} Inner Halo & $2004-2008$ & (112)  & \glsentryshort{hess} & Ann. & \citet{Abramowski:2011:gc}\\ % 112h most exact number given
               & $2010$      & 9.1    &            & Ann. & \citet{HESS:2015cda}\\
               & $2004-2014$ & 254    &            & Ann. & \citet{Abdallah:2016:halo}\\ % 254h most exact number given
               & $2014-2020$ & 546    & \glsentryshort{hess}$^{\dagger}$   & Ann. & \citet{Montanari:2021yic}\\ 
\glsentryshort{mw} Outer Halo  & $2018$ & 10 & \glsentryshort{magic}  & Decay  & \citet{Ninci:2019njk}\\ % 10h most exact number given
\multicolumn{6}{>{\columncolor[gray]{.92}}c}{{\bf Dwarf Satellite Galaxies}} \\
Draco & 2003 & 7.4 & Whipple & Ann. & \citet{Wood:2008hx}\\
      & 2007 & 7.8&  \glsentryshort{magic}$^{\ddag}$ & Ann. & \citet{Albert:2008:dsph}\\
      & 2007 & (18.4) & \glsentryshort{veritas} & Ann. & \citet{Acciari:2010:dwarf}\\
      & $2007-2013$ & (49.8) &   & Ann. & \citet{Archambault:2017wyh}\\
      & $2007-2018$     & 114 & & -- & \citet{KelleyHoskins:2018tevpa}\\
      & $2018$     & 52.6 & \glsentryshort{magic} & Ann. & \citet{MAGIC:2021mog} \\
Ursa Minor & 2003 & 7.9 & Whipple & Ann. & \citet{Wood:2008hx}\\
     & 2007 & (18.9) & \glsentryshort{veritas} & Ann. & \citet{Acciari:2010:dwarf}\\
     & $2007-2013$ & (60.4) &   & Ann. & \citet{Archambault:2017wyh}\\
     & $2007-2018$     & 162 & & -- & \citet{KelleyHoskins:2018tevpa}\\
Sagittarius & 2006 & (11.0) & \glsentryshort{hess} & Ann. & \citet{Aharonian:2007:dwarf}\\
            & $2006-2012$ & 90   &     & Ann. & \citet{Abramowski:2014:dwarf}\\ %90 hours most exact number given
            & $2006-2012$ & (85.5) &     & Ann. & \citet{Abdalla:2018mve}\\
Canis Major & 2006 & 9.6 & \glsentryshort{hess}  & Ann. & \citet{Aharonian:2009:canis}\\
Willman 1 & $2007-2008$ & 13.7 & \glsentryshort{veritas} & Ann. & \citet{Acciari:2010:dwarf}\\
          &             & (13.6)     &          & Ann. & \citet{Archambault:2017wyh}\\
         & 2008 & 15.5 & \glsentryshort{magic}$^{\ddag}$  & Ann. & \citet{Aliu:2009:william}\\
Sculptor & 2008        & (11.8) & \glsentryshort{hess} & Ann. & \citet{Abramowski:2010:carina}\\
         &             &        &            & Ann. & \citet{Abdalla:2018mve}\\
         & $2008-2009$ & 12.5 &           & Ann. & \citet{Abramowski:2014:dwarf}\\
Carina   & $2008-2009$ & (14.8) & \glsentryshort{hess} & Ann. & \citet{Abramowski:2010:carina}\\
         & $2008-2009$ & (12.7) &             & Ann. & \citet{Abramowski:2014:dwarf}\\
         & $2008-2010$ & 22.9 &             & Ann. & \citet{Abdalla:2018mve}\\
Segue 1  & $2008-2009$ & 29.4   & \glsentryshort{magic}$^{\ddag}$ & Ann. & \citet{Aleksic:2011:segue}\\
         & $2010-2011$ & (47.8) & \glsentryshort{veritas} & A.+D. & \citet{Aliu:2012:segue}\\
         & $2010-2013$ & (92.0) &         & Ann. & \citet{Archambault:2017wyh}\\
         & $2010-2013$ & 157.9  & \glsentryshort{magic} & A.+D. & \citet{Aleksic:2013:segue}\\
         &             &        &       & Ann. & \citet{Ahnen:2016:dsph}\\
         & $2010-2018$     & 184 & \glsentryshort{veritas} & -- & \citet{KelleyHoskins:2018tevpa}\\
Bo\"otes 1 & 2009 & 14.3 & \glsentryshort{veritas} & Ann. & \citet{Acciari:2010:dwarf}\\
         &      & (14.0)     &         & Ann. & \citet{Archambault:2017wyh}\\
%        &      & 16        &         & Ann. & \citet{KelleyHoskins:2018tevpa}\\
Coma Berenices & $2010-2013$        & (8.6) & \glsentryshort{hess} & Ann. & \citet{Abramowski:2014:dwarf}\\
               & $2010-2013$        & 10.9 &            & Ann. & \citet{Abdalla:2018mve}\\
      & $<2018$     & 37 & \glsentryshort{veritas} & -- & \citet{KelleyHoskins:2018tevpa}\\
      & $2018$     & 50.2 & \glsentryshort{magic} & Ann. & \citet{MAGIC:2021mog} \\
Fornax   & $2010$        & 6.0 & \glsentryshort{hess} & Ann. & \citet{Abramowski:2014:dwarf} \\
         &      &  &  & Ann. & \citet{Abdalla:2018mve}\\
Ursa Major II & $2014-2016$ &  94.8 & \glsentryshort{magic} & Ann. & \citet{Ahnen:2017pqx}\\
         & $<2018$     & 181 & \glsentryshort{veritas} & -- & \citet{KelleyHoskins:2018tevpa}\\
Triangulum II* & $2014-2016$ &  62.4 & \glsentryshort{magic} & Ann. & \citet{Acciari:2020pno}\\
         & $<2018$     & 15 & \glsentryshort{veritas} & -- & \citet{KelleyHoskins:2018tevpa}\\
Segue II  & $<2018$     & 19 & \glsentryshort{veritas} & -- & \citet{KelleyHoskins:2018tevpa}\\
Canes Ven I  & $<2018$     & 14 & \glsentryshort{veritas} & -- & \citet{KelleyHoskins:2018tevpa}\\
Canes Ven  II  & $<2018$     & 14 & \glsentryshort{veritas} & -- & \citet{KelleyHoskins:2018tevpa}\\
Hercules  & $<2018$     & 13 & \glsentryshort{veritas} & -- & \citet{KelleyHoskins:2018tevpa}\\
Sextans  & $<2018$     & 13 & \glsentryshort{veritas} & -- & \citet{KelleyHoskins:2018tevpa}\\
Draco II  & $<2018$     & 10 & \glsentryshort{veritas} & -- & \citet{KelleyHoskins:2018tevpa}\\
Leo I  & $<2018$     & 7 & \glsentryshort{veritas} & -- & \citet{KelleyHoskins:2018tevpa}\\
Leo  II  & $<2018$     & 16 & \glsentryshort{veritas} & -- & \citet{KelleyHoskins:2018tevpa}\\
Leo IV  & $<2018$     & 3 & \glsentryshort{veritas} & -- & \citet{KelleyHoskins:2018tevpa}\\
Leo V  & $<2018$     & 3 & \glsentryshort{veritas} & -- & \citet{KelleyHoskins:2018tevpa}\\
Reticulum II  & $2017-2018$     & 18.3 & \glsentryshort{hess}$^{\dagger}$ & Ann. & \citet{Abdallah:2020sas}\\
Tucana II  & $2017-2018$     & 16.4 & \glsentryshort{hess}$^{\dagger}$ & Ann. & \citet{Abdallah:2020sas}\\
Tucana III* & $2017-2018$     & 23.6 & \glsentryshort{hess}$^{\dagger}$ & Ann. & \citet{Abdallah:2020sas}\\
Tucana IV*  & $2017-2018$     & 12.4 & \glsentryshort{hess}$^{\dagger}$ & Ann. & \citet{Abdallah:2020sas}\\
Grus II* & $2018$     & 11.3 & \glsentryshort{hess}$^{\dagger}$ & Ann. & \citet{Abdallah:2020sas}\\
\multicolumn{6}{>{\columncolor[gray]{.92}}c}{{\bf Dark satellites}} \\
1FGL J2347.3+0710 & 2010 & 8.3 & \glsentryshort{magic}   & -- & \citet{2011arXiv1109.5935N} \\
1FGL J0338.8+1313 & 2010-2011 & 10.7 & \glsentryshort{magic}   & -- & \citet{2011arXiv1109.5935N} \\
2FGL J0545.6+6018 & 2013-2015 & 8.5 & \glsentryshort{veritas}   & Ann. & \citet{2015arXiv150900085N} \\
2FGL J1115.0-0701 & 2013-2015 & 13.8 & \glsentryshort{veritas}   & Ann.& \citet{2015arXiv150900085N} \\
3FHL J0929.2-4110 & 2018-2019 & 7.8 & \glsentryshort{hess}$^{\dagger}$   & Ann. & \citet{1866411} \\
3FHL J1915.2-1323 & $2018-2019$ & 3.0 & \glsentryshort{hess}$^{\dagger}$   & Ann. & \citet{1866411} \\
3FHL J2030.2-5037 & $2018-2019$ & 8.8 & \glsentryshort{hess}$^{\dagger}$   & Ann. & \citet{1866411} \\
3FHL J2104.5+2117 & $2018-2019$ & 5.5 & \glsentryshort{hess}$^{\dagger}$   & Ann. & \citet{1866411} \\
\multicolumn{6}{>{\columncolor[gray]{.92}}c}{{\bf Intermediate Mass Black Holes}} \\
Galactic Plane Survey & $2004-2007$ & 400 & \glsentryshort{hess} & Ann. & \citet{Aharonian:2008:dm}\\ % 400h most exact number given
& $2005-2006$ & 25 & \glsentryshort{magic}$^{\ddag}$ & Ann. & \citet{Doro:2007icrc}\\
\multicolumn{6}{>{\columncolor[gray]{.92}}c}{{\bf Globular Clusters}} \\
M15 & 2002 & 0.2 & Whipple & Ann. & \citet{Wood:2008hx} \\
    & $2006-2007$ & 15.2 & \glsentryshort{hess} & Ann. & \citet{Abramowski:2011:ngc6388} \\
%    & $2014-2017$ & 115 & \glsentryshort{magic} & -- & In prep. \\
NGC 6388 & $2008-2009$ & 27.2& \glsentryshort{hess} & Ann. & \citet{Abramowski:2011:ngc6388} \\
\multicolumn{6}{>{\columncolor[gray]{.92}}c}{{\bf Other galaxies}} \\
M33 & $2002-2004$ & 7.9 & Whipple & Ann. & \citet{Wood:2008hx} \\
M32 & 2004 & 6.9 & Whipple & Ann. & \citet{Wood:2008hx} \\
WLM & $2018$     & 18.2 & \glsentryshort{hess}$^{\dagger}$ & Ann. & \citet{Abdallah:2021kzs}\\
\multicolumn{6}{>{\columncolor[gray]{.92}}c}{{\bf Galaxy Clusters}} \\
Abell 2029 & $2003-2004$ & 6.1     & Whipple & -- & \citet{Perkins:2006fa}\\
Perseus (Abell 426) & $2004-2005$    & 13.5 & Whipple & -- & \citet{Perkins:2006fa}\\
        & $2008$         & 24.4 & \glsentryshort{magic}$^{\ddag}$   & Ann. & \citet{Aleksic:2010:ngc1275}\\
        & $2009-2017$    & 202.2  & \glsentryshort{magic}   & Decay & \citet{Acciari:2018:perseus}\\
Fornax (Abell S0373)  & $2005$         & 14.5 & \glsentryshort{hess}    & Ann. & \citet{Abramowski:2012:fornax}\\ 
Coma (Abell 1656)    & $2008$         & 18.6 & \glsentryshort{veritas} & Ann. & \citet{2012ApJ...757..123A} \\
\multicolumn{6}{>{\columncolor[gray]{.92}}c}{{\bf Line searches}} \\
\glsentryshort{mw} Inner Halo & $2004-2008$ & (112) & \glsentryshort{hess} & Ann. & \citet{Abramowski:2013:lines}\\
     & $2014$      & 15.2  & \glsentryshort{hess}$^{\dagger}$ & Ann. & \citet{Abdalla:2016olq}\\
     & $2004-2014$ & (254)  & \glsentryshort{hess} & Ann. & \citet{Abdallah:2018:lines}\\
     & $2013-2019$ & 204  & \glsentryshort{magic} & Ann. & \citet{Inada2021}\\
 Segue 1 \glsentryshort{dsph}    & $2010-2013$ & (157.9) & \glsentryshort{magic} & Ann. & \citet{Aleksic:2013:segue}\\
Five \glsentryshort{dsph} galaxies   & $2006-2012$ & (137.1) & \glsentryshort{hess} & Ann. & \citet{Abdalla:2018mve}\\
Five \glsentryshort{dsph} galaxies   & $2007-2013$ & (229.8) & \glsentryshort{veritas} & Ann. & \citet{Archambault:2017wyh}\\
Five \glsentryshort{dsph} galaxies   & $2017-2018$ & 82.0 & \glsentryshort{hess}$^{\dagger}$ & Ann. & \citet{Abdallah:2020sas}\\
%Four \glsentryshort{dsph} galaxies   & $2010-2019$ & (354.4) & \glsentryshort{magic} & Ann. & \citet{MAGIC:2021mog}\\
WLM & $2018$     & (18.2) & \glsentryshort{hess}$^{\dagger}$ & Ann. & \citet{Abdallah:2021kzs}\\
\multicolumn{6}{>{\columncolor[gray]{.92}}c}{{\bf Charged particles}} \\
All-electron & $2004-2007$ & 239 & \glsentryshort{hess} & -- & \citet{Aharonian:2008:electron,Aharonian:2009:electron}\\  % 239h most exact number given
             & $2009-2012$ & 296 & \glsentryshort{veritas} &-- & \citet{Archer:2018chh}\\ % 296h most exact number given
             & $2009-2010$ & 14 & \glsentryshort{magic} & -- & \citet{BorlaTridon:2011dk}\\
Moon shadow & $2010-2011$ & 20 & \glsentryshort{magic} & -- & \citet{Colin:2011wc}\\
            & $2014$ & 1.2 & \glsentryshort{veritas} & -- & \citet{Bird:2016dcv}\\
\end{longtable}
\end{center}

\paragraph{Observations of the \gls{mw} Centre and halo.}

\begin{figure}[h!t]
\centering
 \includegraphics[width=0.47\linewidth]{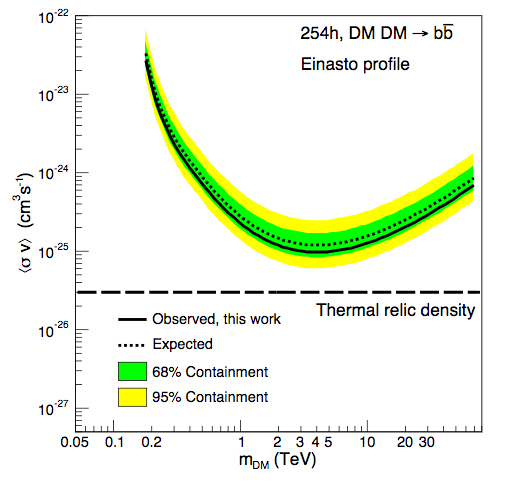} 
  \includegraphics[width=0.47\linewidth]{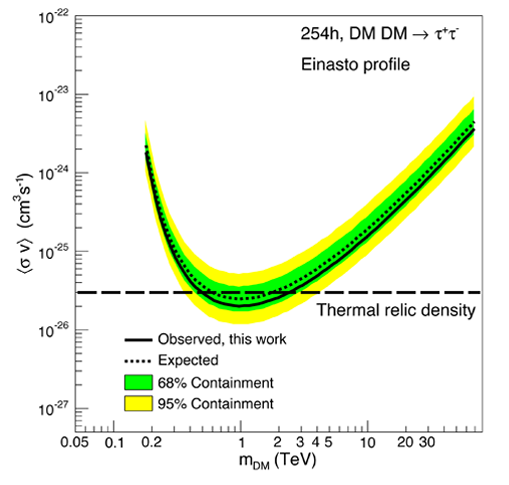} 
  \caption{Constraints  on  the  velocity-weighted  annihilation  cross  section derived  from  10  years  of  \gls{gc}  observations  by \gls{hess} for the $b\bar{b}$ (left panel) and into $\tau^{+}\tau^{-}$ (right panel). The  observed  upper  limit is  plotted as  a
solid black line.  Expected limits are derived from blank-field observations at high Galactic latitudes.  The mean expected limit (black dashed line) is plotted together with the 65\% (dark gray) and 95\% (light gray) confidence bands.  The natural scale for the annihilation cross-section is given as long-dashed black line. Figures from~\citet{Abdallah:2016:halo}, courtesy of the \gls{hess} collaboration.}.
   \label{fig:hess_gc}
\end{figure}

The \glsfirst{gc} is a prime target for \gls{dm} searches with
\glspl{iact}. It is best observed from the Southern Hemisphere, where the \gls{hess} telescopes are located. This guarantees the lowest energy threshold and a longer observability. In 2006, \gls{hess} published the 
observation of a signal from the \gls{gc} above 160~GeV and its interpretation in terms of \gls{dm} after an observation campaign of about
50~h~\citep{Aharonian:2006:gc}. In that work, \gls{dm} density profiles as obtained from N-body cosmological simulations of \gls{mw}-size halos were adopted, namely the \gls{nfw} (\autoref{zhao} with $\alpha,\beta,\gamma=1,3,1$) and Einasto (\autoref{Einasto}). Large differences between the two \gls{dm} profile parametrizations occur only at the very centre of the halo, where the \gls{nfw} profile is more strongly
peaked, while already at distances from the GC larger than $\sim$10 pc the density descriptions become very similar when adopting the same profile normalization at the Solar position. This was the reason for \gls{hess} to focus on an analysis region at a projected Galactocentric distance of $45-150$ pc that corresponds to an angular distance of $0.3-1.0^\circ$ from the \gls{gc}. Furthermore, the contamination from known and expected gamma-ray astrophysical sources at this distance from the \gls{gc} is largely reduced with respect to the very central region. Limits on the annihilation cross-section were first presented in \citet{Abramowski:2011:gc} using 112~h of data and remained for long the most constraining upper limits on \gls{dm} annihilations from \glspl{iact}. More recently,~\citet{Abdallah:2016:halo} gathered 10 years of \gls{hess} datataking, involving a total of 254~h good quality data, which resulted in the \gls{dm} exclusion limits shown in Fig.~\ref{fig:hess_gc} for the $b\bar{b}$ and $\tau^{+}\tau^{-}$ pure annihilation cases. The limits were obtained after an elaborated selection of both the signal and background regions in the proximity of the \gls{gc}, that removes the Galactic plane and known TeV emitters. It can be seen that, while the limits for $b\bar{b}$ approach the thermal relic cross-section value at few TeV, the ones for $\tau^{+}\tau^{-}$ actually already skim the canonical parameter space. These limits represent the strongest ones obtained by \glspl{iact} at present, recently even extended by the announcement of the publication of a data set comprising almost 550 h of data \citep{Montanari:2021yic}. Also the \gls{magic} telescopes \citep{Albert:2006:gc} and the \gls{veritas}  \citep{Beilicke:2012rx} observed the \gls{gc} along the years, yet with more modest performance compared to \gls{hess} due to their location in the Northern Hemisphere as opposed to \gls{hess}. Because of that, neither \gls{magic} nor \gls{veritas} were able to draw conclusions on \gls{dm} physics.

It must be noted that all mentioned \gls{dm} limits are only valid as long as a cuspy profile of the distribution of \gls{dm} toward the \gls{mw} centre is assumed, as it is the case for both \gls{nfw} and Einasto profiles. However, it is still equally plausible that the actual \gls{dm} profile toward the \gls{gc} is  more cored and closer to the so-called Burkert profile, \autoref{burkert}. This would strongly reduce the \gls{dm} sensitivity, not only because of a much smaller  $J-$factor, but also because most \gls{dm} analyses of \gls{iact} data rely on the existence of a \gls{dm} density contrast between a signal and background region. Such cored profiles can appear e.g. when baryonic feedback occurs in connection with strong stellar activity that sweeps outward the matter and dilutes the \gls{dm} concentration in the innermost regions of the Galaxy \citep{DiCintio:2013qxa,Tollet:2015gqa,Read:2018fxs,Lazar:2020pjs}. Also, self-interacting \gls{dm} is expected to form pronounced halo cores \citep [e.g.,][]{Rocha:2012jg,Tulin:2017ara}. Due to the crowded line of sight towards the \gls{gc} and the complex dynamics in the Galactic bulge, an empirical, direct determination of the inner \gls{gc} \gls{dm} density profile from e.g. kinematics data of different tracers in the area is very difficult. Therefore, the inner profile is yet poorly constrained \citep{Benito:2019ngh,Benito:2020lgu}. For these reasons, despite being far more constraining than others, \gls{gc} limits need to be treated with special caution. In 2015, the \gls{hess} collaboration published a study constraining \gls{dm} annihilation from a \gls{dm} core around the \gls{gc}, based on 9~h of dedicated observations with a background subtraction region $\gtrsim 5^\circ$ away from the \gls{gc} \citep{HESS:2015cda}. A similar methodology addressing \gls{dm} decay in the Galactic halo has been pursued by the \gls{magic} collaboration \citep{Ninci:2019njk}. In the conclusion of this section, we will discuss the significant improvement that the \gls{cta} will bring even for this pessimistic scenario of a cored DM profile in the GC. Note that the caveat of potential \gls{dm} cores is also present in the case of \glspl{dsph}. Yet it is considered much less severe compared to the \gls{gc} case. First, because the inner \gls{dsph} \gls{dm} profiles are better constrained, and second, because the \gls{dsph} galaxies are expected to show, at least for standard CDM, cuspier inner profiles due to their comparatively low baryonic content. 
%In addition, when \glspl{dsph} are observed with \glspl{iact}, their poor angular resolution makes cusp or core profiles nearly indistinguishable, as discussed e.g. in \citet{Aleksic:2011:segue}.

\subparagraph{Observation of \glspl{dsph}.}

DSphs are small galaxies, with masses of the order $10^7-10^9$
M$_\odot$,  gravitationally bound to their host galaxy \citep{Strigari:2008ib}. In the \gls{mw}, they are located within the \gls{mw} virial
radius at typical heliocentric distances of
$20-250$~kpc.\footnote{See \citet{Strigari:2018utn} for an excellent  overview on these targets.} They are considered to be in internal dynamical equilibrium and exhibit mass-to-light ratios
of the order of $100-1000$, meaning that they are highy \gls{dm} dominated systems. In addition, the great interest in these objects for indirect \gls{dm} 
searches stems from the fact that \glspl{dsph}
display very reduced stellar activity in the past few Gyr. This turns
out in an unlikely emission of high-energy photons of an astrophysics origin, that is, photons coming from the
depletion of interstellar gas and sources of high-energy cosmic
rays. Thus, any putative gamma-ray signal in these objects would be, most likely, of exotic
origin. As a result, \glspl{dsph} are considered one of the cleanest targets for indirect
\gls{dm} searches. Furthermore, the low stellar activity  may have allowed an efficient \gls{dm} contraction, with reduced
matter redistribution to e.g. supernova explosions and shocks
propagation \citep{Read:2018fxs,Lazar:2020pjs}.  Because of this, such
compact \gls{dm} overdensities result in promising conditions for annihilating \gls{dm} searches according to \autoref{eq:dm_flux}, yet less for
decays~\citep{Geringer-Sameth:2014yza,Bonnivard:2015xpq}. %\citet{Strigari:2018utn} provides an
%excellent  overview on these targets.
%N-body cosmological simulations of \gls{mw}-sized halos~\citep{ViaLactea2008,Springel:2008b} predict that, besides the main ``smooth'' \gls{dm} distribution in the Galactic halo, a wealth of substructures should be present down to scales of $10^{-6}-10^{-12}$~M$_\odot$. Onto these overdensities, baryons could have accreted until starting star and galaxy formation, however, the impact of baryon physics at these targets is probably smaller compared to \gls{mw} sized objects. 
Estimates of the total
number of \gls{mw} dwarf satellites in a $\Lambda$CDM universe predict the
existence of hundreds of these objects within the \gls{mw} virial radius
\citep{Tollerud:2008ze,Hargis:2014kaa}. Discrepancies between
observations and expectations from N-body cosmological simulations
were reported in the past \citep[the so-called ``missing satellite
problem'' discussed in e.g. ][]{Moore:1999nt,Klypin:1999uc,Strigari:2007a,Hargis:2014kaa}. However, not only our
census of \glspl{dsph} is known to be incomplete, but more importantly, the discrepancy
may be related to the fact that a large fraction of  \glspl{dsph}, especially
at the smallest scales, are expected to contain little or no baryonic material in the form of stars and gas. Therefore they would remain 
basically as dark \gls{dm} overdensities \citep{2015MNRAS.448.2941S,Bullock:2017xww}.  
%it is well known that our census is not complete as of today due to
%observational biases and sensitivity limitations of past and current
%stellar surveys. Indeed, when correcting by sky coverage and
%sensitivity, as well as baryonic physics, the predicted number of \gls{mw}
%satellites agrees well with observations,
%e.g.~\citep{2018arXiv180604143G}.
An additional potential issue with these targets is the fact that $\Lambda$CDM predicts a larger number of massive \glspl{dsph} than what it is actually observed. In other words, these massive galaxies should necessarily contain visible counterparts which are nevertheless not seen. This is
the so-called ``too big to fail''
problem~\citep{2006MNRAS.367..387R,BoylanKolchin:2011de}. Possible arguments to mitigate
this issue include a more accurate description of subhalo disruption
and mass loss, and of the role and evolution of the
baryonic contents in these objects 
\citep{10.1093/mnras/sts148,2014ApJ...786...87B,Tomozeiu_2016,2018arXiv180604143G}.  

DSphs are commonly classified in ``classical'' and
``ultra-faint'' galaxies, depending on whether they were discovered before or after the
Sloan Digital Sky Survey \citep[SDSS;][]{2000AJ....120.1579Y}, which
roughly translates into a brightness cut. Yet, this distinction
shadows more subtle differences: ultra-faint \glspl{dsph} possess more
compact sizes, less visible stars, and are typically more \gls{dm} dominated
systems than the classical ones, with mass-to-light ratios sometimes
exceeding 1000 $M_\odot/L_\odot$ \citep{Bonnivard:2015xpq}. In addition, observation \citep{Hayashi:2020jze} and numerical simulations \citep[e.g.,][]{Read:2018fxs,Lazar:2020pjs} suggest that the inner cusp of the \gls{dm} profile increases with decreasing baryonic activity, as expected from classical dwarf systems down to ultra-faint ones. In addition to the nine
classical \glspl{dsph} known, nearly 50 ultra-faint
\gls{dsph} \emph{candidates} have been discovered in photometric surveys since then,
mostly with SDSS itself and, most recently, with Dark Energy Survey (DES)
data~\citep{Drlica-Wagner:2015ufc,Drlica-Wagner:2017tkk}. From this number, about 40 have already
been confirmed to be \glspl{dsph} by means of follow-up spectroscopic
measurements. For each \gls{dsph}, the
$J-$factor is computed via the Jeans hydrostatic equilibrium equation
from the measured 
positions and velocities of member stars. To solve the equation one
needs to input a model for the star phase-space and another one for
the \gls{dm} density profile, whose normalization is left as a free
parameter.\footnote{All the details of these calculations are reviewed, e.g., by
\citet{Strigari:2012gn}.} This technique relies on several assumptions
on the geometry of the \gls{dm} subhalo hosting the \gls{dsph}: triaxiality, anisotropy, and co-rotation
all constitute relevant ansatzes. Another factor is the misreconstruction
of interloper stars, especially that of binary systems, whose intrinsic
high velocities strongly affect the computation of velocity
distribution and, ultimately, of the \gls{dsph} $J-$factor. However, the main limiting factor is
the very small number of stars, down to few tens, observed in the
dimmer ultra-faint \glspl{dsph} compared to the thousands of stars observed in the
larger ones. The interplay between these ansatzes and limitations are
discussed by, e.g., \citet{Bonnivard:2014kza,Bonnivard:2015vua,Chiappo:2016xfs,Klop:2016lug,Ichikawa:2016nbi,Ichikawa:2017rph,Ando:2020yyk}.

\begin{figure}[h!t]
    \centering
    \includegraphics[width=0.45\linewidth]{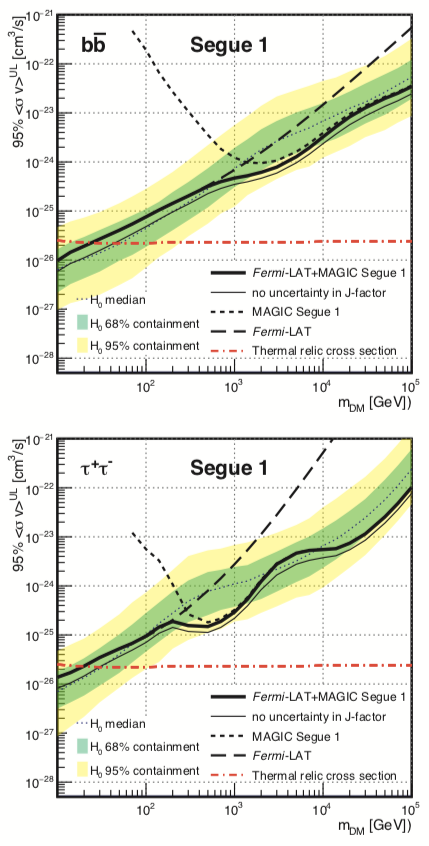}
    \includegraphics[width=0.45\linewidth]{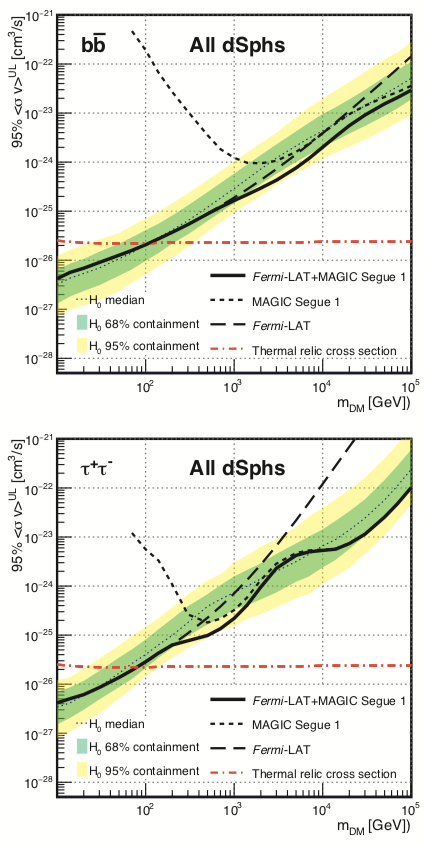}
    \caption{95\% C.L. upper limits on the thermally-averaged
      cross-section for \gls{dm} particles annihilating into $b\bar{b}$ (top
      row) and $\tau^+\tau^-$ (bottom-row). Thick solid lines show the
      limits obtained by combining Segue~1 observations from
      \LAT{} and \gls{magic} (left column), and the observations of 15
      \glspl{dsph} with the \LAT{} and Segue 1 with \gls{magic} (right
      column). Dashed lines show the observed individual \gls{magic} (short
      dashes) and \LAT{} (long dashes) limits. The thin-dotted
      line, dark and light gray bands show, respectively, the median and
      the symmetrical, two-sided 68\% and 95\% containment bands for
      the distribution of limits under the null hypothesis. The
      red-dashed-dotted line shows the thermal relic cross-section
      from \citet{Steigman:2012nb}. Image courtesy of the \gls{magic}
      Collaboration and reproduced from \citet{Ahnen:2016:dsph}.} 
    \label{fig:dsph_combined}
\end{figure}

Since 2004, these targets have been extensively studied by \glspl{iact} in
the context of \gls{dm} searches as shown in
Table~\ref{tab:targets}. Many targets have been sighted with exposures ranging from few hours to more than 150~h. Along almost 20 years we have observed the classical \glspl{dsph} Canis Major, Carina, Draco, Fornax, Leo I-II, Sextans, Sculptor, and Ursa Minor; the tidally-disrupted Sagittarius; and the ultra-faint systems Bo\"otes, Canes Venatici I-II, Coma Berenices, Draco~2, Grus-II, Leo IV-V, Hercules, Reticulum~II, Segue~1-2, Triangulum~II,  Tucana II-III-IV, Ursa Major~2 and Willman~1.\footnote{Note that not all of these ultrafaint systems are already confirmed \glspl{dsph}.} We already noted that in early times only few
hours were devoted to single targets. Part of the reason is that, back then, the prospects of detection were in general more
optimistic than today~\citep{Bergstrom:2005qk}. Also, the large uncertainties
related to the estimation of the \gls{dm} content in \gls{dsph} galaxies asked
for a diversification strategy, such that a larger pool  of different \glspl{dsph} was targeted. As a matter of fact, the exclusion \gls{dm} limits have significantly improved since the first publications for the classical \glspl{dsph} to the latest ones on both classical and ultra-faint dwarf systems. This was possible thanks to extended observation times
and improved analysis techniques, as reviewed in detail by  \citet{Rico:2020vlg}. \citet{Aleksic:2012cp} were the first to present a dedicated maximum likelihood approach for \gls{dm} analyses with \glspl{iact}, which allows to account for both the spectral and morphological shape of putative \gls{dm} emission, this way significantly enhancing the sensitivity to \gls{dm}. This work was also the first one to combine data from different sources and instruments. The same approach was consequently applied to analyses
of the deep observations performed with the \gls{magic} stereo
system~\citep{Aleksic:2013:segue,Ahnen:2016:dsph,Ahnen:2017pqx,Acciari:2020pno}. As used in \citet{Ahnen:2017pqx,Acciari:2020pno}, the likelihood binned in energy is built as follows:
\begin{equation}
\begin{array}{rl}
\mathcal{L}(\langle\sigma v\rangle;\nu_i|\mathcal{D}_i) =&\mathcal{L}\left(\langle\sigma v\rangle;s_{ij},J,\tau_i|N^{\text{ON}}_{ij},N^{\text{OFF}}_{ij}\right)  \\
 = &{\displaystyle \prod_{j=1}^{\text{bins}}\left[\frac{\left(s_{ij}(\langle\sigma v\rangle+b_{ij})\right)^{N^{\text{ON}}_{ij}}}{N^{\text{ON}}_{ij}!}\;e^{-((s_{ij}(\langle\sigma v\rangle+b_{ij})))} \times  \frac{(\tau_ib_{ij})^{N^{\text{OFF}}_{ij}}}{N^{\text{OFF}}_{ij}!}e^{-(\tau_i\,b_{ij})}\right]} \\
     \times & {\displaystyle \prod_{j=1}^{\text{bins}}
     \left[\mathcal{G}(\tau_i|\tau_{\text{obs},i},\sigma_{\tau_{\text{obs},i}})\right]} \quad  \times \quad\mathcal{G}(J|J_{\text{obs}},\sigma_J) \\
\end{array}
\label{eq:binned_iact_likelihood}
\end{equation}

where the likelihood is the product of likelihoods over the number of $N_{\text{bins}}$ energy intervals that have $\langle\sigma v\rangle$ as parameter of interest, the astrophysical $J-$factor, the normalization between signal and control regions $\tau_i$ and the estimated number of signal and background events $s_{ij},b_{ij}$  as nuisance parameters. The number of measured events in the signal and background extraction regions are $N^{\text{ON}}_{ij}$ and $N^{\text{OFF}}_{ij}$, respectively; The terms $\mathcal{G}(J)$ and $\mathcal{G}(\tau_i)$ are the Gaussian likelihoods for the $J-$factor and the ratio of the signal- to background-region  exposure. The estimated signal $s_{ij}$ depends on the parameter of interest  $\langle\sigma v\rangle$ as:

\begin{equation}
    s_{ij}(\langle\sigma v\rangle)\,=\,T^{\text{obs}}_j\,\int_{\text{bin},j}dE'_{\text{bin},j}\;\int dE\;\frac{d\phi(\langle\sigma v\rangle)}{dE}\;A_{\text{eff}}(E)\,G(E'|E)\,,
\end{equation}
where $T^{\text{\text{obs}}}_j$ is the observation time, $\phi(\langle\sigma v\rangle)$ is the intrinsic flux, $A_{\text{eff}}(E)$ the effective area and $G(E'|E)$ the migration matrix between true and reconstructed energy.

Marginalizing the instrument response functions and
the $J-$factor uncertainties in a full likelihood approach
not only allows to stack results of the same instrument, but also, as said, to combine limits from different instruments. In \citet{Ahnen:2016:dsph}, the unbinned approach from \citet{Aleksic:2012cp} was used to combine for the first time \gls{iact} data from \gls{magic} on Segue 1 with multi-\gls{dsph} data from \LAT{}. These limits are shown in
Fig.~\ref{fig:dsph_combined}, and still provide the most constraining results for indirect \gls{dm} searches at
\glspl{dsph} with \glspl{iact}.\footnote{Note, however, that these limits could be deteriorated by some yet unaccounted uncertainty on the Segue I \gls{dm} content and its corresponding $J-$factor \citep{Bonnivard:2015vua}.} 
A similar unbinned likelihood analysis method based
on {\it event weighting} has been developed by the \gls{veritas}
collaboration using individual event
information to  improve the sensitivity from a joint
analysis of five individual \gls{dsph} targets into a single limit~\citep{Archambault:2017wyh}. Also, the \gls{hess} collaboration had adopted a simplified version of \autoref{eq:binned_iact_likelihood} to combine data from five \glspl{dsph}, to search both for continuum \citep{Abramowski:2014:dwarf} and line \citep{Abdalla:2018mve} emission. Currently, a collaborative effort is ongoing
among the \gls{hess}, \gls{magic}, 
\gls{veritas}, \LAT{}, and \gls{hawc} collaborations to provide combined \gls{dm} limits
using a vast amount of  data collected from \glspl{dsph} so far \citep{Fermi-LAT:2021hze}. This study
shall include more than 500~h of \gls{iact} data plus 10 years
\LAT{} integrated time on 20 \glspl{dsph} and 1038 days of \gls{hawc} exposure on 12 \glspl{dsph}. % using the method of
%\citet{Aleksic:2012cp}. 

In summary, limits from \glspl{dsph} now reach values of cross-section
of the order of $10^{-24}$~\svunits{} and are  perceived by the
community as the most robust obtained with \glspl{iact}.  In the outlook of this section, we will  shortly discuss the expectation for \gls{cta} on \glspl{dsph}.

\subparagraph{Observation of dark satellites.}
In the standard cosmological framework, small dense \gls{dm} structures form first in the early Universe and later merge to larger halos. A natural consequence of this scenario of hierarchical bottom-up structure formation is that it predicts abundant substructures, or ``subhalos'', inside larger halos like our own Galaxy. The most massive of these subhalos in our own Galaxy host the known dwarf satellite galaxies of the Milky Way, while smaller subhalos with masses below $\sim10^7~M_{\odot}$ may harbour no stars or gas at all and thus may remain completely dark. These less massive \gls{dm} subhalos, with no astrophysical counterparts at other wavelengths, can represent excellent targets for indirect \gls{dm} searches  as well \citep{2013PhR...531....1S}: they may be located close and thus ease the detection of gamma radiation from \gls{dm} annihilation or decay, while background processes are absent by definition. Indeed,  N-body simulations of Milky-Way-sized halos predict the existence of thousands of them \citep{ViaLactea2008,Springel:2008b,2009MNRAS.398L..21S} and some are expected to yield large \gls{dm} annihilation fluxes at Earth given their typical distances and masses \citep[e.g.,][]{ 2012A&A...538A..93Z, 2012ApJ...747..121A,2016JCAP...05..028S, 2016JCAP...09..047H, Coronado-Blazquez:2019pny}. 

Since we do not know a priori the exact location of these ``dark subhalos'', current \glspl{iact} are not particularly suited to discover them given their narrow field of view of a few degrees wide. This is in contrast with the \LAT{}, which continually surveys the entire gamma-ray sky and thus naturally arises as the ideal instrument to search for dark subhalo candidates in the GeV energy range. Recently, the \gls{hawc} observatory, with a broad field of view as well, came into operation and enabled such searches also in the TeV regime. In most cases, the dark satellites' search strategy is based on looking for them among the pool of unidentified gamma-ray sources (unIDs) in current point-source catalogs. The key point is to identify those unIDs that exhibit spectral or spatial properties compatible with that expected from \gls{dm} annihilation. Indeed, it is intriguing that about one third of the sources in all available gamma-ray catalogs are still unidentified and some of them may be dark satellites just awaiting for a proper classification. In the absence of an unequivocal \gls{dm} signal, this strategy allows to set competitive \gls{dm} constraints by comparing the number of \gls{dm} subhalo candidates among the list of unIDs to that expected from a combination of N-body simulation results and flux instrumental sensitivity \citep{2010PhRvD..82f3501B,2012PhRvD..86d3504B,2012MNRAS.424L..64M,2012A&A...538A..93Z, 2012JCAP...11..050Z,2012ApJ...747..121A,2014PhRvD..89a6014B,2015JCAP...12..035B,2016JCAP...05..028S,2016ApJ...825...69M,2016JCAP...05..049B,2017PhRvD..96f3009C,2017JCAP...04..018H,2018arXiv181111732A,Calore:2019lks,Coronado-Blazquez:2019puc,Coronado-Blazquez:2019pny,2020Galax...8....5C}. Traditionally, however, these \gls{dm} constraints have been subject to large uncertainties given the lack of a precise characterization of the \gls{dm} subhalo population, such as their structural properties and radial distribution in the Galaxy. A statistical approach on the computation of the number of expected dark subhalos and their prospects of detection with \gls{cta} were reported by~\citet{Coronado-Blazquez:2021avx}.

\begin{figure}[h!t]
\centering
\includegraphics[width=0.45\linewidth,valign=b]{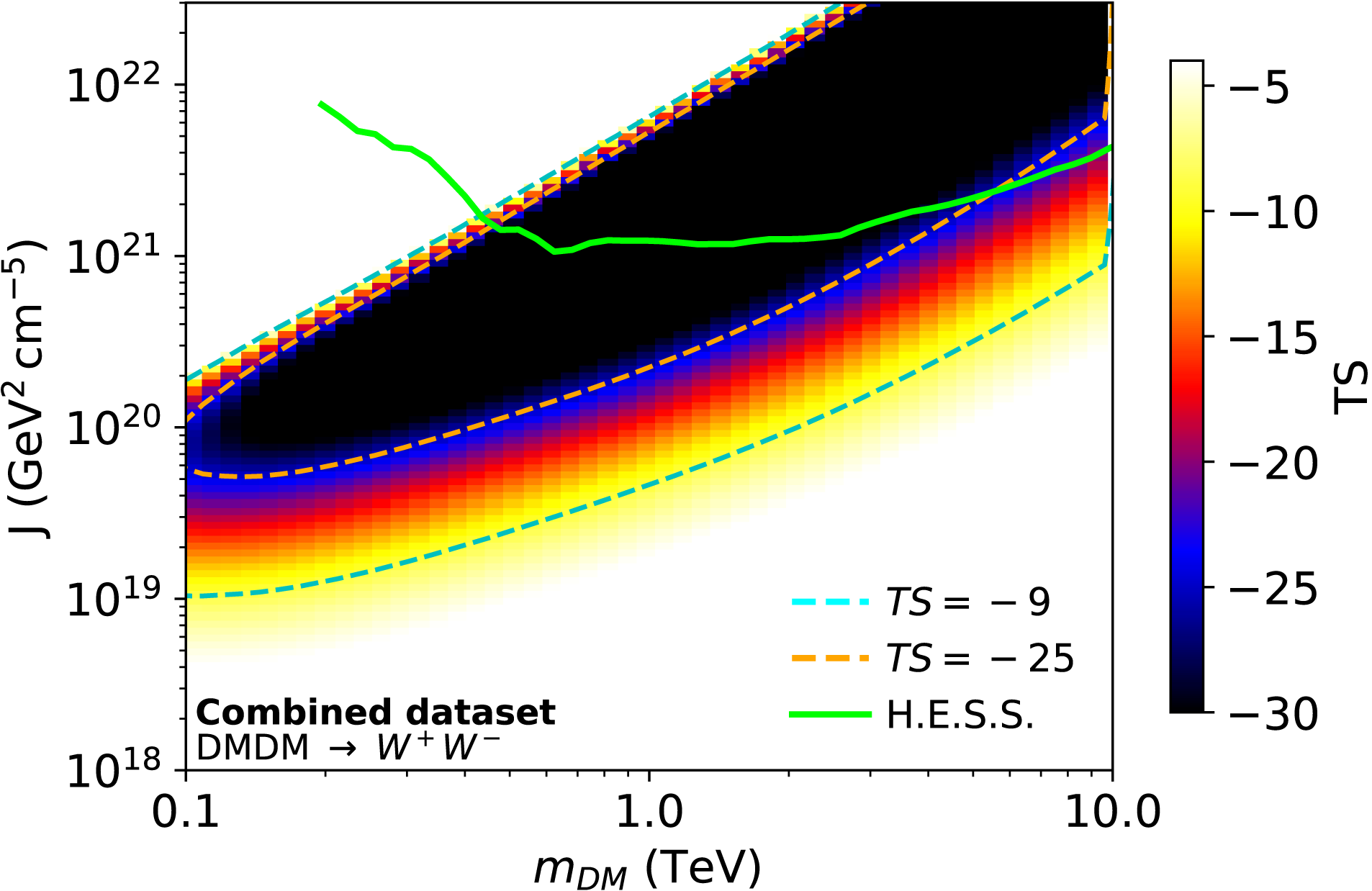}\hfill
\includegraphics[width=0.52\linewidth,valign=b]{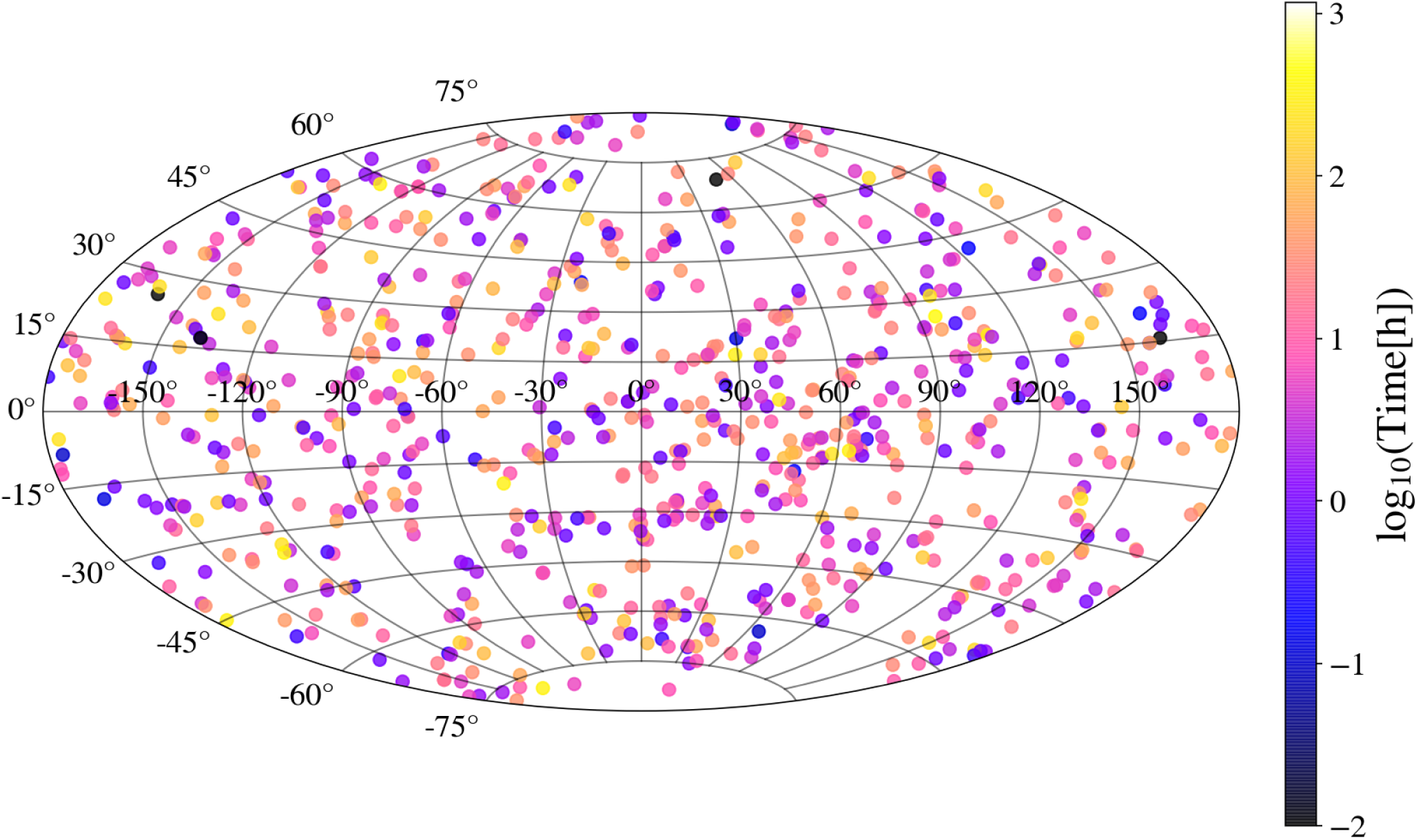}
\caption{The two approaches to search for dark satellites with \glspl{iact}. First, by following up observations by suitable unidentified gamma-ray sources detected by \LAT{}, which is the most viable strategy with current \glspl{iact}: the left panel, taken from \citet{1866411}, shows the upper limit on the sum of the $J$-factors in a combined analysis of three unIDs observed by \gls{hess} under the assumption of a thermal \gls{dm} annihilation cross-section (solid light gray line). Under the assumption that the signals detected by \LAT{} in fact stem from \gls{dm} subhalos and annihilation into $W^+W^-$ bosons (likelihood contours in the background), the \gls{hess} null results significantly restrict the possible $J$-factor range. Second, with the future \gls{cta}, also the serendipitous discovery in any observation's field of view becomes a feasible search strategy: the right figure shows a random realization of sky coverage with the \gls{cta} within the first decade, extrapolated from \gls{magic} stereo observations. Figure from \citet{Coronado-Blazquez:2021avx}, see page~\pageref{sec:subhalos_cta} for further discussion.} 
\label{fig:subhalo_searches_iacts}
\end{figure}

While current \glspl{iact} are not the best-suited instruments to discover dark satellites, they have been used to scrutinize in further detail some of the best \gls{dm} subhalo candidates among the pool of unIDs identified in the above mentioned works, given their superior sensitivity at TeV energies and angular resolution. In ~\citet{2011arXiv1109.5935N,2011arXiv1110.4744N}, the \gls{magic} Collaboration reported on the observation of two objects in 2010 and 2011, namely 1FGL J2347.3+0710 and 1FGL J0338.8+1313, among the list of 630 unIDs present in the First \LAT{} catalog of point sources  \citep[1FGL;][]{2010ApJS..188..405A}. No \gls{vhe} gamma rays were detected after an exposure of around 10~h on each source. These two objects were selected as the most promising out of ten after having survived several selection criteria that were based on the expected \gls{dm} signal properties. In particular, they selected only non-variable sources located at high ($\lvert${b}$\rvert>10^\circ$) Galactic latitudes, to avoid contamination from Galactic objects like pulsars, and whose spectra are well described by a power law in the \LAT{} energy band and are particularly {\it hard} (spectral index $\Gamma>-2$). They also performed a careful study to search for counterparts of all unIDs at other (X-ray, optical, radio) wavelengths. Later, the \gls{veritas} Collaboration reported on the observation of two other unIDs in a \gls{dm} context, namely 2FGL J0545.6+6018 and 2FGL J1115.0-0701, with again around 10 h exposure on each target \citep{2013arXiv1303.1406G,2015arXiv150900085N}. These objects were identified as the best candidates among the 576 unIDs in the Second \LAT{} catalog of point sources  \citep[2FGL;][]{2012ApJS..199...31N}. As summarized by \citet{2015arXiv150900085N}, \gls{veritas} adopted the same selection criteria proposed by \gls{magic} in \citet{2011arXiv1109.5935N,2011arXiv1110.4744N}, the only difference with respect to \gls{magic} being the \gls{veritas} peculiarities in terms of observation site and instrumental sensitivity. Once again, the null detection of \gls{vhe} gamma rays from the location of these two unIDs only led to set upper limits to the flux from these objects. In \citet{1866411},\footnote{Note that the recent publication by \citet{1866411}   is the only \gls{iact} study on this topic so far published in a peer-review journal.}  the \gls{hess} collaboration recently reported the follow-up observation of dark-satellite candidates among unIDs in the Third Catalog of Hard \LAT{} sources \citep[3FHL;][]{TheFermi-LAT:2017pvy}. Applying similar selection criteria as by the \gls{magic} and \gls{veritas} works to 177 unIDs in the 3FHL, they obtained six subhalo candidates out of which four were observed for between three and nine hours (see \autoref{tab:targets}). As among these four, the source 3FHL J2104.5.2117 was recently associated with high probability to an AGN in the 4FGL catalog \citep{Fermi-LAT:2019yla}.   No signal was detected from either of the four targets. As no robust assumption on the $J$-factor of the putative \gls{dm} subhalos can be made for this source class, the \gls{hess} collaboration adopted a reverse conclusion: on the assumption of a thermal \gls{dm} annihilation cross-section in the order of \autoref{freezeout-xsection},  upper limits on the dark-satellite $J-$factor were derived. This is illustrated in \autoref{fig:subhalo_searches_iacts} (left).

Yet, we note that although no detection was obtained in neither \gls{magic}, \gls{veritas}, or \gls{hess} observations of unIDs, the resulting flux or $J-$factor upper limits in all cases may be still useful to test the \gls{dm} subhalo hypothesis and to shed further light on the nature of these sources.
%\footnote{One word of caution: all mentioned \gls{iact} works on unIDs have not been yet published in refereed journals as of today.} 
So far, no \gls{iact} dark satellite observations have been yet performed of any of the unIDs in the most recent \LAT{} 4FGL catalog \citep{Fermi-LAT:2019yla}, containing more than one thousand unIDs.  Also, \citet{Coronado-Blazquez:2019puc,Coronado-Blazquez:2019pny} have recently highlighted potential \gls{dm} subhalo candidates among the 2FHL \citep{Ackermann:2015uya}, 3FGL \citep{Acero:2015gva}, and 3FHL catalogs which also have not yet been observed by \glspl{iact}. 

Finally, thanks to a significantly larger field of view and foreseen large-scale survey observations, also dark satellite discovery searches may be feasible with the next-generation \gls{iact}, the \gls{cta}. This is illustrated in \autoref{fig:subhalo_searches_iacts} (right) and further discussed on page~\pageref{sec:subhalos_cta} in the outlook of this section.

\subparagraph{Observation of \glspl{imbh}.}
There are theories that predict the existence of black holes (BHs) of masses between $10^2-10^6$~M$_\odot$, sometimes
referred to as \glspl{imbh}: they could form as
remnants of collapse of Population III stars, resulting in $10^2$~M$_\odot$-mass BHs \citep[scenario A, ][]{Madau:2001sc,Heger:2002by} or from direct collapse of primordial huge gas overdensities in early-forming halos, resulting in $10^6$~M$_\odot$-mass BHs \citep[scenario B, ][]{Koushiappas:2003zn}.
Many of them
could reside in the \gls{mw} halo.  The average number of \glspl{imbh} in the \gls{mw} was estimated by numerical simulations to be of the order of a thousand (scenario A) to a hundred (scenario B) \citep{Bertone:2005xz}. As a result of the increased gravitational potential due to infalling baryons on a central accreting system, the \gls{dm}
could have readjusted and shrunk, giving rise to the formation of
so-called ``mini-spikes'' \citep{Bertone:2005xz,Bertone:2009kj}. On the other hand, mini-spikes are rapidly disrupted as a result of dynamical processes like BH formation or merging events. The interesting fact is that, starting
from a typical \gls{nfw} distribution for the \gls{dm}, the adiabatic
growth of the spike leads to a final \gls{dm} density profile even cuspier than \gls{nfw}, with a central slope of index $-7/3$ as opposed to -1 for NFW \citep{Bertone:2009kj}. For scenario B, which foresees more luminous objects than scenario A, the corresponding gamma-ray luminosity would be of the order of the gamma-ray luminosity of the entire \gls{mw} halo, which made \glspl{imbh} very
interesting targets for \gls{dm} searches~\citep{Bertone:2005xz}. 
Similarly to the dark satellites' case discussed above, it is possible that some \glspl{imbh} are hidden in unidentified EGRET or \LAT{} point-like sources. Before the advent of \LAT{}, and after having performed a selection of the best EGRET unidentified sources, \gls{magic} observed the brightest of them in 2006 for 25~h,  with no detection~\citep{Doro:2007icrc}. The source was later on associated with a bright pulsar using \LAT{} data. Also, using 400~h of data collected from 2004 to 2007 in the region between $-30$ and $+60$ degrees in Galactic longitude, and between $-3$ and $+3$ degrees in Galactic latitude, \gls{hess} could
exclude the \gls{imbh} formation scenario B at a 90\% confidence level for \gls{dm} particles with annihilation cross-section values $\langle \sigma \mathit{v} \rangle$ above $10^{-28}$~\svunits and masses between 800 GeV and 10 TeV \citep{Aharonian:2008:dm}.

\subparagraph{$\qquad$Observation of globular clusters.}
Globular clusters are star compounds that share properties
with \glspl{dsph}. They are in general more compact and with narrower metallicity distributions than \glspl{dsph}. There is a
large debate on how these clusters formed, considering that it is
possible to find globular clusters already 1~Gyr after the Big
Bang. Despite they are generally thought to contain low amounts of \gls{dm}, the debate on whether or not they reside in a \gls{dm} halo is still
open \citep{Mashchenko:2004hj,Mashchenko:2004hk,Lane:2010rd,Bradford:2011aq,Ibata:2012eq,2014MNRAS.438..487D,2017MNRAS.464.2174B,Creasey:2018bgv}. From the cosmological point of view, there is no conflict in telling that they were formed in \gls{dm} overdensities, as the 
\glspl{dsph}, however, there is no strong observational evidence for the
presence of a \gls{dm} halo as of today. One possibility is in fact that the luminous
part of a globular cluster is just located in the inner region of a flat
\gls{dm} overdensity, which would not modulate significantly the
stellar velocities for different radii \citep{Ardi:2020hlx}. 
The interest in globular clusters from the \gls{dm} search perspective has
recently been revived by the observational evidence of a $\sim30$ GeV DM
signature in the gamma-ray spectra of 47 Tuc \citep{Brown:2018pwq} and
Omega Cen~\cite{Brown:2019whs}; though a pure DM interpretation of this
observational evidence is still a matter of debate \citet{Bartels:2018qgr,Brown:2019teh}.

\begin{figure}[h!t]
\centering
\includegraphics[width=0.45\linewidth]{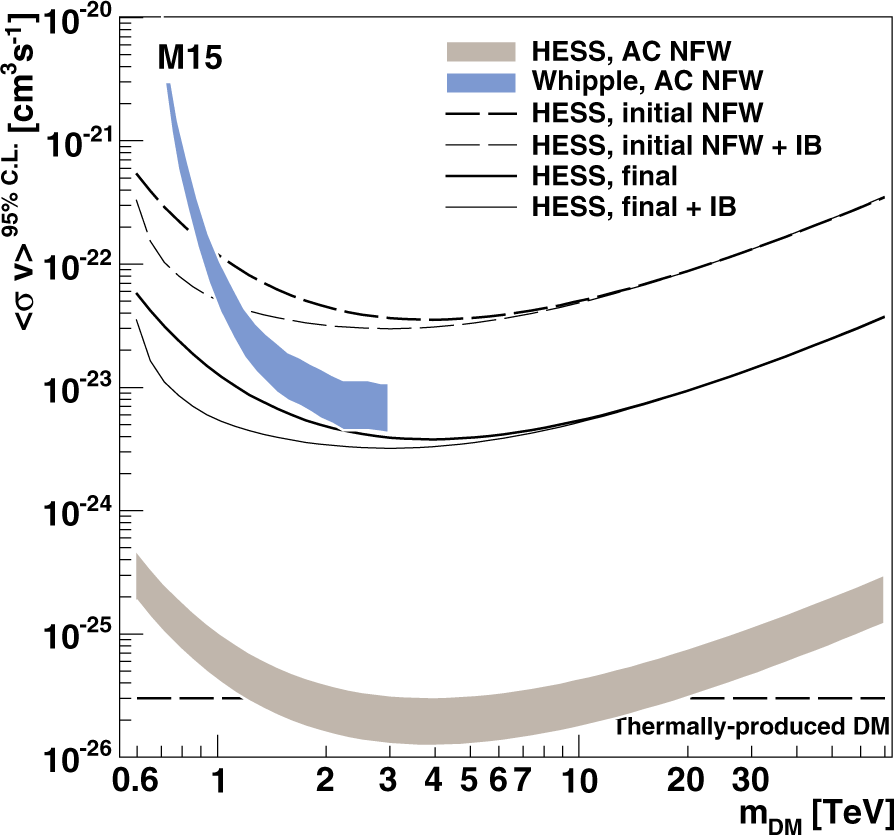}
\caption{95\% C.L. upper limits on the \gls{dm} annihilation cross-section by \gls{hess} \citep{Abramowski:2011:ngc6388} and Whipple \citep{Wood:2008hx} observations of the globular cluster M15. Limits were calculated for Whipple (blue band) and \gls{hess} (brown band)  assuming an \gls{nfw} density profile, adiabatically contracted (AC) due to the presence of an \gls{imbh}. Additionally taking into account for kinetic heating by stars results into the ``final'' profile assumed by \gls{hess} (solid lines). The dashed lines denote an initial \gls{dm} halo model assuming $M_{\rm vir}=10^7\,M_{\odot}$. The final-state gamma-ray spectrum is modeled in this work according to \citet{Bergstrom:1998a}. IB refers to accounting for internal bremsstrahlung effects. Figure from \citet{Abramowski:2011:ngc6388}.} 
\label{fig:globclusters_hess}
\end{figure}

For what regards the observation of globular clusters with
\glspl{iact}, M15 was observed from 2002 to 2004 
by Whipple~\citep{Wood:2008hx} and later re-observed together
with NGC 6388 by \gls{hess} in
2006--2009~\citep{Abramowski:2011:ngc6388}, shown in Fig.~\ref{fig:globclusters_hess}. \gls{magic} also invested
more than 150~h on M15, with which only upper limits on the gamma-ray emission were obtained, and discussed in the context of astrophysics~\citep{Acciari:2019ysf}. The interpretation in terms of \gls{dm} is under preparation. The best
exclusion curves for annihilating \gls{dm} come from the observation of NGC 6388 by \gls{hess}, and
are of the order of $10^{-24}-10^{-25}$~\svunits. However it must be noted that these results rely on strong
assumptions about the amount of \gls{dm} in this particular object.

\subparagraph{Observation of galaxy clusters.}
Within the standard $\Lambda$CDM scenario, 
galaxy clusters, with masses around $10^{14}-10^{15}$~M$_\odot$, are
the largest gravitationally bound objects in the Universe and the most recent structures to
form~\citep{Voit:2004ah}. They are relevant for what concerns indirect \gls{dm}
searches \citep[e.g.,][]{Jeltema:2008vu,Colafrancesco:2010kx,SanchezConde:2011ap,Pinzke:2011ek,Gao:2011rf}, because \gls{dm} is the 
dominant mass component within this type of systems, accounting for up to 80\%
of the total virial mass.\footnote{The remainder component is dominated by 
intra-cluster medium (ICM) gas.} However, galaxy clusters are extragalactic objects, many of them located at cosmological distances, and therefore any radiative signal from \gls{dm} annihilation or decay would be significantly diluted or attenuated by the extragalactic background light. 
Given their large halo masses, the presence of a large amount of substructures in clusters is expected to be particularly important for \gls{dm} searches, as they intrinsically boost the total $J$-factor for annihilating \gls{dm} by a factor of $30-50$~\citep{SanchezConde:2011ap,Sanchez-Conde:2013yxa,Zavala:2015ura,Moline:2016pbm,Hiroshima:2018kfv}. Because of that, galaxy clusters are targets competitive to \glspl{dsph} for indirect \gls{dm} searches in term of expected $J-$factors \citep{SanchezConde:2011ap}. Although this {\it subhalo boost} is not present in a decaying \gls{dm} scenario, for which the ``clumpiness'' of \gls{dm} is not relevant, galaxy clusters are even more competitive compared to \glspl{dsph} in the case of  decaying \gls{dm}: this is to be understood by the huge total mass budget of a cluster integrated within a small opening angle.  Note also that, as in other targets, in galaxy clusters any potential \gls{dm}-induced gamma-ray emission is not only expected as prompt emission, but also as secondary emission mostly generated by synchrotron and inverse Compton processes from high energy electrons and positrons produced in \gls{dm} interactions~\citep{Colafrancesco:2005ji}. 

Clusters of galaxies also host strong astrophysical gamma-ray
emitters in many cases~\citep{Blasi:1999aj}. Thus, any careful search for \gls{dm}
signatures in galaxy clusters must disentangle these from more ``conventional'' astrophysical contributions. For instance, AGNs sitting in individual cluster
galaxies or a sizeable interaction of cosmic rays (CRs) with the ICM may potentially yield gamma-ray fluxes larger than the one expected from \gls{dm} interactions. Fortunately, a different spatial morphology of the gamma-ray signal is expected in each case: while the signal from \gls{dm} is more extended, especially in the case of decay, the signal from CR interactions and radiative energy losses is more compact, and individual galaxies are point-like. This can be used as a powerful tool for discriminating between the different components, as discussed by \citet{Doro:2012xx}. 

\begin{figure}[h!t]
\centering
\includegraphics[width=0.45\linewidth]{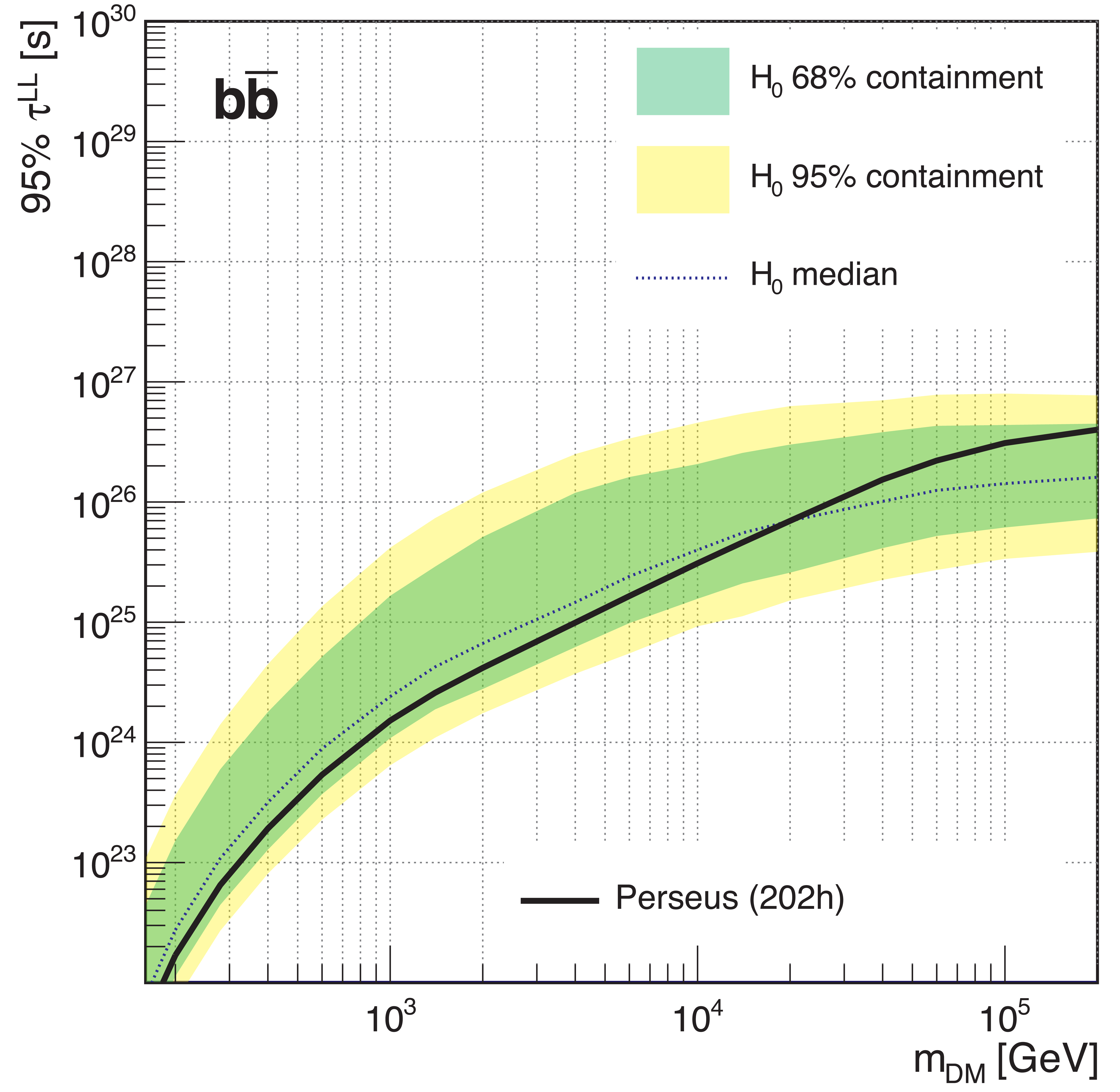}
\includegraphics[width=0.45\linewidth]{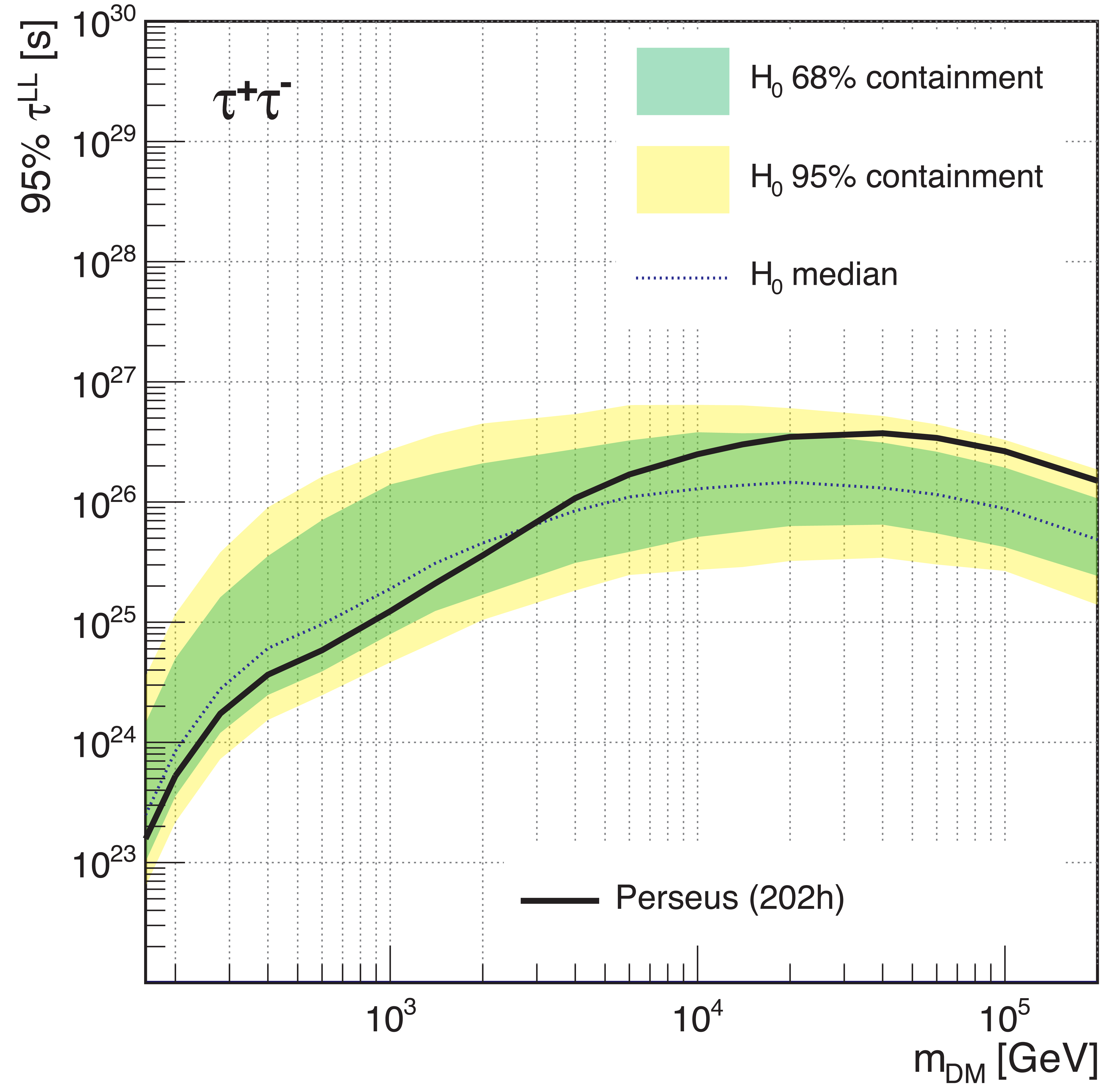}
\caption{95\% C.L. lower limits on the \gls{dm} decay lifetime $\tau^{\text{LL}}$ (solid black lines) as obtained from the 202~h observation of the Perseus cluster of galaxies with the \gls{magic} telescopes for the $b\bar{b}$ (left panel) and  $\tau^{+}\tau^{-}$ (right) decay channels. The expected limits (dashed lines) and the two sided 68\% and 95\% containment bands are also shown. These are the most constraining decay limits obtained so far above few hundreds GeV. Figure from \citet{Acciari:2018:perseus}.} 
\label{fig:decay_dm}
\end{figure}

To date, the deepest scrutiny of a galaxy cluster in gamma rays is the one performed with the \gls{magic} telescopes for the Perseus cluster~\citep{Acciari:2018:perseus}. \gls{magic} observed this object for more than 400~h over several years. This deep observation of the Perseus cluster allowed to set the strongest constraints to date on decay \gls{dm} using a subset of 202~h of data \citep{Acciari:2018:perseus}. Such constraints are reported in Fig.~\ref{fig:decay_dm} for two pure decay channels that roughly bracket the minimum and maximum number of photons expected from decay, i.e., by decays into purely
$b\bar{b}$ quarks or  $\tau^{+}\tau^{-}$ leptons. These \gls{magic} limits are of the order of
$10^{26}$~s, that is, more constraining than any other limit set to decaying \gls{dm}  in the \gls{dm} mass range above few hundreds of GeV. At lower energies, stronger constraints are found using \LAT{} data of the diffuse emission~\citep{Ackermann_2012}.

For \gls{dm} annihilation, the strongest \gls{dm} constraints from clusters as of today were set  by \gls{hess} by the observation of the Fornax cluster \citep{Abramowski:2012:fornax}, which is one of the clusters with the highest expected \gls{dm} annihilation signal. Apart from Perseus and Fornax, also the Coma cluster was observed for 18.6h by the \gls{veritas} telescopes \citep{2012ApJ...757..123A}, however, yielding less stringent constraints on \gls{dm} annihilation.

\paragraph{Constraints on line-like spectral signatures.} Self-annihilation or
decay of \gls{dm} particles may create unique
line-like gamma-ray spectral features not expected at TeV energies from any other astrophysical process~\citep{Bertone:2005xz}. In particular, narrow spectral lines can be expected from events such as $\chi\chi\rightarrow\gamma\gamma$ and $\chi\chi\rightarrow X\gamma$, where $X=Z^0,H,...$ are found at loop or enhanced level \citep{Bergstrom:1994mg,Jungman:1994cg,Bergstrom:1997fh,Arina:2009uq,Rinchiuso:2018ajn,Beneke:2018ssm}. Furthermore, in case the annihilation is mediated by a heavy particle, also peculiar features such as `box'-like features in the gamma-ray spectral energy distribution may occur~\citep{Ibarra:2012dw}. Also, emission from virtual internal
states can happen for specific particle models via bremsstrahlung, providing a pronounced component
peaked towards the spectral cutoff at the highest
energies \citep{Bringmann:2007nk,Bringmann:2008kj,Bringmann:2013oja}. 

Such a signal would be
readily distinguishable from astrophysical high-energy gamma-ray sources that are expected to only produce continuous spectra, and may constitute a robust "smoking-gun" identification of \gls{dm}. For \gls{wimp} masses above several hundreds of GeV and above the energy reach of space-based detectors, Earth-bound gamma-ray detectors are moreover the only instruments to search for such features. 
%\footnote{There are only few theories to explain bump features in the spectral energy distribution of TeV gamma rays with known physics.}. 
Also, for \glspl{iact}, line-like features are even easier to be searched for than continuous spectra: the large residual cosmic-ray background is a generic challenge for \glspl{iact}, yet line-like gamma-ray spectra are very distinct from the background spectral energy distribution and, thus, a powerful discrimination between background and signal is possible in terms of their spectral properties.
Nevertheless, specific analysis algorithms must be developed   in order to look for such signatures in \gls{iact} data. In particular, \glspl{iact} are not only, like any other instrument, affected by a limited energy resolution, in this case of the order of 10\%, but may also show specific systematic uncertainties at the energy scale as a result of using the atmosphere as calorimeter.

%There have been  claims of hint of a line-like signal at about 130~GeV in the Fermi data of the Galactic Centre region~\citep{S_VIB_Line, S_Line} which received a huge attention. If confirmed, the \gls{wimp} particle should  have a mass
%of about $m_\chi\sim130$ GeV  and annihilation rate
%(assuming Einasto profile) of  $\langle \sigma \mathit{v} \rangle_{\gamma\gamma} =
%1.27\times 10^{-27}$~\svunits \citep{Bringmann:2012ez}. 

\begin{figure}[h!t]
\centering
 \includegraphics[width=0.45\linewidth]{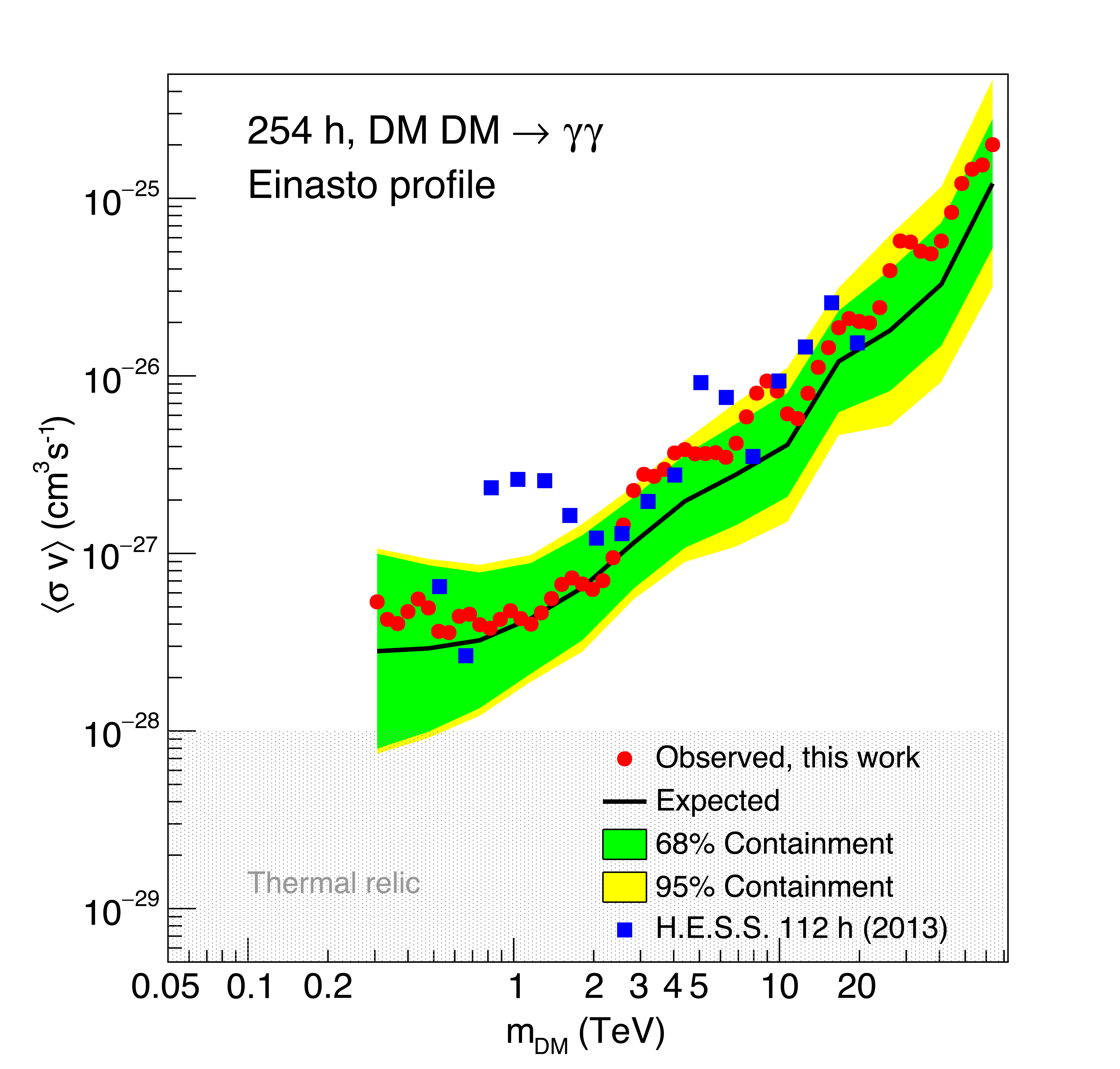}\hspace{0.2cm}
 \includegraphics[width=0.45\linewidth]{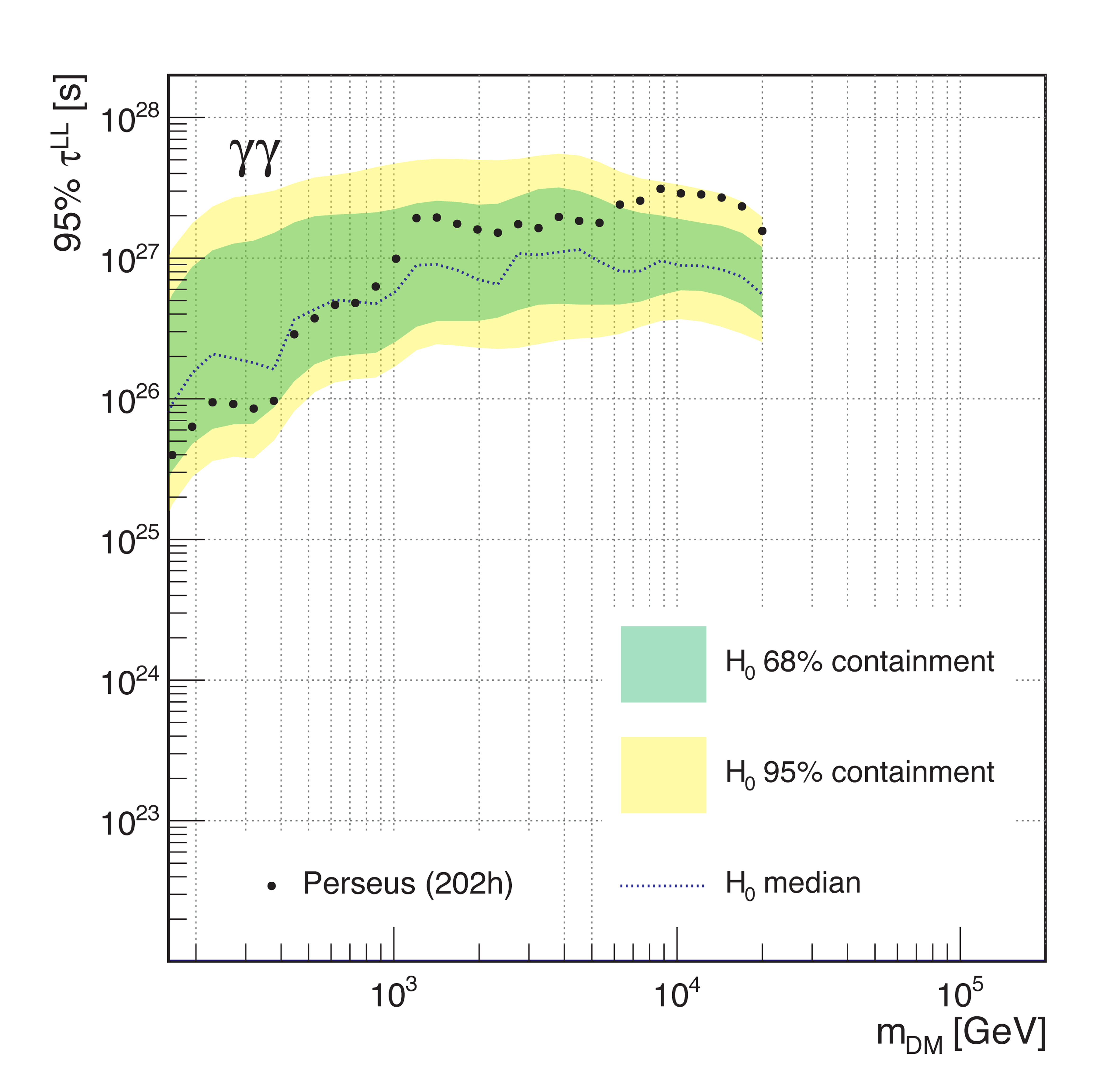} 
  \caption{Constraints on the velocity-averaged annihilation cross-section $\langle \sigma v \rangle$ for the prompt annihilation into two photons derived from \gls{hess} observations of the inner \gls{gc} region (left), and of the lifetime from prompt \gls{dm} decay into two photons derived from \gls{magic} observations of the Perseus galaxy cluster (right). Figures from~\citet{Abdallah:2018:lines,Acciari:2018:perseus}, respectively.}  
   \label{fig:hess_lines}
\end{figure}

Using data collected with \gls{hess}, upper limits on line-like emission have been obtained in the energy range between 100 GeV and 70 TeV from the central part of the \gls{mw} halo \citep{Abdalla:2016olq,Abdallah:2018:lines}. For a \gls{dm} particle mass of 1 TeV, limits on the velocity-averaged \gls{dm} annihilation cross-section $\sv$ reach the level of $10^{-27}$ \svunits{} when adopting an Einasto \gls{dm} density profile for the Galaxy. 
Using combined \gls{dsph} observations, limits were derived between 200 GeV and 40 TeV by \gls{hess}~\citep{Abdalla:2018mve,Abdallah:2020sas} and between 100 GeV and 10 TeV by \gls{veritas}~\citep{Archambault:2017wyh}.   
Additional limits were obtained by \gls{magic} after a 158~h observation of the Segue~1 \gls{dsph} \citep{Aleksic:2013:segue}; also an analysis of the \gls{gc} region was recently presented by \gls{magic} \citep{Inada2021}. Line emission can be likewise emitted in \gls{dm} decay processes, and this was scrutinized by \citet{Acciari:2018:perseus} in the analysis of 202~h of data towards the Perseus galaxy cluster. The current best limits on line-like emission from \gls{dm} annihilation (left) and decay (right) at TeV energies are shown in Fig.~\ref{fig:hess_lines}. Note that these limits correspond to a pure annihilation or decay process into two photons. For this process only constituting  a small fraction among other annihilation or decay channels, as expected for a loop-suppressed process, the total constrained annihilation cross-section may be much larger (and correspondingly, the lifetime be shorter) than suggested by the shown data. In Fig.~\ref{fig:hess_lines} (left), this is considered by the grey-shaded region of an expected signal for thermal \glspl{wimp} downshifted to $\langle \sigma \mathit{v} \rangle \lesssim 10^{-28}$~\svunits{} already implying annihilations into two photons happening with a branching ratio smaller than $10^{-2}$. 

%These limits complement those obtained with the \LAT{} at lower energies~\citep{Abdo:2010nc} and are shown in 

\paragraph{Searches with charged particles. \label{pageref:antiparticles}} In the annihilation or decay process of \gls{dm} particles, hadrons and leptons are abundantly produced~\citep[see, e.g.,][]{Donato:2003xg}. Below these, antiparticles would constitute a species that could be easier discriminated against the astrophysical cosmic-ray background, as, e.g., the antiparticles produced after the spallation of CRs in the interstellar medium. This possibility gained a great interest a decade ago after the accumulation of evidence of a positron excess from satellite-borne particle detector experiments such as PAMELA \citep{Adriani:2008zr}, \LAT{}~\citep{2012PhRvL.108a1103A} and subsequently AMS~\citep{Aguilar:2013}. As no antiproton excess was found simultaneously, an explanation  for this positron excess that generated particular interest in the community was to invoke the existence of  \emph{leptophilic} \gls{dm}, i.e., \gls{dm} particles that prefer leptonic annihilations or decays over hadronic channels. Also, the ballon-borne ATIC reported an excess in the electron+positron spectrum around~500 GeV \citep{Chang:2008aa}. However, the sensitivity of these balloon or space-borne measurements was limited to energies below a few TeV. In turn, \glspl{iact} are sensitive to Cherenkov light generated in atmospheric showers initiated by sufficiently energetic photons or charged particles. In this sense, they are also suitable detectors for cosmic electrons, positrons and antiprotons beyond TeV energies. 

Leptonic showers from cosmic electrons or positrons are currently indistinguishable from gamma-ray induced showers. Yet, in contrast to gamma-ray induced showers, which are always aligned to a very localized cosmic gamma-ray source,  cosmic leptons reach the Earth from perfectly isotropical directions. As such, in a data sample without a gamma-ray source in the field of view, electrons or positrons  can be discriminated from the residual hadronic cosmic-ray background in \gls{iact} data by comparing the whole data sample with corresponding \gls{mc} simulations of hadron events, via the definition of discriminator parameters specifically designed for that purpose. Consequently, the systematic uncertainties need to be carefully controlled, because even a small mismatch between \gls{mc} and real data would result in a bias in the estimated electron flux. This search was pioneered by \gls{hess} back in 2008, which used 239 hrs of data,  extending the measured electron+positron spectrum deeper into the TeV range~\citep{Aharonian:2008:electron}. They found evidence for a steepening in the all-electron spectrum above 600 GeV, while they excluded the spectral peak previously claimed by ATIC \citep{Aharonian:2009:electron}. An analogous measurement by \gls{veritas}  confirmed a spectral break in the energy spectrum at around ~700 GeV \citep{Archer:2018chh}, as did a preliminary measurement presented by \gls{magic}   \citep{BorlaTridon:2011dk}. The DAMPE satellite detector published an all electron+positron spectrum up to 4.6 TeV, confirming a spectral break in the all-electron spectrum at around 900~GeV \citep{Ambrosi:2017wek}, as so did the CALET experiment measuring the spectrum to 4.8~TeV on the International Space Station \citep{Adriani:2018ktz}.

%\mhc{Can we discuss the paragraph on Moon shadow? Also only proceedings, no results, and I think interpreting shadow for electrons even more tricky when still not clear results published on "simple" all-electron flux with \gls{magic}? Here some shorter alternative formulation of the paragraph:} 
Unfortunately, the \gls{iact} detection technique does not directly allow to discriminate electrons from positrons, to investigate the excess of positrons at TeV energies. However, the Earth' magnetic field is separating electrons from positrons, and this separation could be observable where the isotropic flux is shadowed by the Earth \citep[as exploited by the \LAT{} measurement;][]{2012PhRvL.108a1103A} or the Moon. In fact, the feasibility of disentangling electrons and positrons by observing a cosmic-ray {\it Moon shadow} with \glspl{iact} has already been investigated  and some data been collected with the \gls{magic} and \gls{veritas} telescopes despite the big observational challenges \citep{Colin:2011wc,Bird:2015opa}.

%A drawback of the above technique is that positrons cannot be discriminated from electrons. A different experimental approach may allow \glspl{iact} to disentangle all electrons by charge by making use of the so-called {\it Moon shadow}. In fact, the all-electrons incoming flux gets diverted by the Earth magnetic field in a way that some particles are absorbed by the Moon which creates a sort of shadow, at one side for positively charged particles, and at the other side for negatively charged particles. Additionally, this shadow is cast at a distance from the Moon that depends on the particle rigidity: the higher the energy of the electrons the closer the shadow to the Moon. This search was first pursued by \gls{magic} and preliminary results were discussed at conferences \citep{Colin:2011wc}.

%%%%%%%%%%%%%%%%%%%%
%
%
%
%
%
%
%
\subsection*{Conclusions and outlook}
\label{sec:dm-conclu}
\begin{figure}[h!t]
  \centering
  \includegraphics[width=\linewidth]{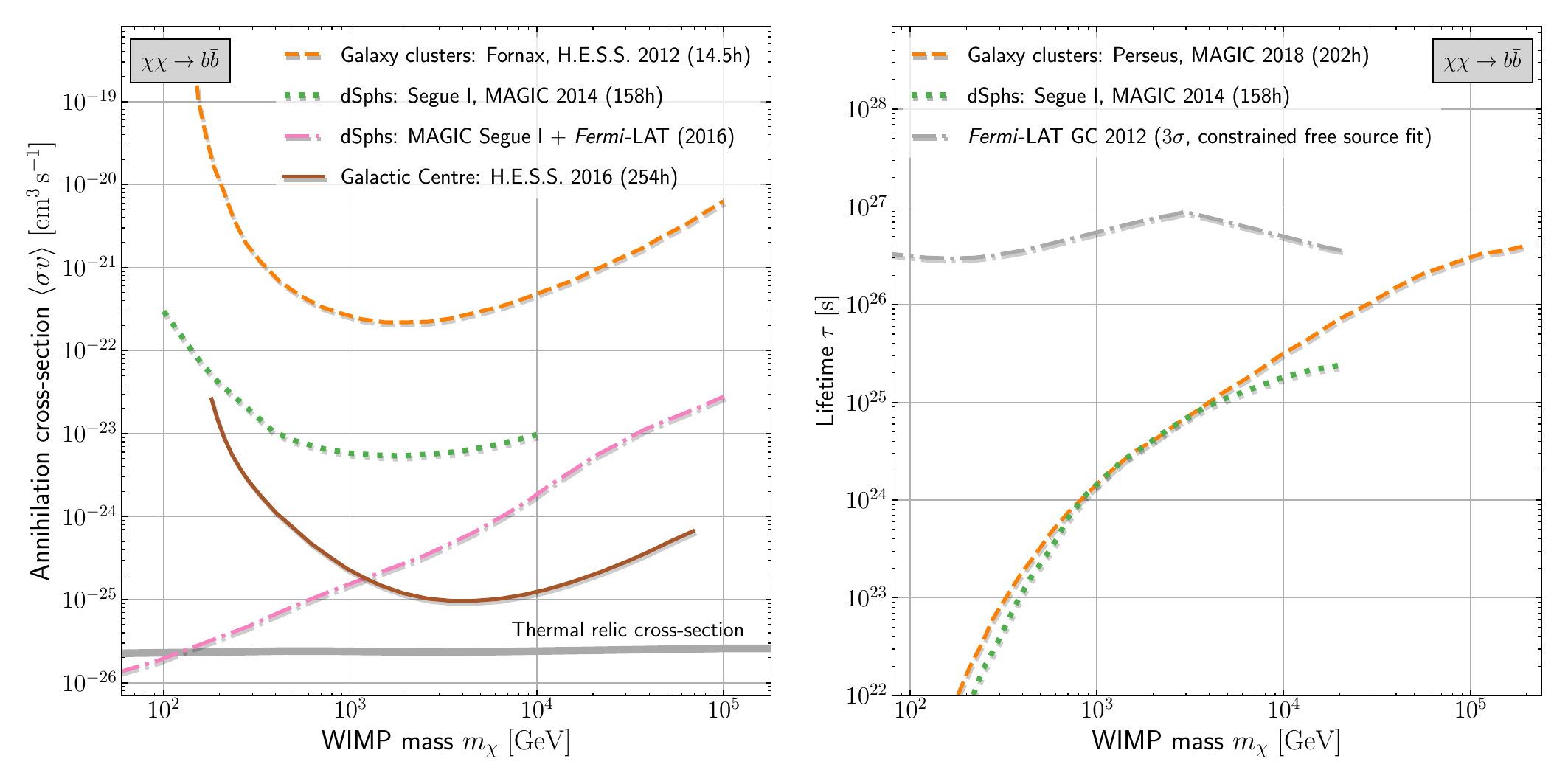}
  \caption{\label{fig:compare} Selection of some of the most representative \gls{wimp} limits by \glspl{iact} for the $b\bar{b}$ annihilation channel (left) and the $b\bar{b}$ decay channel (right).} 
\end{figure}

The length of Tab.~\ref{tab:targets} substantiates the huge effort in the pursuit of \gls{dm}  made by \glspl{iact} over the past decade. Not only target classes have been diversified, but also novel analyses and algorithms have been developed specifically for such searches. In Fig.~\ref{fig:compare} we report some of the most important limits produced by \glspl{iact} so far. It is important to comprehensively discuss this effort.

On the left panel of Fig.~\ref{fig:compare} we report some of the most relevant limits obtained on annihilating \gls{dm} for the $b\bar{b}$ channel. We show the \gls{hess} observation of the \gls{gc} halo adopting an \gls{nfw} profile as a solid line~\citep{Abdallah:2016:halo}, the \gls{magic} stereo observations of the Segue~1 \gls{dsph} in dotted~\citep{Aleksic:2013:segue}, the \LAT{} combined limits from the observation of 15 \glspl{dsph} with 6 years of data together with \gls{magic} Segue I with a dot-dashed line line \citep{Ahnen:2016:dsph}, and the 14.5~h limits obtained with \gls{hess} observation of the Fornax galaxy cluster in dashed line~\citep{Abramowski:2012:fornax}. This figure is emblematic. We see that only searches in an object with a large \gls{dm} concentration and close distance such as the GC ($J-$factor of the order of $10^{21}$ Gev$^2$ cm$^{-5}$) have a chance to skim the thermal relic value at masses about $1-10$~TeV. The second best limits, although possibly less subject to model uncertainties, are those obtained with Segue~1, a very \gls{dm}-dominated \gls{dsph} ($J-$factor of the order of $10^{19}$ Gev$^2$ cm$^{-5}$). Galaxy clusters provide weaker constraints, albeit it must be noted that the corresponding limits were obtained with a much more reduced exposure than the others shown in the same figure. On the other hand, at lower masses, \LAT{} dominates the sensitivity. 

This is not necessarily the end of the game. For instance, there are several mechanisms that can provide a cross-section larger than the thermal one, such as Sommerfeld effects~\citep{Nagayama:2021cwd}, which are activated resonantly for some \gls{dm} masses due to the particles' extremely non-relativistic velocities today.  These resonant phenomena predict that there are some values of mass for which the cross-section may be much larger than what predicted by the thermal limit scale, still consistent with today's observed \gls{dm} density. However, such resonances can appear at virtually any mass. Therefore, indirect detection of \gls{dm} could be `around the corner' for such models. Furthermore, it is important to remember that, should more than one \gls{dm} particle exist, the thermal value would need to be re-discussed. In any case, the \glspl{iact} \gls{dm} annihilation limits obtained so far are the strongest for \gls{dm} masses in the TeV scale.

On the right panel of Fig.~\ref{fig:compare} we report some of the most relevant limits obtained on decaying \gls{dm} with \glspl{iact}. We report the results from 202~h observation of the Perseus galaxy cluster (dashed line) and those obtained with 158~h exposure on the Segue~1 \gls{dsph} galaxy (dotted line) with \gls{magic}~\citep{Aleksic:2011:segue,Aleksic:2010:ngc1275}. We also show, for comparison, the \LAT{} decay limits from the GC~\citep{Ackermann_2012}. One can see that for energies larger than 10~TeV, lower limits at the order of $10^{26}$~s are obtained with \glspl{iact}, and they are competitive with those obtained at lower energies with \LAT{}. Here, there is no reference value to be reached, in contrast to the annihilation case, and in line of principle, any value above the exclusion curve is equally probable. We are not in the position to assess the likelihood of a model.
\bigskip

\begin{figure}[h!t]
  \centering
  \includegraphics[width=0.7\linewidth]{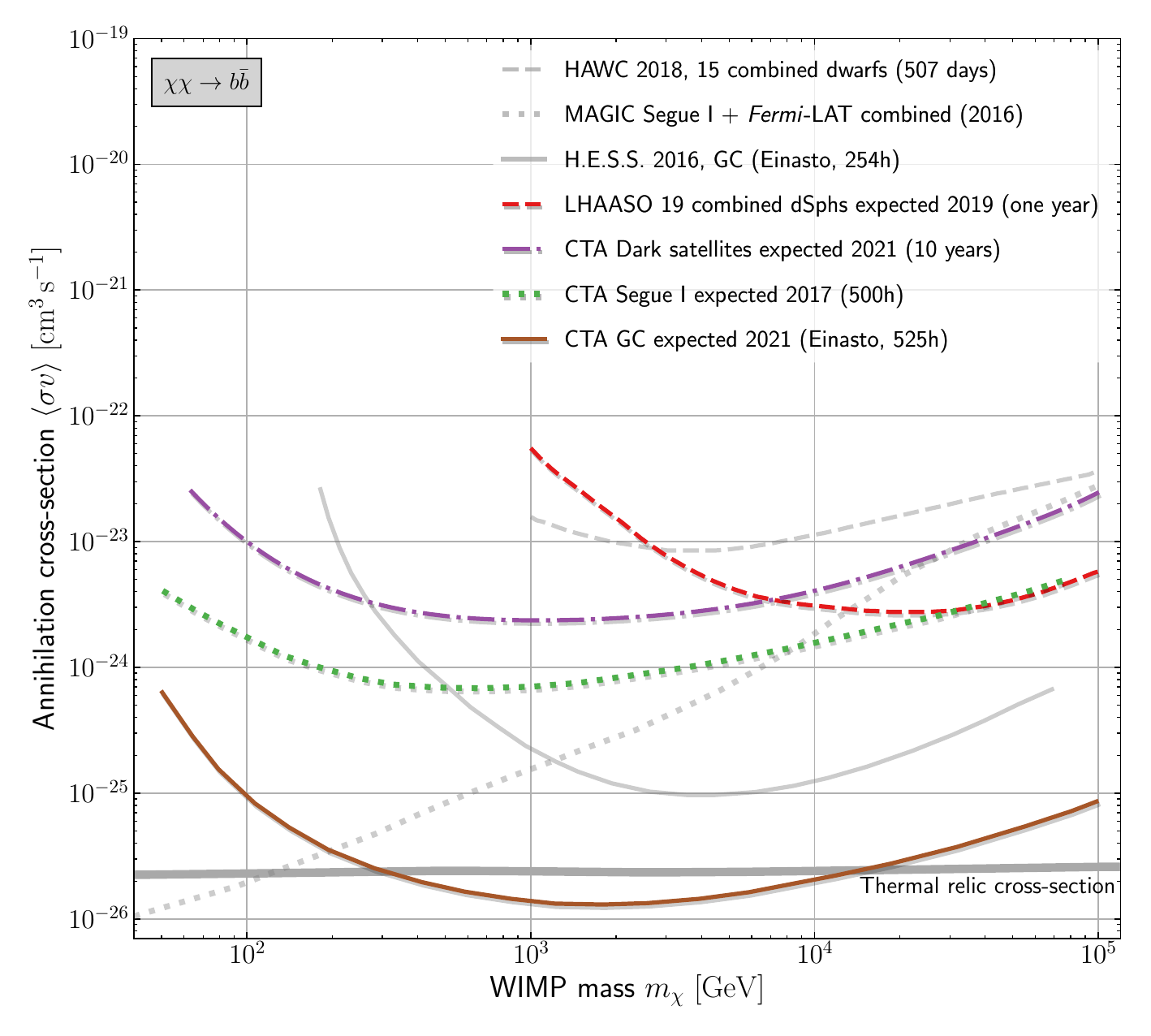}
  \caption{\label{fig:compare2} Selection of some of the most representative sensitivity prospects for \gls{wimp} searches with future gamma-ray detectors, namely, \gls{cta} \citep{Acharya:2017ttl,Acharyya:2020sbj,Coronado-Blazquez:2021avx} and \gls{lhaaso} \citep{He:2019dya}. In shaded colors, a selection of current limits by \glspl{iact} (as shown in Figure~\ref{fig:compare}) and \gls{hawc} \citep{HAWC:2017mfa} are overlaid.} 
\end{figure}

\paragraph{Outlook}
\glspl{iact} have certainly produced unique information about \gls{dm} candidates at the TeV mass scale, with a sensitivity unmatched by any other technique or experiment. Yet, these results should be put into context with the other gamma-ray detectors in the field, operating or planned. Throughout this Chapter, we already presented results from \LAT{}, a pair-production detector hosted aboard a satellite, sensitive to gamma rays from hundreds MeV to hundreds GeV. Data from \LAT{} dominate the \gls{dm} limits shown in Fig.~\ref{fig:compare} at the lowest \gls{dm} masses. Towards the high energy end of the \gls{iact} sensitivity range, we find a different kind of detector class, that of the shower-front detectors that sample the very particles in the extended atmospheric showers. Instruments of this class have been the decommissioned ARGO-YBJ and MILAGRO experiments, now followed by the \gls{hawc} and \gls{lhaaso} \citep{He:2019dya} detectors. All these were or are hosted in the Northern Hemisphere. A project for a Southern Hemisphere shower front detector goes under the name of \gls{swgo}~\citep{BarresdeAlmeida:2020hkh}. The sensitivity of these instruments for \gls{dm} particles  is competitive with currebnt \glspl{iact}, as shown in Fig.~\ref{fig:compare2} for \gls{lhaaso} (dashed line). 
%We also include in the same figure \gls{dm} detection prospects for a 525~h observation of the \gls{gc} halo with \gls{cta} \citep[dotted line,][]{Acharyya:2020sbj} and 500~h observation of the Segue I \gls{dsph} with \gls{cta} \citep[dashed line,][]{Acharya:2017ttl}. 
At lower energies, direct detection experiments hardly reach the sensitivity needed to test \gls{dm} candidates with masses of few hundred GeV, and particle accelerators similarly cannnot access the energy scale of multi-TeV particles. Therefore, \glspl{iact} are crucial to probe the parameter space of heavy \gls{wimp} candidates between a few TeV and the unitary upper bound for \gls{wimp} masses expected at around few hundreds of TeV \citep{Griest:1989wd}.

A huge leap will be made with the \glsfirst{cta} \citep{Acharya:2013sxa}, which will supersede in the coming decade the \glspl{iact} we have discussed in this chapter, \gls{hess}, \gls{magic}, and \gls{veritas}. \gls{cta} will not only have an improved sensitivity, but also both a better energy and angular resolution  compared to current \glspl{iact}. A better sensitivity will obviously boost the performance obtained for the same observation time. In practice, this will significantly increase our chances for detection and, in the absence of it, will translate into more constraining \gls{dm} limits. The improved energy and angular resolutions will allow to provide more robust constraints: the peculiarity of \gls{dm} spectra can be assessed only if the energy resolution is good enough not to wash out their unique properties. An improved angular resolution will allow for a better investigation of the \gls{dm} signal region, especially in those cases for which abundant and complicated astrophysical emission is expected to be found, such as the \gls{gc} or galaxy clusters, or in cases where the spatial extension itself can be a \gls{dm} smoking gun, as expected in dark satellites \citep{Coronado-Blazquez:2019puc}. The envisioned \gls{cta} strategy for \gls{dm} searches is discussed in \citet{Acharya:2017ttl}: a deep survey of the \gls{gc} region,  a dedicated pointing on the best available \gls{dsph} at the time,  and deep surveys of both the Perseus galaxy cluster and the Large Magellanic Cloud are all parts of the so-called \gls{cta} Key Science Programs for the first decade. 

For the \gls{gc} region, a deep scan over 525 hours will be performed in the first three years of operation primarily with \gls{cta} South, covering a circular area of ten degrees in diameter around the \gls{gc}. Additionally, an extended survey with over 300 more hours on a large area North or South of the \gls{gc} region is envisaged \citep{Acharya:2017ttl}. \citet{Acharyya:2020sbj} studied in detail how this wealth of data can be used to probe \gls{wimp} candidates with annihilation cross-sections below the thermal relic value. In \autoref{fig:compare2}, it is displayed how this expectation compares to current limits on \gls{dm} annihilation in the TeV range, with a sensitivity exceeding the best current limit by \gls{hess} by one order of magnitude (solid line). Importantly, \citet{Acharyya:2020sbj} shows that this sensitivity is even guaranteed when accounting for the uncertainty of the instrumental background and astrophysical emissions being present in the data. Furthermore, they found that, even for a cored inner Galactic \gls{dm} density profile, a sensitivity closely reaching the thermal value can be achieved thanks to the large covered area and a template-fitting analysis approach. As a result, \gls{cta} may robustly allow the discovery of TeV \gls{dm} at the canonically expected thermal cross-section. 
In parallel to the observation of the \gls{gc} region, 300 hours of the most promising \gls{dsph} target are guaranteed in the first three years of operation of \gls{cta}. In addition to the interest that such deep observation possesses by itself to search for \gls{dm} in this type of objects, there is one more reason that makes it particularly useful. Indeed, in case of a putative detection of a \gls{dm} signal in the \gls{gc} region, this additional time on a \gls{dsph} may allow to further scrutinize such discovery in a \gls{dm} context by means of other type of target, and to test or exclude possible alternative explanations. In \autoref{fig:compare2}, \gls{dm} prospects are shown assuming the Segue I \gls{dsph} with its currently best estimated $J-$factor and for a total of 500~h exposure time (dotted line). Independently of the success of these initial \gls{dm} searches with CTA, 700 more hours will be dedicated in the first decade of \gls{cta} operations to hunt for \gls{dm} either at the \gls{gc} or at the position of the best \gls{dsph} target \citep{Acharya:2017ttl}.

Not only the \gls{gc} and \glspl{dsph} are viable targets for \gls{cta}.  \citet{2011PhRvD..83a5003B,2016JCAP...09..047H,Coronado-Blazquez:2021avx} also discussed \gls{cta}'s capabilities to search for dark satellites. In contrast to current \glspl{iact}, \gls{cta} may also have a high chance to realistically \textit{discover} these objects. This is not only due to the improved sensitivity, but also to its larger field of view and planned dedicated large-area Galactic-plane and extragalactic surveys \citep{2013APh....43..317D}.  
%Interestingly, they also showed that \gls{cta} could be more sensitive than the LAT to less massive subhalos should the integration angle be wisely chosen according to both the \gls{cta}'s angular resolution and expected subhalo spatial extension. 
\label{sec:subhalos_cta}
While \citet{2011PhRvD..83a5003B,2016JCAP...09..047H} only investigated the individual surveys for such discovery potential, \citet{Coronado-Blazquez:2021avx} recently scrutinized the chances for a serendipitous discovery of a dark satellite in all \gls{cta} data taken within the first decade of \gls{cta} operations. As shown in \autoref{fig:compare2} (dot-dashed line), this strategy could yield a sensitivity competitive to those obtained by a dedicated deep \gls{cta} observation of the most promising \gls{dsph} galaxy to date.

\bigskip
Can we expect novelties beyond the advent of \gls{cta}? Certainly new \glspl{dsph} have been discovered these years thanks to the Dark Energy Survey \citep{Abbott:2018jhe}, and opportunities to identify promising targets for indirect \gls{dm} searches will be greatly increased with the Very Rubin Observatory in the near future \citep{Drlica-Wagner:2019xan}. Possibly, in a decade from now, all or the majority of \gls{mw} \gls{dsph} satellites will be detected \citep{Hargis:2014kaa}. Improved spectroscopic capabilities, like DESI \citep{2016arXiv161100036D} or the Prime Focus Spectrograph recently put in operation at the Subaru telescope \citep{2018SPIE10702E..1CT}  will also allow for more precise member star velocity estimations and, thus, for a more accurate \gls{dm} modeling. Data from the GAIA satellite will also enable a better removal of interloper stars \citep{Helmi:2018okc}. All together, this information should lead to a boost in accuracy of the determination of \gls{dsph} $J-$factors. It is also still possible to find a \gls{dsph} exhibiting a higher $J-$factor than any other \gls{dsph} currently known. It was shown in \citet{Doro:2017vjf} that current \glspl{iact} may reach the thermal cross-section value if they were able to observe for 500~h a \gls{dsph} with a $J-$factor five times the one of Segue 1. %Such a \gls{dsph} could indeed exist, at least as of predictions from N-body simulations. 
Another possibility is that this $J-$factor was reached by a dark satellite serendipitously detected, both massive  and close enough to the Earth.\footnote{In this case, however, it may be difficult to assess the underlying \gls{dm} particle physics model until we have some knowledge on its exact distance and mass by other means.} 
Also, multi-wavelength approaches complementing \gls{iact} \gls{dm} searches from the radio band to MeV gamma rays are more and more identified as very useful. Data from other wavelengths either may set strong constraints on heavy \gls{dm} candidates on their own \citep{Regis:2021glv}, or allow to improve the  \gls{vhe} gamma-ray analysis by providing additional information \citep[as e.g. for \LAT{} in][]{DiMauro:2021qcf}.

%Below the GeV energy range, some predictions have also been made for future MeV missions like e-ASTROGAM \citep{2018PDU....21....1C}. Above all, these predictions highlight the important role that an enhanced angular resolution could play in this kind of searches: the dark satellites with the highest annihilation fluxes are expected to be extended, in contrast to what happens with common gamma-ray sources, e.g., AGNs. Thus, any unID showing a significant spatial extension may be a particularly good \gls{dm} subhalo candidate.

\bigskip
All in all, the next decade will gather great expectations for the discovery of TeV \gls{wimp} \gls{dm} or in challenging the paradigm of massive particle \gls{dm}. Even if particle \gls{dm} may be first discovered in colliders or direct detection experiments, an accurate \gls{dm} \emph{identification}
and \emph{characterization} will only be possible by combining
signatures from different detectors and methods~\citep{Cahill-Rowley:2014boa}. A signature in telescopes is furthermore necessary to establish the definitive link between a potential \gls{dm} signal at the lab and the long sought \gls{dm} in the Universe. To this aim, \glspl{iact} to study the properties and budgets of massive particle \gls{dm} in astrophysical reservoirs play a crucial role, already today and all the more in the future.

\section*{Verification}
\begin{enumerate}[start=1,label={Q8.\arabic*:},wide=0pt, leftmargin=3em]
    %\item Q1. How can the observed apparent lack of \gls{dm} in some galaxies be an argument in favor of \gls{dm} in contrast to modified gravitational theories to explain the ``\gls{dm} problem''?
    \item Why the expected gamma-ray spectrum of \gls{dm} annihilation is \textit{exactly} cut-off at the rest mass of the \gls{dm} particle (at half the rest mass for decay)?
    \item Why does one not search for \gls{wimp} relic annihilations or decays of local \gls{dm} particles in a detector?
    \item According to the canonic scenario of thermal \gls{dm} freeze-out in the early Universe, would there be less or more \gls{dm} left over today if the annihilation cross-section was larger than $\langle \sigma \mathit{v} \rangle\sim 10^{-26}$~\svunits?
    \item After first results of \gls{iact} \gls{dm} searches were published with few hours of data, why improved limits from the same instruments were published not before hundreds of more hours had been accumulated?
\end{enumerate}

%\newpage
\section{Axion-like Particles}
\label{sec:alp}

%\paragraph{Introduction: axions and the gamma-ray connection.}
As of today, the so-called Peccei-Quinn (PQ) mechanism -- proposed in the late '70s of the past century \citep{1977PhRvL..38.1440P} -- represents the most convincing solution to the strong Charge-Parity (CP) problem in Quantum Chromodynamics (QCD), i.e., the absence of CP violation in strong interactions. It was soon realized that a by-product of the proposed PQ mechanism is the existence of a pseudo-Nambu-Goldstone boson called {\it axion} \citep{1978PhRvL..40..223W,1978PhRvL..40..279W}.\footnote{The name {\it axion} is typically attributed to Frank Wilczek, who supposedly took it from a laundry detergent alluding to the fact that the new particle would ``remove a stain''.} Very interestingly, these QCD axions may account for a fraction or the whole (cold) DM content in the Universe and, indeed, constitute one of the most valid alternatives to \glspl{wimp} at present. Contrary to \glspl{wimp} though, the axion would be very light, as its mass has already been experimentally constrained to be $\ll$eV. In fact, axions represent the most famous exponent of the whole class of so-called weakly interacting slim particles (WISPs) (see e.g. \citet{Arias2012} for a recent review). 

One interesting property of these axions, particularly relevant for the purposes of this book, is a predicted interaction with photons in the presence of external magnetic or electric fields described by the following Lagrangian \citep{1978PhRvD..18.1829D,1983PhRvL..51.1415S}:
\begin{equation}
{\cal L}_{a \gamma} = -\frac{1}{4}~g_{a\gamma}~F_{\mu \nu}\tilde{F}^{\mu \nu}a = {g_{a\gamma}}~{\bf E \cdot B}~a, 
\label{eq:lagrangian}
\end{equation}

where $a$ is the axion field, $g_{a\gamma}$ is the coupling strength, $F$ is the electromagnetic field-strength tensor, $\tilde{F}$ its dual, ${\bf E}$ the electric field, and ${\bf B}$ the magnetic field. For the QCD axion, the mass is proportional to the coupling strength. Yet, there are other, more general states arising from extensions of the SM, for which this relation does not hold, frequently known as \glspl{alp} because they still share many similarities with the QCD axion \citep{1978PThPh..59..274C,1986MPLA....1..541L,1982PhRvL..49.1549W,1984PhLB..149..351W,2006JHEP...05..078C}. In string theory, for example, it is possible to properly formulate the entire PQ mechanism, which leads to predict the existence of many of these ``generalized'' axions or ALPs, the QCD axion being just a particular case \citep{1984PhLB..149..351W,2006JHEP...05..078C,2010PhRvD..81l3530A,2012JHEP...10..146C}.
%many of these ``generalized'' axions, frequently called axion-like particles (\glspl{alp}) because they still share many similarities with QCD axions, are predicted by string theories once the Peccei-Quinn mechanism is properly formulated within the boundaries of these theories, the QCD axion being just one of them \citep{19rev,20rev,21rev,22rev}.

\begin{figure}[!ht] 
    \centering
    \includegraphics[width=0.3\linewidth]{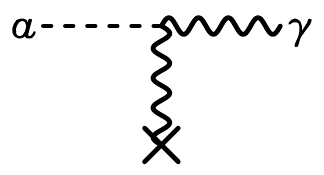}
    \caption{Feynman diagram showing the interaction of photons and \glspl{alp} in the presence of an external electric or magnetic field. This interaction obeys the Lagrangian in Eq.~(\ref{eq:lagrangian}). }
    \label{fig:feynman}
\end{figure}

The interaction with photons described by Eq.~(\ref{eq:lagrangian}) represents the main vehicle used to experimentally search for axions and \glspl{alp} up to present \citep{2005PhRvL..94l1301Z,2006PhRvD..74a2006D,2007JCAP...04..010A,2010PhRvL.105q1801W,2017NatPh..13..584A,2018PrPNP.102...89I}. This interaction results in conversions or {\it oscillations} of photons into \glspl{alp} and vice-versa; see Fig.~\ref{fig:feynman}. This same property of axions and \glspl{alp} can be used to look for them by means of astrophysical observations as well, whose main observable is precisely photons. Indeed, should \glspl{alp} and photons couple to each other, significant distortions in the spectra of astrophysical sources may be induced that could be measured, e.g., in the gamma-ray energy range with current or future IACTs.\footnote{Note that we only refer to \glspl{alp} here: the search for QCD axions is in principle not possible in gamma rays given the too small couplings expected at the axion masses that can be probed by gamma-ray telescopes (see, however, \citet{2017JHEP...01..095F} for ``boosted'' photon/axion couplings). This will be further discussed later in the section.} Thus, photon-ALP conversions may potentially have very important implications for gamma-ray astrophysics and, very excitingly, \glspl{alp} offer the possibility to search for new physics with gamma rays. Even in the absence of ALP-induced spectral features in gamma-ray data, a null result might be used to provide valuable insight, i.e, competitive experimental constraints, on the precise ALP nature.

%

%\md{Yo quitaria este parrafo de abajo, que iria bien en un paper en si, pero aqui interrumpe el hilo del tema. Ademas, quizas la estructura se hace en la introduccion, pero aqui no.}
%\strike{In the following sections, we will describe in further detail the main characteristics of these ALP-induced spectral features and what astrophysical environments/objects are, and why, the best targets to look for ALP/photon mixing in the energy range covered by IACTs. We will then review the IACT efforts already done in this direction and how these IACT observations have been used, in the beginning to claim hints of ALP imprints in gamma-ray data, later to set limits on the relevant ALP parameter space (ALP mass versus coupling constant). We will continue with a discussion of these limits in a broader axion/ALP context, and will end up by looking at the future of ALP searches with current and future IACTs.}

%\subsection*{ALP/photon conversions in the gamma-ray Universe}
%\paragraph{ALP/photon conversions in the gamma-ray Universe.}
ALP-induced imprints in the spectra of gamma-ray sources are expected to happen around and above a particular critical energy for conversion that, in convenient units for gamma-ray experiments, can be expressed as \citep{Hooper2007}\

\begin{equation} 
E_{\text{crit}} \equiv \frac{m_a^2}{2~g_{a\gamma}~B} \sim 2.5~{\rm GeV}~\frac{m^2_{\mu {\rm eV}}}{g_{11}~B_{\text{G}}},
\label{eq:ecrit}
\end{equation}

\noindent where the sub-indices refer to dimensionless quantities: $m_{\mu eV} \equiv m_a/ \mu {\rm eV}$, $g_{11} \equiv g_{a\gamma}/10^{-11}$ GeV 
%$^{-1}$
and $B_{\text{G}} \equiv  B/{\rm Gauss}$; $m$ is an effective ALP mass $m^2 \equiv |m_a^2-\omega_{\text{pl}}^2|$, with $m_a$ the ALP mass and $\omega_{\text{pl}}=0.37 \times 10^{-4} \mu {\rm eV} \sqrt{n_e/{\rm cm}^{-3}}$ the plasma frequency of the medium under consideration ($n_e$ being its electron density). Recent results from the CAST experiment \citep{2017NatPh..13..584A} provide an upper limit of $g_{11} \leq 6.6$ for ALP masses $m_a \leq 0.02$~eV. At present, this bound represents the most general and stringent limit in this ALP mass range. 

Just around this $E_{\text{crit}}$ an oscillatory spectral pattern is expected whose exact properties will depend on both the morphology and orientation of the magnetic field along the direction of propagation of the ALP. The mixing between photons and \glspl{alp} becomes maximal above $E_{\text{crit}}$ and independent of the energy, the probability for conversion being just proportional to $(g_{a\gamma}\times B)^2$. This is the so-called {\it strong mixing regime}. A modification of the observed photon flux is then expected within this energy interval, because photons convert into \glspl{alp} and vice-versa. The mixing remains maximum up to an energy $E_{\text{max}} \sim 2.2 \times 10^6$~GeV~g$_{11}$~B$_{\mu G}$, where birefringence effects heavily start to suppress the conversions. We note that other effects might significantly alter this general picture. For instance, at TeV energies, \citet{2017JCAP...01..024K} recently noticed that photon-photon refraction is expected to diminish the expected photon-ALP maximal mixing, this attenuation being more and more efficient as the energy increases.

The probability for photon-ALP conversion is proportional to the modulus of the magnetic field {\it times} the physical size, $s$, where this field is confined. More precisely, it is \citep{Hooper2007,2009PhRvD..79l3511S}:

\begin{equation}\label{eq:prob2}
P_{a\gamma} =\frac{1}{1+(E_{\text{crit}}/E_{\gamma})^2}~
\sin^2\left[\frac{B~s~g_{a\gamma}}{2}\sqrt{1+\left(\frac{E_{\text{crit}}}{E_{\gamma}}\right)^2}\right].
\end{equation}

\begin{figure}[!ht] 
    \centering
    \includegraphics[width=0.6\linewidth]{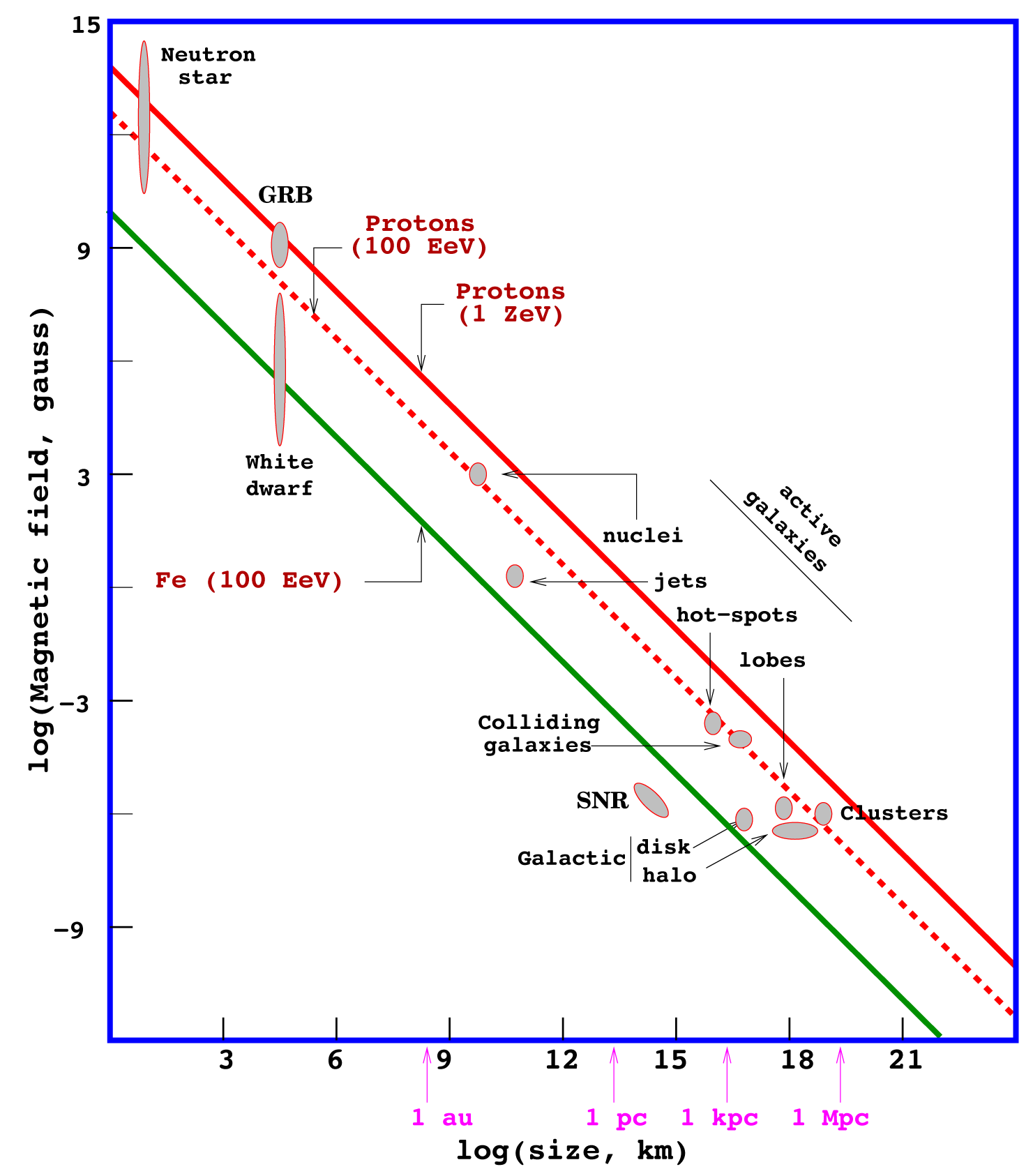}
    \caption{The Hillas criterium establishes a relation between physical sizes and magnetic fields in astrophysical objects/environments so that an efficient acceleration of ultra-high energy cosmic rays is possible. This correspondence can also be used as a very good proxy to understand where in the Universe an efficient conversion between photons and axions/\glspl{alp} can be expected. Figure taken from \citet{Hooper2007}.}
    \label{fig:hillas}
\end{figure}

This means that an efficient conversion can happen not only in relatively small regions of the Universe exhibiting very strong magnetic fields (e.g., AGNs, neutron stars) but also along large physical volumes filled with weak magnetic fields (e.g., the intergalactic medium between galaxies, the intra-cluster medium in galaxy clusters or the interstellar medium in a galaxy like our own). Very interestingly, this particular ``property'' is very closely connected to the so-called {\it Hillas criterion} which is required to accelerate high-energy cosmic rays in astrophysical objects (see Fig.~\ref{fig:hillas}), in such a way that this criterion can be actually used as a good proxy for the identification of the best astrophysical targets for ALP searches. 

The precise effect of photon-ALP conversions on gamma-ray spectra can be very diverse, as diverse are the possible astrophysical scenarios that potentially allow for efficient conversions; thus, a detailed, dedicated analysis must be done for each particular physical case. At first sight, it might seem that photon-ALP conversions would always lead to an attenuation of the gamma-ray flux at Earth, as a fraction of the emitted photons would convert into \glspl{alp} and therefore would never reach the Earth. Yet, reconversions of \glspl{alp} into gamma-ray photons along the line of sight are  possible, which would alter the level of attenuated flux. An additional ingredient needed for an accurate prediction is the precise role of the so-called extragalactic background light (EBL). This EBL -- the result of the light emitted by galaxies over the entire cosmic history -- permeates the whole Universe and induces an attenuation of the gamma-ray flux by pair production, i.e., $\gamma \gamma^{EBL} \rightarrow e^-e^+$. 
The attenuation is maximal for a background photon energy $\epsilon \sim~(500$ GeV/E) eV, which means that infrared/optical background photons are the ones responsible for the attenuation in the energy range specific to IACTs. The inclusion of the EBL in the computations of photon-ALP conversion leads to very interesting predictions. Indeed, for distant VHE sources, ALP conversions in the intergalactic medium can lead to either an attenuation or an enhancement of the observed flux at Earth, depending on the exact configuration of the magnetic fields, energies considered, and source distance. An enhancement is possible because \glspl{alp} may travel unimpeded through the EBL and, only at later times before reaching the observer, some of them would convert back into gamma rays, for instance in the magnetic field of our galaxy. Intriguingly, this might potentially allow us to observe gamma-ray emitters located at distances larger than those expected from conventional EBL models. Additionally, it is interesting to note that photon-ALP conversion might also happen at the sources themselves, in this way providing an alternative scenario to avoid absorption of $\gamma$-rays in the dense ultraviolet radiation field of AGNs, see e.g.~\citet{2012PhRvD..86h5036T}.

Note that none of the above given arguments explain the reason to perform the astrophysical ALP search particularly in the gamma-ray energy band and not at any other photon wavelengths. Indeed, the latter is certainly possible and has already been done in some works (see e.g. \citet{ 2013ApJ...772...44W,2017ApJ...847..101B} for X-ray astrophysical searches). When typical values of the astrophysical magnetic fields strengths are considered and the CAST upper bound is taken into account, photon-ALP conversions are expected to show up in the $\sim$ GeV-TeV range provided that the ALP mass is ultra light, i.e. $neV \leq m_a \leq \mu eV$; see Eq.~(\ref{eq:ecrit}). %Namely, gamma-ray observations allow us to probe ultra-light \glspl{alp} down to $\sim .

\paragraph{Current status of ALP searches with IACTs.} 
From an historical perspective, the event that triggered a rapidly increasing interest for \glspl{alp} within the gamma-ray astrophysics community was, very likely, the claim by the PVLAS collaboration back in 2006 of the observation of optical rotation generated in vacuum by a magnetic field -- possibly associated to a neutral, light boson \citep{2006PhRvL..96k0406Z}. Certainly, should such a light boson exist, it might have important astrophysical and cosmological consequences that could be tested \citep{1995PhRvL..75.2077F,2008LNP...741...19S,2008LNP...741..115M,2010PhRvD..81f3508V,2010PhRvD..82l3508W,2015JCAP...05..050A,2008PhLB..659..847D}. The PVLAS claim was excluded soon after an upgraded setup \citep{2008PhRvD..77c2006Z}; however the whole effort aimed at looking for ALP-induced features in the spectra of gamma-ray sources had already been boosted. The idea of using astrophysical observations to set constraints on ALP parameters was certainly not new and it had been already exploited well before 2006, e.g.~\cite{1996PhRvL..77.2372G,1996PhLB..383..439B,2002PhRvL..88p1302C,2003JCAP...05..005C,2005PhRvD..72b3501M}; yet the mentioned PVLAS claim first and, later, some astrophysical (gamma-ray) hints of photon-ALP conversions, as described below, generated a momentum in the community that stands until present time.

As said previously, efficient photon-ALP conversions may happen in those astrophysical environments that exhibit the right combination of magnetic field strength and size of the magnetized region. Indeed, prospects studies have been published that predict the size and shape of the expected ALP-induced spectral features due to conversions in or around the gamma-ray emitters themselves \citep{Hooper2007,2007PhRvD..76l3011H}, within galaxy clusters \citep{2012PhRvD..86g5024H}, the intergalactic medium along cosmological distances \citep{2007PhRvD..76l1301D,2008LNP...741..115M,2009JCAP...12..004M}, our own galaxy \citep{2008PhRvD..77f3001S}, or a combination of these scenarios \citep{2009PhRvD..79l3511S}. With these predictions at hand, different approaches to look for \glspl{alp} with IACTs are possible; however all of them will be most likely based on the search and analysis of a systematic residual after applying the best-fit, conventional model to the available gamma-ray data. Only if we found significant departures with respect to this best-fit model, and after having discarded other possible conventional sources of error for such residuals, one might seriously consider it as a signal of exotic physics. On the contrary, the absence of any significant deviations with respect to the conventional baseline model can be used to set constraints on ALP parameters in such a way that all those combinations of ALP parameters that should have yielded a measurable effect in the data can now be excluded. 

\begin{figure}[!ht] 
    \centering
    \includegraphics[width=0.8\linewidth]{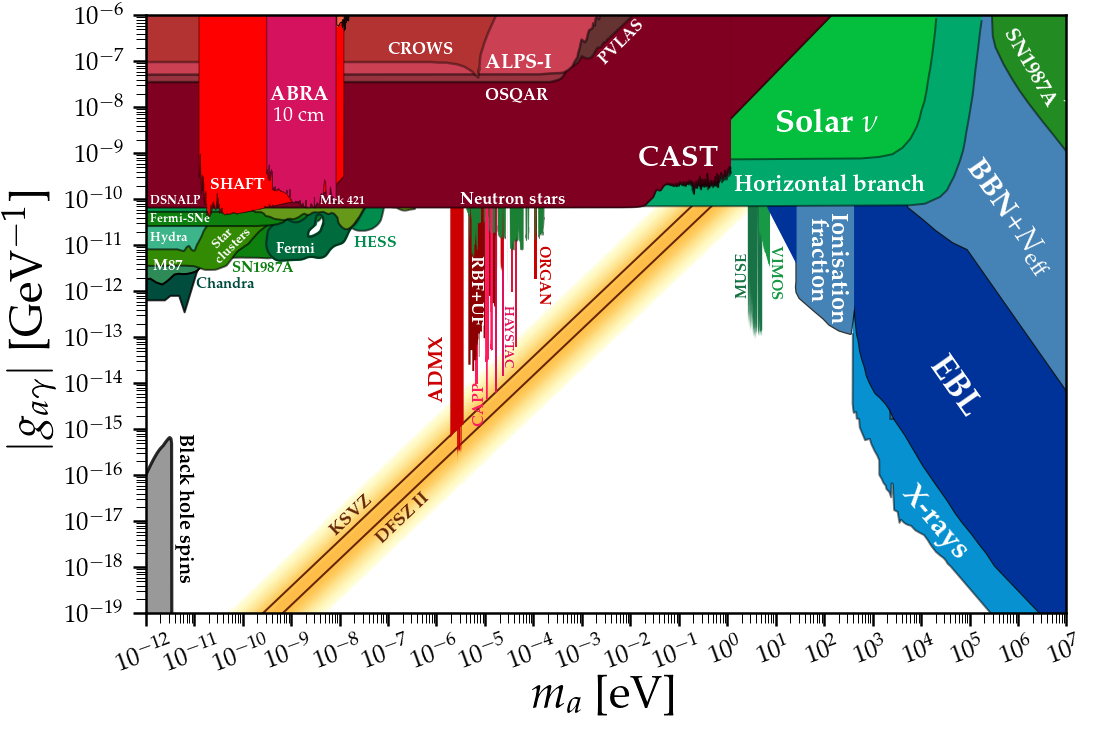}
    \caption{Summary of current constraints to the ALP-photon coupling. Results obtained by H.E.S.S. are displayed in the upper left regime of the displayed parameter space, for ALP masses around $0.1~\mu$eV. Figure taken from \url{https://cajohare.github.io/AxionLimits/} (version as of June 11, 2021).}
    \label{fig:ALPsummaryIACT}
\end{figure}

Along the years, two competing but complementary data analysis strategies were slowly arising as a consequence of both the predicted features and the matureness of the field: one aimed at statistically detecting the predicted spectral oscillations or {\it irregularities} around the critical energy for photon-ALP conversion; the other focused on the strong mixing regime, where the ALP-induced effect on the gamma-ray spectra is expected to be maximal. In most cases, the involved gamma-ray analyses were used to set limits on the ALP parameter space in the absence of any clear ALP-induced spectral trace in the data.\footnote{See, however, the claims made and briefly described below in this same section.} The obtained limits are nicely complementary to those achieved in the lab by other means, both in terms of the masses and of the couplings being tested. This is illustrated in Figure \ref{fig:ALPsummaryIACT}, which shows the current summary of limits obtained in the region of the ALP parameter space relevant to IACTs. In particular, the H.E.S.S. Collaboration searched for ALP-induced irregularities in the spectrum of PKS 2155-304, an AGN located at $z=0.116$ \citep{2013PhRvD..88j2003A}. The authors of this work considered photon-ALP conversions happening both in the intergalactic medium between us and the source, and those in the intra-cluster magnetic field where the AGN resides. Since the magnetic field strength in each of these astrophysical scenarios is different by orders of magnitude (nG versus $\mu$G, respectively), according to Eq.~(\ref{eq:ecrit}) the data analysis will lead to two different sets of limits in the ALP parameter space. This can be seen in Fig.~\ref{fig:ALPsummaryIACT}. The most competitive upper limits are obtained for the case of the intra-cluster medium, and are as good as $g_{11} < 2.1$ for $15 < m_{neV} < 60$ (at 95\% C.L.). 
%More recently, \citet{Liang:2018mqm} compiled the spectra of ten bright Galactic TeV sources observed by H.E.S.S. (mostly pulsar wind nebulae and supernova remnants) and searched for ALP-imprinted spectral oscillations in the data as well. In the absence of these oscillations in a combined analysis of these sources, their results exclude $g_{11} < 2.5$ at ALP masses around $0.1-0.2~m_{\mu eV}$ (at 95\% C.L.). 

Perhaps surprisingly, the H.E.S.S. collaboration so far is the only IACT experiment who has published ALP limits from their proprietary data, namely, on PKS 2155-304 \citep{2013PhRvD..88j2003A}. In addition, using public H.E.S.S. data, \citet{Liang:2018mqm} have set limits on the ALP parameter space from a compilation of ten bright Galactic sources. Similarly, \citet{Guo:2020kiq} have combined public data from H.E.S.S. and \LAT{} on PKS 2155-304 and another extragalactic object (PG 1553+113) for a study of ALP implications.\footnote{A similar work by \citet{2018arXiv180504388M} focuses on the AGN labeled as NGC1275 in the Perseus galaxy cluster by a joint analysis of \LAT{} and public MAGIC data. 
%Indeed, very recently a work appeared that makes use of both {\it Fermi}-LAT and MAGIC (public) data to perform a joint ALP search in the energy range covered by both instruments \citep{2018arXiv180504388M}. 
The latter result 
%-- although preliminary and not yet formally published in a peer-review journal --  
suggests that those results in \citet{2016PhRvL.116p1101A} can be improved by roughly an order of magnitude under an optimistic, yet uncertain, configuration of the magnetic field. 
%We note that this work could be easily improved by the MAGIC collaboration by performing a dedicated analysis at high energies using their own, private data containing precise information on every detected event.
Yet, we note that this work has not been accepted as refereed article as of today.} \cite{2012PhRvD..86h5036T,Galanti:2015rda} have studied an ample sample of blazar spectra at VHE energies seen by \LAT{}, H.E.S.S., MAGIC, and VERITAS, as did similarly \cite{Xia:2019yud} with combined public data from \LAT{} and H.E.S.S., MAGIC, and VERITAS from supernova remnants to search for irregularities in the gamma-ray spectra.
Although not strictly in the VHE band, other limits exist that were obtained with gamma-ray data as well, namely those using {\it Fermi}-LAT observations at lower, but partially overlapping energies than those covered by IACTs. In particular, a search for spectral oscillations in the spectrum of NGC1275, an AGN located at the center of the Perseus galaxy cluster, was used to set the most competitive ALP constraints derived from gamma rays as of today \citep{2016PhRvL.116p1101A}. These limits exclude couplings $g_{11} > 0.5$ at ALP masses $0.5 \lesssim m_a \lesssim 5$ neV (95\% C.L.) under the default (conservative) scenario considered by the authors to describe the configuration of the involved magnetic fields: the one in the galaxy cluster hosting the AGN and the one in our galaxy. %A subsequent, very similar work focused on the same object and using the same exact data analysis procedure to set their limits but using $\sim$25\% more LAT observation time, ended up with slightly better limits.
Additional works using Fermi-LAT data to set constraints on the ALP parameter space are the ones by \citet{2018PhRvD..97f3003X} and \citet{2018PhRvD..97f3009Z}. In the former, the authors used three supernova remnants as targets and reported a suspicious oscillatory behaviour of their spectra at small ($\sim4.2\sigma$) significance. The uncertainties in the measurement do not allow extract robust conclusions and, instead, the authors speculated that this result could also be caused by different parts of the remnant contributing to the observed emission. In \citet{2018PhRvD..97f3009Z}, the authors focused on PKS 2155-304 and set competitive limits with respect to those in \citet{2016PhRvL.116p1101A}. Yet, their results are largely affected by important uncertainties in the modelling of the involved magnetic fields and, thus, the derived results should be taken with some caution. More recently, \cite{2021PhRvD.103h3003L} and \cite{2021arXiv211013636L} searched for ALP-photon oscillations in the gamma-ray spectra of blazars to set competitive ALP constraints as well, in the former work using Fermi-LAT and ARGO-YBJ data of Mrk~421; in the latter combining Fermi-LAT and MAGIC observations of both Mrk~421 and PG~1553+113. Their constraints turn out to be nicely complementary to those in \citet{2016PhRvL.116p1101A} in the ALP mass range between a few $\mu$eV to a few neV (95\% C.L.).

In addition to setting meaningful limits on the ALP parameter space, some hints of the presence of \glspl{alp} were claimed along the years using gamma-ray data. Already back in 2007, \glspl{alp} with masses $<<10^{-10}~eV$ were invoked to explain alleged anomalies in the flux of distant AGNs as measured by IACTs \citep{2007PhRvD..76l1301D}. At that time, the Universe seemed less opaque to gamma rays than predicted by the latest available EBL models; e.g.~\citet{2006Natur.440.1018A,2008Sci...320.1752M,2009ApJ...691L..91S,2012ApJ...758L..13H}. This so-called ``transparency hint'' within the ALP and VHE communities, pointing to ultra-light \glspl{alp} as a possible (exotic) solution to the transparency problem, was reinforced a few years later by a statistical analysis of half a hundred of VHE sources \citep{2012JCAP...02..033H}. This work found an apparent suppression of the pair production process at a $4.2\sigma$ level; in other words, that there was a systematic overcorrection of the intrinsic VHE spectra of cosmological sources by EBL absorption, especially at the highest attenuations. The same study showed that ultra-light \glspl{alp} could represent a viable explanation for such anomaly as well. In a subsequent work \citep{2013PhRvD..87c5027M} the same authors calculated the ALP parameter values that were needed to account for the apparent VHE flux anomalies measured by IACTs, i.e., they shaped a specific contour in the ALP parameter space for the mentioned ``ALP transparency hint''. Despite these early and subsequent works supporting both a too low level of EBL density and an ALP explanation for it (e.g., \citet{2009MNRAS.394L..21D,2011PhRvD..84j5030D}), this hint of the existence of \glspl{alp} has not survived over time. Nowadays, it is generally accepted that state-of-the-art EBL models are in good agreement with gamma-ray data, i.e., that the measured transparency of the Universe to VHE gamma rays is as expected \citep{2011PhRvD..83f3005B,2012Sci...338.1190A,2013A&A...550A...4H,2013A&A...554A..75S,biteauwilliam,2015ApJ...813L..34D} and, consequently, that there is no need to invoke the existence of \glspl{alp} in this context. In parallel, also the ``transparency hint'' region itself in the ALP parameter space has considerably shrunk since it was first proposed as a consequence of more recent ALP searches, which have, as already described above, imposed tight and robust constraints in this same region.

But not only the ``ALP transparency hint''; also other potential issues with conventional physics were reported in the VHE regime that might point to \glspl{alp}. To cite some, the hard spectra exhibited by some AGNs seemed difficult to explain by conventional emission mechanisms \citep{2011PhRvD..84j5030D}; the intrinsic spectrum of a few AGNs showed a marginal deviation at the highest measured energies with respect to the expected power-law behavior (the so-called ``pile-up problem'';  \citet{2011JCAP...11..020D}); the existence of spectral hardening precisely at those energies where the EBL absorption is essential as well as its significant ($12\sigma$) unexpected dependence with redshift \citep{2014JETPL.100..355R}; the existence of spectral breaks at a few GeV \citep{2013JCAP...11..023M}, the energy spectrum of some Galactic pulsars \citep{2018JCAP...04..048M}, etc. These apparent puzzles -- although inconclusive and possible to explain within the boundaries of conventional VHE physics, e.g., particles accelerated at relativistic shocks \citep{2007ApJ...667L..29S}, internal photon-photon absorption \citep{2008MNRAS.387.1206A}, secondary gamma rays produced along the line of sight by cosmic-ray protons interaction with background photons \citep{2010APh....33...81E,2011ApJ...731...51E,2012ApJ...751L..11E,2013PhRvD..87f3002A} -- gradually stacked up to offer the ultra-light ALP as a single, viable and very appealing explanation able to simultaneously resolve all of them. As a result, the momentum to perform the ALP quest in gamma rays persisted along the years and, indeed, continues intensively at present.

\paragraph{Future of ALP searches with IACTs.}  There is definitely room for further work on ALP searches in gamma rays, even with current IACTs. As described above, up to now only the H.E.S.S. Collaboration have published IACT data to search for \glspl{alp} and, in the absence of a signal, to set constraints on the ALP parameter space. Yet, both MAGIC and VERITAS are equally suited to perform similar studies in the Northern Hemisphere. Even more, archival data taken by these telescopes over the last years could be easily used as well to search for ALP-induced spectral features. A good example is the deep 400h observation of the Perseus galaxy cluster recently accomplished by the MAGIC telescopes. The initial goal of this observational campaign -- the longest  ever done by an IACT for a single object -- was to look for the long-awaited first detection of cosmic-ray induced gamma-ray emission from clusters \citep{2016A&A...589A..33A} and, more recently, to set limits on DM decay \citep{2018PDU....22...38A}. Since these observations include NGC1275 and IC310 in the field of view, two AGNs sitting in the Perseus cluster and observed at a very high significance with MAGIC, the data possess all that it is necessary to search for spectral irregularities in a similar way than what was done by the {\it Fermi}-LAT collaboration in \citet{2016PhRvL.116p1101A}. We remind that this LAT work, which also adopted Perseus as target, provided the best ALP limits derived from gamma ray observations so far. Another potential line of action to be further exploited by current IACTs can be provided by the measurement of new photons exhibiting very high opacities.\footnote{These may be possible to obtain from both new discoveries of sources located at cosmological distances and/or for already known sources observed at even higher energies.} A statistical analysis combining this hypothetical new data with what has been already observed, in a similar way to that done by \citet{2012PhRvD..86g5024H,2013PhRvD..87c5027M,biteauwilliam}, could provide further insight on the role of \glspl{alp} on photon propagation through cosmological distances.

Beyond current IACTs, CTA, with its $\sim$10 times better sensitivity with respect to current IACTs, should make it possible to test the photon-ALP conversion scenario at an unprecedented sensitivity in the energy range between a few tens of GeV up to a few dozen TeV \citep{2013arXiv1305.0252S}. Indeed, a 300h CTA observation of the Perseus galaxy cluster is foreseen during the first five years of CTA operations \citep{2019scta.book.....C}, which could be used to improve upon existing limits over a wider ALP mass range. A dedicated sensitivity estimate for gamma rays from NGC1275 inside the Perseus cluster  shows that CTA could reach couplings as low as $g_{11} \gtrsim 0.1$ for masses around 10 neV \citep{Abdalla:2020gea}. In this way, \gls{cta} may become the most sensitive instrument to \glspl{alp} at these masses, largely exceeding laboratory-based dedicated experiments such as CAST \citep{2005PhRvL..94l1301Z} or ALPS II   \citep{Bahre:2013ywa} and even competing with the future IAXO \citep{2013JPhCS.460a2002I} at masses 5 neV $\lesssim m_a \lesssim$ 100 neV. These prospects were based on the search for ALP-induced spectral irregularities around the critical energy for conversion. In addition to the already discussed overlap and complementarity with \LAT{}, other next-generation instruments will complement CTA particularly in its low energy end as well, such as LHAASO \citep{Bai:2019khm,Long:2021udi}, or the future eASTROGAM \citep{2017arXiv171101265D} or AMEGO \citep{2018AAS...23141902H} missions. By design, these next-generation experiments are expected to possess superb spectral capabilities in the search for ALP-induced spectral features. Indeed, preliminary results focusing on photon-ALP conversions in the Perseus galaxy cluster \citep{2018AAS...23135549C} show that these instruments could be sensitive to couplings $g_{11} > 0.3$ in the approximated ALP mass range $0.5 \lesssim m_a \lesssim 500$ neV. 
Preliminary prospects work was also done that focused on using the spectral hardening of distant gamma-ray sources instead \citep{2013arXiv1305.0252S,2013APh....43..189D}. This was further explored in detail for CTA by \citet{2014JCAP...12..016M} by adopting a scenario consisting of four bright gamma-ray sources that are combined under the same consistent likelihood analysis. From this work, a significant detection of an ALP signal could be possible for ALP masses below 100 neV and couplings $g_{11} > 2$. At the highest energies covered by CTA, not only CTA but also HAWC \citep{2017ApJ...843...39A} and HiSCORE \citep{2014APh....56...42T}, working in a complementary (even higher) energy range, will be much more sensitive to high energy photons from distant gamma-ray emitters compared to current IACTs. This increased sensitivity will allow to test at a higher confidence if the once claimed higher transparency of the Universe to gamma rays than expected from current EBL models is on solid grounds or not and, if so, whether \glspl{alp} could still represent a viable explanation. These and other instruments covering the VHE range up to tens to hundreds of TeV could also be used to search for spectral oscillations as well \citep{2018JHEAp..20....1G}.
%2017arXiv171201839V

\begin{figure}[!ht] 
    \centering
    \includegraphics[width=0.8\linewidth]{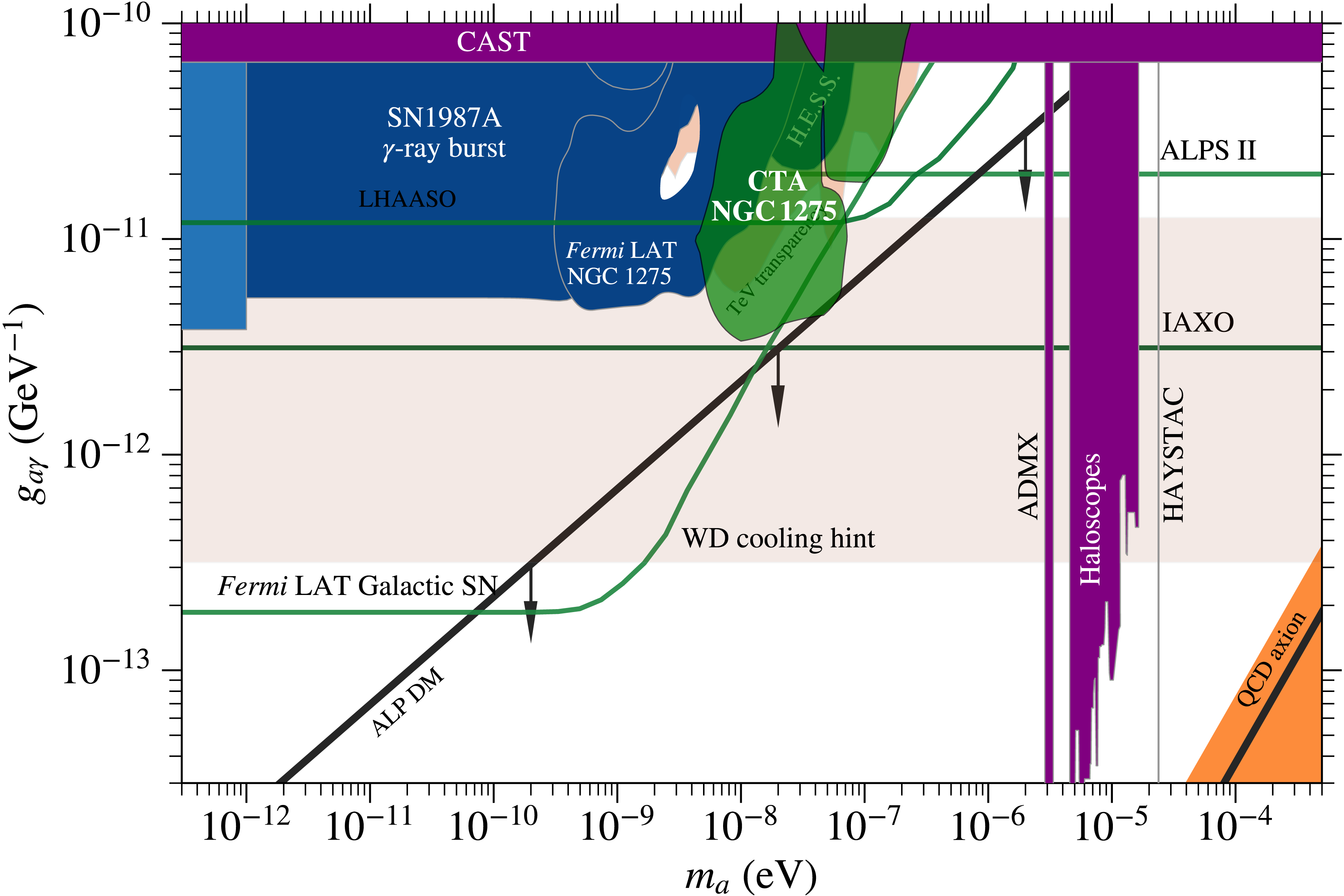}
    \caption{Future ALP search prospects for different experiments and astrophysical targets (depicted as horizontal lines and by the CTA excluding region). CTA should be able to provide very competitive limits around a few tens of neV, skimming the part of the parameter space where ALP could constitute the entirety of the DM in the Universe (region below the diagonal line). Figure taken from \citet{Abdalla:2020gea}.}
    \label{fig:ALPsummary}
\end{figure}

This whole picture of the near future of ALP searches with planned gamma-ray and other instruments is summarized in Fig.~\ref{fig:ALPsummary}. In order to put this particular effort into the more general ALP search context, as done for Fig.~\ref{fig:ALPsummaryIACT}, the figure also shows a broader energy range than the one that can be actually probed by gamma rays. Very interestingly, gamma rays may have the potential to look for \glspl{alp} in a region of the parameter space where they could constitute a portion or all of the DM content in the Universe, and which it is difficult to explore with any other current or planned detection technique or instruments. We  refer to the interested reader the recently published review by \citet{Batkovic:2021fzr} for a further reading of the methods and technical challenges of ALP searches with IACTs.

\section*{Verification}
\begin{enumerate}[start=5,label={Q8.\arabic*:},wide=0pt, leftmargin=3em]
    \item If all the DM in the Universe was made out of \glspl{alp}, would this scenario be in agreement with all gravitational evidence for DM (CMB, large-scale structure, mass distribution in galaxies and clusters)?
    \item Why is an enhancement of the gamma-ray flux at Earth possible for cosmological sources if photons oscillate into ALPs?
    \item Could there also be a way to probe the ALP DM particles present in the Universe?
\end{enumerate}

%\newpage
\section{Primordial Black Holes\label{sec:pbh}}
\label{sec:pbhs}

%\paragraph{Formation, evolution, and current limits on PBHs.}
The search for \glspl{pbh} traces back to several decades ago. \citep{Zeldovich:1967,Hawking:1971ei}. \gls{pbh}s are different from  astrophysical
\glspl{bh}. The latter form by
gravitational collapse of dying heavy stars thus retrieving masses close to that of the progenitors. \gls{pbh}s instead formed in the radiation dominated era, via different mechanisms that bounds their mass scale to the moment of generation. Because of their origin, they are not contributing to the count of
baryonic matter. In half a
century, numerous theoretical
studies and experimental constraints have accumulated in search of
evidence for \glspl{pbh}, especially as \gls{dm} candidate \citep[see, e.g.][for recent reviews]{Carr:2016drx,Sasaki:2018dmp,Green:2020jor}. There are several theoretically proposed formation
mechanisms, which allow the mass to range from the Planck mass scale to that of the
supermassive \glspl{bh}~\citep{Carr:2005zd,Carr:2009jm}. An exact
mapping of the \gls{pbh} formation time and evolution combines constraints from 
cosmology, thermodynamics, general relativity and quantum mechanisms, at the extremes of their validity. The search for \gls{pbh} gained significant momentum  after the
discovery  of the GW150914 and  merger
event of two stellar mass \glspl{bh} by the Ligo-Virgo collaboration~\citep[LVC;][]{Abbott:2016aoc}. Since then, a large number of further confirmed GW events from merging BHs has been reported~\citep{Bailes2021}. It was soon realized that \glspl{bh}
compatible with such merger events could constitute a large fraction
or even the totality of \gls{dm} (see Sec.~\ref{sec:dm_foundations}) and be of
primordial nature, thus opening new very intriguing possibilities~\citep{Bird:2016dcv}. 
Clearly, a
discovery of \glspl{pbh} would constitute a major breakthrough, shedding
light onto the early Universe, quantum mechanics mechanisms, particle
physics, and \gls{bh} thermodynamics. 

Most of the fate of a \gls{pbh} relates to the time of formation,
which foremost determines its mass. By relating the Universe density at
time $t$ with the critical density to generate a \gls{bh}, one finds a 
relation between the \gls{pbh} mass and the time of formation \citep{Hawking:1971ei}: 
\begin{equation}\label{eq:pbh1}
M(t)\sim \frac{c^3\;t}{G} \;\sim\; 10^{38}\;\frac{t}{1s}\; [g] \;=\; 0.5\times10^5\;\frac{t}{1s} \; [M_\odot] \,,
\end{equation}
where $t$ is the Universe age at the time of creation, $M$ the \gls{pbh} mass, $c$ the speed of
light, $G$ the Cavendish constant. Different mechanisms allow
the generation of \glspl{pbh}. One general possibility is the collapse of
regions with higher density than the surroundings. This could be the
case during sudden pressure reduction, as possibly happened
during the quark-hadron era, or during dust-like phases or slow
reheating after inflation. Another possibility is are phase
transitions, for example connected to bubble collisions, collapse of
cosmic strings or domain walls. It is beyond the purpose of this text
to properly summarize all theories and the reader is referred to
specific review \citep[such as][and references therein]{Carr:2005zd,Khlopov:2008qy}.  

The mass of the \gls{pbh} has
important consequences in the \gls{pbh} evolution and affects several
observable quantities. If the \glspl{pbh} were formed at a precise moment,
their mass distribution could be extremely narrow, however, this may well not be the case: if  \glspl{pbh} were
generated through scale-invariant mechanisms, like generic quantum
fluctuations of the space time, or during a wide time elapse, or have
been subject to significant further mass accretion, they could have a
wider mass distribution~\citep{Carr:2009jm}. The number density of \glspl{pbh} is primarily constrained to be smaller than
the \gls{dm} at current times: $\Omega_{\text{PBH}}<\Omega_{DM}=0.25$. However,
their initial number could have  been larger but constantly reducing
over time due to the Hawking
mechanism~\citep{Hawking:1974rv}. According to Hawking, \glspl{bh} radiate
over time, thus reducing their entropy and increasing their
temperature, with faster pace as time goes by, eventually reaching a
final stage of complete disruption called \emph{evaporation}. The mass
loss rate during the \gls{bh} evolution in quantum decay scenarios is
governed by: 
\begin{equation}\label{eq:pbh_massloss}
    \frac{dM_{\text{BH}}}{dt}=-\frac{\alpha(M_{\text{BH}})}{M_{\text{BH}}^2}\,,
\end{equation}
 where $\alpha(M)$ is a parameter that depends on the number of
 degrees of freedom (d.o.f.) available at any given time of the
 evaporation process.  Clearly, the larger the temperature and the
 larger the number of d.o.f., the faster the pace of mass loss and the
 shortest the fate to evaporation. In particular, when the temperature
 exceeds the chromodynamics scale of $200-300$~MeV, quarks and gluons
 start to be directly radiated, which subsequently fragment and
 generate hadrons, photons and leptons, which in turn produce a strong
 non-thermal emission. Above this value, considering the number of
 d.o.f. for quarks and gluons is larger than that for leptons and
 photons, the former dominate the observed gamma-ray spectrum~\citep[see for
 example Fig~1 of][]{Halzen:1991uw}. While below the
 chromodynamics scale, the number of d.o.f. is more accurately known,
 one should mention that there are also models like that of \citet{Hagedorn:1968zz} that foresee a strong multiplication
 of the number of d.o.f. during the final stages of evaporation, which
 would imply an event faster final stage evolution before the
 evaporation, with complete disruption in nanoseconds to
 microseconds. It should be noted that the Hawking mechanism in its
 simpler description may not be accurate. Recently it was found for
 example that small mass \glspl{bh} may become unstable due to
 quantum-gravitational effects earlier than predicted, with relevant
 consequences on observables like galaxy
 clustering~\citep{Raccanelli:2017xee}. A \gls{pbh} with mass M has a temperature~\citep{Halzen:1991uw}:
\begin{equation}\label{eq:pbh_temperature}
    T_{\text{BH}}(M)=\frac{\hbar\,c^3}{8\pi\,G\,k_{\text{b}}}\frac{1}{M}
    %\sim 10^{-7}\left(\frac{M_\odot}{M}}\right) [K]
    \sim 100 \,\left(\frac{10^{15}\,\text{g}}{M}\right) \;[\text{MeV}]\,,
\end{equation}
which increases over time as long as mass is lost, so that the life
expectation for a \gls{bh}, also called \emph{evaporation time} is: 
\begin{equation}\label{eq:pbh_lifetime}
    \tau_{\text{BH}}(M)=\frac{G^2M^3}{\hbar\,c^4}
    %\sim 10^{64}\left(\frac{M}{M_\odot}}\right)^{64} [yr]
    \sim 10^{10} \left(\frac{M}{10^{15}\,\text{g}}\right)^3 \;[\text{yr}]
\end{equation}
where $\hbar$ is the reduced Planck constant and $k_{\text{b}}$ the Boltzmann
constant. The temperature can be connected to the lifetime through the
following formula~\citep{MacGibbon:2015mya}: 
\begin{equation}\label{eq:pbh_temp_tau}
    T_{\text{BH}}(\tau)\approx \left[4.8\times10^{11}\frac{1\,\text{s}}{\tau_{\text{BH}}(M)}\right]^{1/3}\;[\text{GeV}] \,.
\end{equation}

The reason to use a reference mass of $M_\dagger=10^{15}$g
($\sim10^{-18}$M$_\odot$)   is that Eq.~\ref{eq:pbh_lifetime} shows
that all \glspl{pbh} roughly lighter than this value would have already
evaporated at present time. \glspl{pbh} at about M$_\dagger$ would instead
be evaporating now, possibly associated with cataclysmic events such
as long-duration \glspl{grb}.   Heavier \glspl{pbh} would be only mildly
emitting radiation and almost stable, constituting a sort of
non-baryonic collisionless cold fluid, and thus a possible
contributor to the \gls{dm}~\citep{Feng:2013pwa,Barack:2018yly}.  Evaporated,
evaporating, and non-evaporating \glspl{pbh} imply therefore rather different
observables and means for investigation, and quite different chances
for detection. A collection of constraints can be found online\footnote{\url{https://github.com/bradkav/PBHbounds}}. 

%Fig.~\ref{fig:pbh_limits}, reproduced from this collection  provides a collection of such limits. 
%\begin{figure}[h!t]
%    \centering
%    \includegraphics[width=0.9\linewidth]{./figures_ch8_color/pbh_bounds}
%    \includegraphics[width=0.9\linewidth]{./figures_ch8_color/pbh_bw}
%    \caption{Collection of limits for the \gls{pbh} density fraction as a function of the mass. Reproduced from \url{http://doi.org/10.5281/zenodo.3538999}.} 
%    \label{fig:pbh_limits}
%\end{figure}

%From Fig.~\ref{fig:pbh_limits} one can see a wide collection of experimental limits to \glspl{pbh}. 
For already evaporated \glspl{pbh}, the limits are obtained with cosmological probes such as the CMB or theoretical arguments. 
%They are not shown because the scale would not be appropriate. 
In
the range $<10^{17}$~M$_{\odot}$, evaporated \glspl{pbh} would have
filled the Universe with particles and radiation that would have
affected baryogenesis, for example through injection of substantial
amount of antiparticles, or through interaction with photons of the
cosmic microwave background.  In the range
$10^{-17}$~M$_{\odot}$ instead, \glspl{pbh} would have strongly radiated MeV
photons in their final phase, as shown in Eq.~\ref{eq:pbh1}. This
radiation would have filled the Universe, a prediction that is not
observed on top of a standard astrophysical MeV diffuse background,
thus providing very robust constraints too~\citep{Laha_2020}.  
\gls{pbh} with masses around M$_\dagger$ would be instead evaporating at
present time. We could  observe such events for example as gamma--
or X--ray bursts or even through emission lines such as those observed
from the \gls{gc}. \glspl{iact} are sensitive in the TeV range, and
would be therefore valid  probes for \gls{pbh} evaporating now, as we will
discuss in more depth in the following. Finally, for larger initial masses, \glspl{pbh}
would appear as massive non-baryonic objects. As such, they would
naturally act like weakly interacting massive objects, a standard
requisite for a \gls{dm} particle theory.  For such
objects, we could expect  electromagnetic or hadronic signatures
besides the gravitational ones. 
The detection of gravitational waves from black holes
with estimated mass of few tens of
GeV~\citep{Abbott:2016blz,Abbott:2016nmj,Abbott:2016pea}
revived the interest of \glspl{pbh} as \gls{dm} candidates of similar
mass~\citep{Barack:2018yly,Bird:2016dcv}.

%\bigskip
\paragraph{Search for \glspl{pbh} with \glspl{iact}.} How well do \glspl{iact} fit into
this wide world of \gls{pbh} observables? %Unfortunately, the MeV diffuse
                                %emission from already evaporated \gls{pbh}
                                %is at energies unreachable for
                                %\glspl{iact}. Additionally,  
\glspl{iact} have narrow field of view (FOV) which makes any attempt of
detection of diffuse emission hard to attain. Probably the main avenue
for \glspl{iact} to search for signatures of ongoing  \gls{pbh}
disruptions. As discussed, in the final phase, evaporating \glspl{pbh} would
deliver an increasing emission, culminating with a disruption and a
burst of very high energy (GeV to TeV) radiation, lasting however short
time, from fractions of seconds \citep{Hagedorn:1968zz} to several seconds~\citep[see Fig.~3
  of][]{MacGibbon:2015mya}. The expectation for the gamma-ray
spectrum a superposition of two components: A primary component comes
directly from Hawking radiation and is thus peaked at around the \gls{pbh} mass.
A secondary component comes from the decay of hadrons produced
in the fragmentation of primary quarks and gluons and peaks at somewhat
lower energies. Due to available number of d.o.f., a large fraction of
Hawking radiation from evaporating \gls{pbh} would be in the form of quark
and gluon jets. The secondary components would provide a  photon
contribution larger that the primary one.  

Despite presumable intense radiation during evaporation, such
disrupting events would not be easy to catch with \glspl{iact}: their short
duration would in fact reduce the number of external triggers such as
those provided by the Fermi GMB or LAT instruments, in reason of the
small sensitivity for seconds-elapsing \gls{pbh} evaporation. Evaporating
\glspl{pbh} could alternatively  appear serendipitously as extremely
ephemeral but bright emission anywhere in the FOV of \glspl{iact} during
regular observations. Such events would likely not be automatically
recognized by the standard trigger and data reconstruction
mechanisms. A specific search for evaporating \gls{pbh} must be developed on
\gls{iact} data.  

\begin{figure}[h!t]
    \centering
    \includegraphics[width=0.7\linewidth]{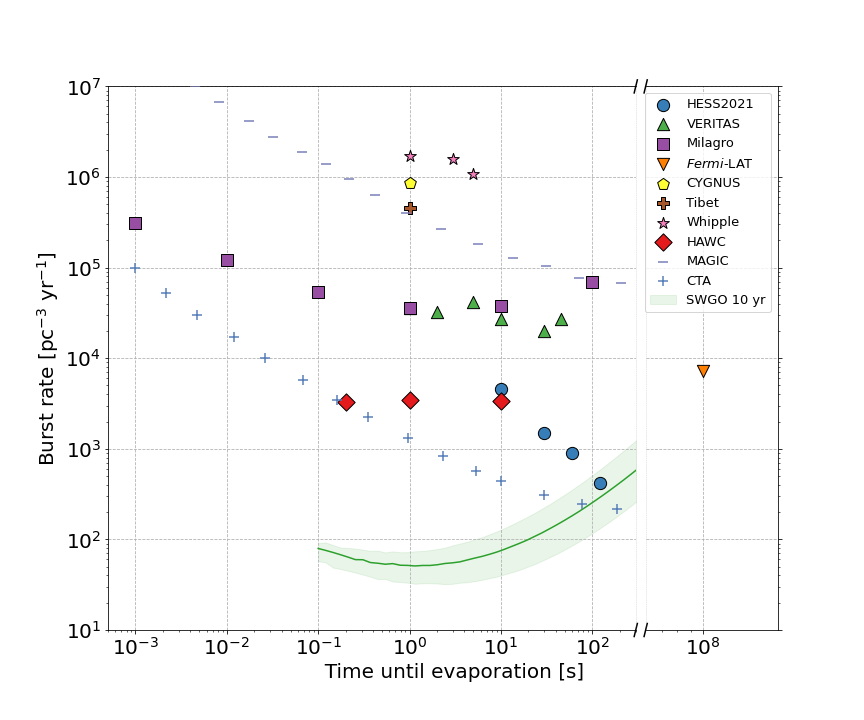}
    \caption{A collection of limits on the \gls{pbh} volume density as a function of the time to evaporation for \gls{pbh} of mass of the order of $10^{15}$g. The image is an update of Fig. 4 of \citet{Lopez-Coto:2021lxh} after inclusion recent \gls{hess} data from~\cite{Tavernier:2021} and the predictions for \gls{magic} and \gls{cta} by \citet{Cassanyes2015}.}
    \label{fig:pbh_limits_iacts}
\end{figure}

\paragraph{Results.}
The idea of using Cherenkov light to detect \glspl{pbh} was first developed
by \citet{Porter:1978,Porter:1979} and further
implemented with the realization of a specific trigger system called
SGARFACE~\citep{Krennrich:2000,LeBohec:2005wt}. The SGARFACE detector,
mounted on the single-dish Whipple telescope, was a specific trigger,
based on analogue pixel summation, sensitive to gamma-ray 'glows'
developing in the atmosphere after short and intense bursts of gamma
rays above 100~MeV. The Cherenkov light at the ground from sub-GeV
showers would be extremely spread, due to the increased Coulomb
scattering of the atmospheric shower particles, much more than for TeV
showers, almost impeding a directional reconstruction. Individually,
they would not be detectable, but in case of a strong burst, the sum
of all these 'glows' was simulated and shown to be detectable with
SGARFACE. The motivation for SGARFACE was two-fold. First, it was to
constrain the short time scale bursts from \glspl{pbh} in the sub-microsecond
to 100 microsecond scale, as predicted by~\citet{Hagedorn:1968zz}. But second and foremost, it was to be used to
look for other short time scale bursts in the sub-GeV to GeV
regime, including counterparts to giant radio pulses from pulsars.  
%SGARFACE made use of both the signal and time structure of the shower for gamma/hadron separation: gamma-ray showers are generically more compact in time. 
The intensity peaks were searched in windows of 60, 180, 540, 1620,
4860~ns and 15$\mu$s, with digitization bin of 20~ns. The predicted
sensitivity of SGARFACE was ranging from $0.8$ to $50$~ph
cm$^{-2}$s$^{-1}$ for primaries above 200~MeV and pulse width of 60~ns
and 15$\mu$s respectively and for a burst modelled as a power law with
spectral index of $-2.5$. Results obtained with this instrument were
presented by \citet{Schroedter:2009} using 2.2 million
events taken over 6 years of operation. That study also presents in
details the instrument, the reconstruction and the analysis. The
sensitivity limits were computed for \gls{pbh} of mass $10^{13}$~g, for
different burst durations, and ranged between 1 and $10^3$ bursts per
cubic parsec per year for duration from ns to $\mu$s if assuming the
Hagedorn mechanism. Larger duration bursts, of the order of 
seconds, were constrained also using the Whipple telescope standard
trigger~\citep{Connaughton:1998,Linton:2006yu}, providing additional
results on \gls{pbh} evaporation from \glspl{iact}. In particular the latest result
with Whipple~\citep{Linton:2006yu} made use of about 2200~h of good
quality data taken between 1998 and 2003. The search was made over
time windows of 1, 3 and 5 seconds.  
A peak in emission was determined when a certain number of consecutive
events were recorded within a certain window of time. It is clear that
the larger the window, the larger the probability of contamination
from background events. The background contamination was attenuated by
scrambling all the time of events. Statistically generated burst
surviving the search criteria on the scramble sample were thus
removed. The probability of observing a burst event of intensity $b$
in a time window $\Delta t$ at an angular direction $\theta$ and an
astrophysical distance $r$ can be written as~\citep{Linton:2006yu}: 
\begin{equation}
    n_{\text{BH}}(b,\Delta t) = \rho_{\text{BH}}\,\tau\int_{\Delta\Omega} d\omega \int r^2\,P(b,N_{\text{d}}(r,\theta,\Delta t))\;dr\,,
\end{equation}
where $\rho_{\text{BH}},\tau$ are the \gls{bh} numerical volume density and $\tau$ the lifetime, $\omega$ is the solid angle of the signal search, and $P(b,N_{\text{d}}(r,\theta,\Delta t)$ is the Poisson distribution of a \gls{bh} emitting $N_{\text{d}}$ gamma-rays and producing an event of intensity $b$ where the number of emitted gamma rays is:
\begin{equation}
    N_{\text{d}}(r,\theta,\Delta t)= \frac{A(\theta)}{4\pi\,r^2}\int_{\Delta t}dt\int \frac{d^2N(E,t)}{dE\,dt}\,A_0(E)\;dE\,,
\end{equation}

where $A(\theta)$ is the effective area. These formulas can be used to
provide upper limits on the \gls{pbh} density using the number of detected
bursts, once the initial gamma-ray spectrum is specified. Of course,
this depends on the exact knowledge of the quantum process at work
during the evaporation phase, as well as with \gls{bh} properties
(e.g. spin) and its surroundings (for example if a chromosphere is
formed). This search provided the first constraints to the \gls{pbh} density
at $1.08\times10^6$ \gls{pbh} pc$^{-3}$ yr$^{-1}$ (see
Fig.~\ref{fig:pbh_limits_iacts}):

\begin{equation}\label{eq:pbh_dnde}
    \frac{dN_\gamma}{dE}=9\times10^{35}\times\begin{cases}
               \left(\frac{T_\tau}{1\,\text{GeV}}\right)^{-3/2}\left(\frac{E}{1\,\text{GeV}}\right)^{-3/2}\; \text{GeV}^{-1} \quad \text{for } E<k_{\text{b}}\,T_\tau\\
               \left(\frac{E}{1\,\text{GeV}}\right)^{-3}\; \text{GeV}^{-1} \quad \text{for } E>k_{\text{b}}\,T_\tau
            \end{cases}
\end{equation}
where $T_\tau=7.8\,k_{\text{b}}^{-1}(\tau/1\,\text{s})^{-1/3}$ [TeV] is the \gls{pbh} temperature with a lifetime of $\tau$.

In the thesis of \citet{Cassanyes2015},  the potential of \gls{magic} to search for \glspl{pbh} was investigated for the first time. Cassanyes assumed $1\div10,000$~h of good
quality data, and photons from 158~GeV to 10~TeV. He found that the
largest sensitivity comes for integration times of about several
minutes, a value larger than the few seconds found by \gls{veritas} and
\gls{hess} This is probably explained by a different emission model than
Eq.~\ref{eq:pbh_dnde}. Within this window, upper limits of the order
of $\sim10^4$ pc$^{-3}$ yr$^{-1}$ can be achieved. However, \gls{magic} has
not published any results on actual data with such technique. 

Recently, there is increased activity in the field to analyze accumulated existing \gls{iact} data for \gls{pbh} signatures: \cite{Tavernier:2019exh, Tavernier:2021} are analyzing so far almost 5000 hours of \gls{hess} data in search of \gls{pbh} events, reaching upper limits on the evaporation rate to $<527$~pc$^{-3}$yr$^{-1}$ (see Fig.~\ref{fig:pbh_limits_iacts}. Also the \gls{veritas} collaboration is pursuing such reanalysis of their data \citep{Pfrang:2021iky}, with also looking for short-term optical fluctuations by \gls{pbh} microlensing events \citep{Pfrang:2021pfy}.

\paragraph{Outlook.}
Are the current limits interesting? 
%By looking at
%Fig.~\ref{fig:pbh_limits} and 
%Fig.~\ref{fig:pbh_limits_iacts}, one can
%see that 
These limits are several orders of magnitude away from the
limits obtained with e.g. the extragalactic background (EGB), in the
range of \gls{pbh} evaporating at present times. However, the
theoretical mapping of \gls{pbh} formation and evolution still allows
for a wide range of possibilities to significantly make these limits
stronger. There are at least two direct ways to improve the
limits. First, the mass loss of the \gls{bh} is governed by the number of
d.o.f. As such, if one relaxes the assumption that only Standard Model
d.o.f. are to be used, boosted emission can
occur as a result of multiplication of
d.o.f. during the final stage of evaporation~\citep{Hagedorn:1968zz}. Secondly, those limits
assume an homogeneous \gls{pbh} distribution in the Universe. However,
clustering in galaxies is plausible~\citep{Raccanelli:2017xee}, which
would significantly change these limits. Finally, one should not
forget that possibly \glspl{pbh} have a narrow mass range and therefore
a limit in one mass region does not necessarily affect limits in other
part of the mass spectrum. 

There are instruments sensitive to \glspl{pbh} in the TeV range,\footnote{Note that there are also efforts to search for \glspl{pbh} at lower, MeV gamma-ray energies \citep{Ray:2021mxu, LeyVa2021,Ghosh:2021gfa}.}
specially the wide FOV air shower detectors such as MILAGRO and HAWC.
MILAGRO~\citep{Abdo2015} and HAWC \citep{2020JCAP...04..026A} provided results with 4.6 and 2.6 years of data, respectively.
Such large observation times compared to \glspl{iact} is justified by the fact
that MILAGRO and HAWC observe particles within the showers, and thus have a
daily duty cycle.  The optimal search time for MILAGRO and HAWC was around 1
second.  At smaller intervals, the sensitivity gets low, at
larger intervals, the background contamination becomes
dominant. With HAWC, limits on the burst rate were obtained of the order of $\sim3\times10^3$ pc$^{-3}$
yr$^{-1}$.  %In the same work by~\citet{Abdo2015}, an extrapolation for
%the performance of HAWC was attempted, resulting again with limits in
%the ballpark of $\sim10^4$ pc$^{-3}$ yr$^{-1}$. 
More recently, \citet{Lopez-Coto:2021lxh} investigated the prospects for this search with the planned instrument \gls{swgo}, showing how this instrument detector can improve of more than one order of magnitude the limits from HAWC as shown in Fig.~\ref{fig:pbh_limits_iacts}. An endeavor to
compute the prospects for \gls{cta} was made by
\citet{Cassanyes2015}. Cassanyes used the same method to estimate the
performance of \gls{magic} with \gls{cta}'s figure of merit.  He found that the
sensitive \gls{pbh} volume of \gls{cta} would be a factor of about 790
larger than in \gls{magic}: a factor of 25 due to the larger FOV, and a
factor of $10^{3/2}$ because the 10-times better sensitivity allows
for a  $\sqrt{10}$ farther distance. With this performance, \gls{cta} could
set constraints of the order of $\sim10^2$ pc$^{-3}$ yr$^{-1}$. The
advantage is that no specific time is required to \glspl{iact}: \glspl{pbh}
could appear serendipitously as bright and short bursts, lasting from
sub-second to seconds, anywhere in the field of view.  
Detecting these events require to build a specific off-line
reconstruction pipeline. As discussed by~\citet{Doro:2017vjf},
considering that \gls{cta} will surely strongly compress and reduce the
global information from its hundreds of cameras in order to save disk
space, these pipelines should be developed and investigated as soon as
possible in order to avoid losing important data for \gls{pbh}
searches.

 \section*{Verification}
\begin{enumerate}[start=8,label={Q8.\arabic*:},wide=0pt, leftmargin=3em]
\item Can the imprint of non-baryonic matter in the CMB anisotropies be caused by \glspl{pbh}?
\item What gamma-ray instruments are best suited to search for \glspl{pbh} evaporation?
\item What gamma-ray instruments are best suited to search for evaporated \glspl{pbh}?
\item Why a tailored analysis is needed by \glspl{iact} in search for \glspl{pbh}?
\item Why CTA may be less competitive than other gamma-ray instruments in the same energy band?
\end{enumerate}

%\newpage
\section{Tau-Neutrinos}
\label{sec:tau}
The nature of the astrophysical emission of neutrinos is for many decades one of
the hottest topics in astrophysics. The interest has been boosted in recent times by the
measurement of an astrophysical diffuse flux of high-energy
neutrinos by IceCube~\citep{Aartsen:2014gkd} and the significant
spatial and temporal coincidence observed between a 290~TeV neutrino
event by IceCube and a flaring blazar by Fermi-LAT and
MAGIC~\citep{IceCube:2018dnn}. Several unknowns are yet to be understood: What is the full spectrum and composition of the diffuse neutrino flux? What astrophysical targets are neutrino emitters and where does the neutrino emission takes place? Are neutrinos produced via leptonic or hadronic processes? Are neutrinos also the result of interaction of high energy cosmic rays with local ambient fields? These and others questions are effectively tackled by large-scale ground-based instruments such as the mentioned IceCube, Antares~\citep{AdrianMartinez:2011uh}, the planned KM3NeT~\cite{AdrianMartinez:2016} at TeV to PeV energies, and the \gls{pao}~\citep{Aab:2015kma}, the \gls{ashra} \citep{Asaoka20137}, or the future \gls{grand}~\citep{Alvarez-Muniz:2018bhp} and \gls{poemma} observatory~\cite{Olinto:2020oky} at PeV energies and above.
Also, although not primarily designed for neutrino detection, \gls{iact}s are instruments capable not only to
detect electromagnetic counterparts to neutrino astrophysical targets, but also possibly
 to directly detect signatures of neutrino events crossing the
Earth's atmosphere or neutrinos emerging from the Earth's crust in a technique called \textit{Earth-skimming} detection technique~\cite{1999ICRC....2..396F}.

\bigskip
Extraterrestrial neutrinos are generated \citep[see, e.g.,][for recent
reviews]{Becker:2007sv,Baret:2011zz} through general production mechanisms such as nuclear decay (proton-proton) or photopion (proton-photon) interactions eventually
generating neutrinos through reactions such as:   
\begin{equation}\label{eq:pi_dec}
\begin{array}{rl}
    p\;p\rightarrow  & \pi^+\;\pi^-\;\pi^0  \mbox{, followed by:}\\
    & \pi^+\rightarrow \mu^+\; \nu_\mu \rightarrow e^+\; \nu_e\; \overline{\nu_\mu}\;\nu_\mu \quad \mbox{or}\\
    & \pi^-\rightarrow \mu^-\; \overline{\nu_\mu} \rightarrow e^-\; \overline{\nu_e}\; \overline{\nu_\mu}\;\nu_\mu 
     \end{array}
\end{equation}
\noindent and
\begin{equation}\label{eq:nu2}
\begin{array}{rl}
    p\;\gamma\rightarrow  &  \Delta^+\ \mbox{, followed by:}\\
    & \Delta^+\rightarrow p\;\pi^0 \; (2/3)\quad \mbox{or} \\
    & \Delta^+\rightarrow n\;\pi^+ \; (1/3)
\end{array}
\end{equation}

\noindent
Therefore, depending on the ambient matter and radiation fields, the
above processes can give rise to neutrino emissions in a wide range of
energies and fluxes~\citep{Ackermann:2019cxh}. At MeV energies, a cosmic neutrino background is
expected from solar
MeV neutrinos generated within our Sun as well as from farther 
stars. Above 0.1~GeV the most abundant population is that of
atmospheric neutrinos, generated by the interaction of cosmic rays
with the Earth's atmosphere. Atmospheric neutrinos constitute the by-far
larger background in large-scale neutrino detectors. In this energy range one can also
find the rarer interesting extraterrestrial neutrinos
 produced e.g., in an \gls{agn} or a \gls{grb} and possibly in all other non-thermal emission regions in which high energy cosmic rays are efficiently generated. Finally,
above $5\cdot10^{19}$~eV, neutrinos are found
after $\Delta$ resonances (Eq.~\ref{eq:nu2}) of ultra high energy cosmic rays with the cosmic
microwave background~\citep{Beresinsky:1969qj} (cosmogenic neutrinos). Therefore, the phenomenology of neutrino production is wide, and often accompanied to high energy cosmic ray reservoirs.

During hadronic processes in Galactic and extragalactic environments,
the expected neutrino generation rate ratio is
$(\nu_e,\nu_\mu,\nu_\tau)=(1,2,0)$, due to the fact that positive,
negative and neutral pions are produced equally during proton
interactions. Therefore, \nutaus are not expected to be produced
significantly through these mechanisms at the source. However, during propagation,
neutrino families mix because of the non vanishing mass
eigenstates. Already for distances larger than few astronomical units,
the mixing is total and a uniform neutrino family mix is expected at
Earth: $(\nu_e,\nu_\mu,\nu_\tau)=(1,1,1)$ (for the computation of
mixing, see~\citet{Becker:2007sv}). It is important to recall that the
family mixing depends on terms proportional to the difference of the
squared masses of two neutrino flavors. Therefore, the detection of
\nutaus is very relevant not only to test astrophysical cosmic
ray processes and unveil new emission mechanisms at the source, but
also to test neutrino physics models.  

\begin{figure}[h!t]
    \centering
    \includegraphics[width=0.9\linewidth]{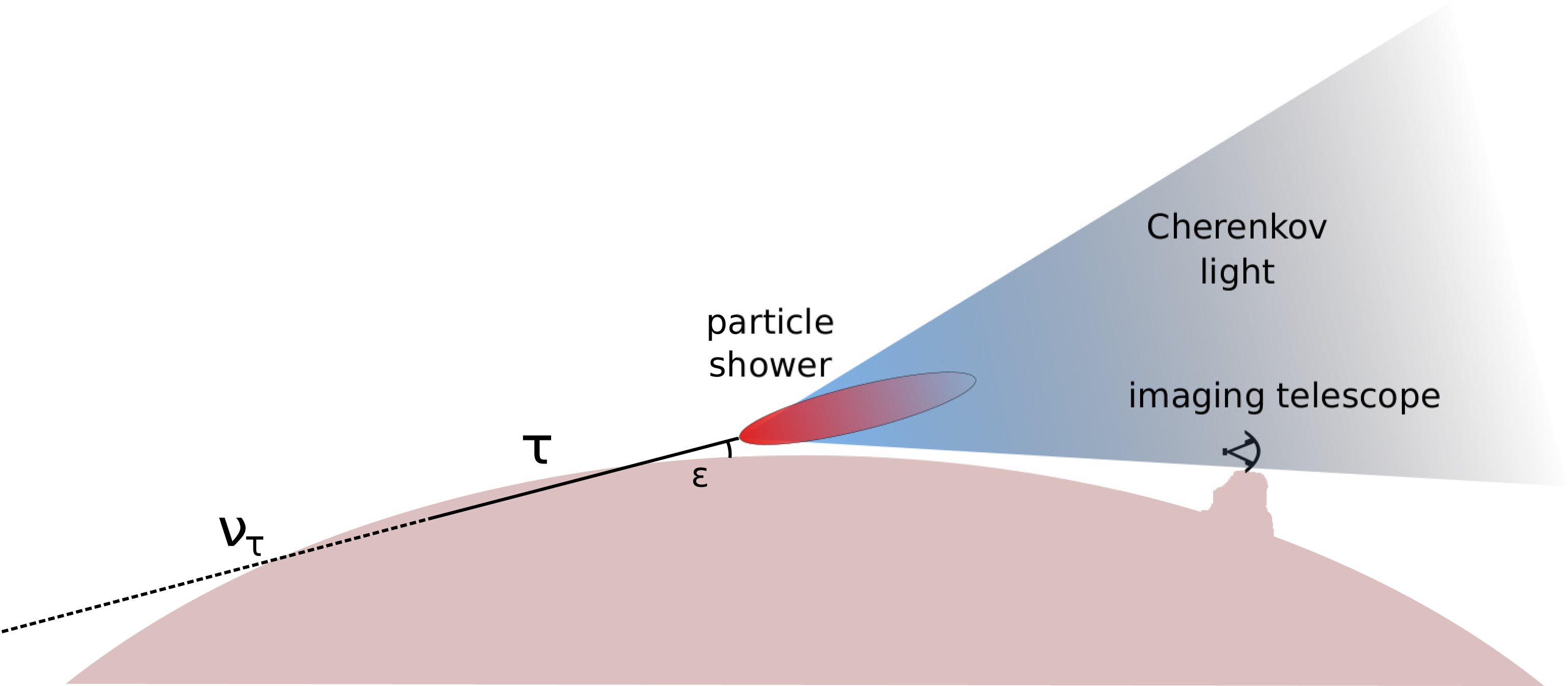}   
    \caption{Artistic sketch of the Earth-skimming \nutau-technique, with the basic scheme of detection for \gls{iact}s. The Cherenkov light cone is produced by charged shower particles emerging from the direction of the \nutau. Courtesy of N. Otte, reproduced from \citet{Otte:2018uxj}.}
    \label{fig:tau}
\end{figure}

By ice or water detectors with a huge instrumented volume (such as IceCube, Antares, or KM3Net), TeV-PeV neutrinos are observed through the Cherenkov light generated in
particle showers produced by the neutrino interactions in the detector~\citep{IceCube:2015vkp}. Above the PeV range, the
ice or the water are increasingly opaque to the crossing of electron or muon
neutrinos, and only \nutaus can be detected from below. Above PeV energies, the  technique  to detect Earth-skimming cosmic \nutaus  comes into play: When neutrinos hit the Earth at its rim and cross a certain amount of the Earth's crust, they have a non negligible chance of converting to a lepton which subsequently may emerge into the air and induce an upwards-directed atmospheric shower. The advantage of this approach is very similar to seeking for signatures from cosmic primaries interacting with the atmosphere: One uses the Earth itself as suitable large interaction volume, and only needs to place a detector in or close to the induced secondary particle showers. Also, while \nutaus are the only candidates for the Earth-skimming technique, they probe indirectly the whole neutrino family of astrophysical origin due to the neutrino mixing. The Earth-skimming approach is pursued by the \gls{pao}~\citep{Fargion:2000iz,Fargion:2003kn}, and detectors like \gls{ashra} and \gls{grand} are primarily designed for it. 
% GRAND is instead a planned array of thousand of radio antennas to be placed over a steep mountain side that are sensitive to the radio emission from  atmospheric showers generated with the same skimming neutrinos at extremely high energies (EHE).
Finally, this technique can also be adopted by \gls{iact}s, detecting the Cherenkov light of the ultra-relativistic particles in the neutrino-induced atmospheric showers. The setup for detecting Earth-skimming \nutaus with \gls{iact}s is illustrated in Fig.~\ref{fig:tau}. 

\bigskip
Neutrino interactions with matter (N) happen with charged current (CC) and neutral current (NC) mechanisms:
\begin{eqnarray}
    \mbox{CC:}\quad N+\nu_l &\rightarrow\ldots\rightarrow& X+l\quad \mbox{where}\quad l=e,\mu,\tau \\
    \mbox{NC:}\quad N+\nu_l &\rightarrow& X+\nu_l\quad \mbox{where}\quad l=e,\mu,\tau 
\end{eqnarray}
where $X$ is a generic product that leads to a hadronic
cascade. Of the leptons found in the reaction products, electrons are rapidly absorbed in matter and the chance of exiting the Earth's crust are low. When muons are created in these interactions, they also lose most of their energy inside the Earth, but with longer radiation length than electrons and occasionally pass through the crust and decay in the atmosphere. However, due to its rather large lifetime, a muon emerging from the Earth would rarely decay into a particle shower within the telescope \gls{fov} and only generate direct Cherenkov light, which mostly would go undetected. The situation drastically changes for tau neutrinos. A PeV tau particle created after a CC \nutau interaction in the crust  can travel through a rather large amount of matter with a significant chance to decay shortly after leaving the crust at the right distance from the telescopes and generate extensive air showers.\footnote{At the highest  $\sim$ EeV energies, the   \nutau $\rightarrow \tau \rightarrow$ \nutau $\rightarrow \tau \rightarrow \ldots$, recreation chain increases the chance of a high-energy tau leaving the crust and, consequently, the telescopes' acceptance \citep{Alvarez-Muniz:2017mpk}.} 
Table~\ref{tab:tau} \citep[extracted from][]{Fargion:2000iz}, summarizes the main tau decay modes: a large cross-section is found into hadronic-induced showers (65\%), electromagnetic showers (18\%) or muons (17\%). The first two components are detectable if generated at the right distance range and within the telescopes \gls{fov}

\begin{table}[h!t]
\centering
\begin{tabular}{llll}                                                                       
\hline
 Decay        &    Secondaries         &  Probability  & Air-shower  \\
\hline
$\tau \rightarrow \mu^{-}\bar{\nu}_{\mu} \nu_{\tau} $&  $\mu^{-}$  &  17.4\%& -- \\
\hline
$\tau \rightarrow  e^{-}\bar{\nu}_{e} \nu_{\tau} $&  $e^{-}$  &  17.8\%  &Pure e.m.    \\
\hline
$\tau \rightarrow \pi^{-} \nu_{\tau} $&  $\pi^{-}$  &  11.8\%  & Pure had \\
$\tau \rightarrow \pi^{-}\pi^{-} \pi^{+}\nu_{\tau} $&  $2\pi^{-}$,$\pi^{+}$ & 10\%  & Pure had  \\
\hline
$\tau \rightarrow \pi^{-}\pi^{0} \nu_{\tau} $&  $\pi^{-}$, $\pi^{0}\rightarrow
2\gamma$  &  25.8\%  & Mixed had/e.m.    \\
$\tau \rightarrow \pi^{-}2\pi^{0} \nu_{\tau} $&  $\pi^{-}$,
$2\pi^{0}\rightarrow 4\gamma$  &  10.79\%  &Mixed had/e.m.   \\
$\tau \rightarrow \pi^{-}3\pi^{0} \nu_{\tau} $&  $\pi^{-}$,
$3\pi^{0}\rightarrow 6\gamma$  &  1.23\%  &Mixed had/e.m.   \\

$\tau \rightarrow \pi^{-}\pi^{+} \pi^{-} \pi^{0} \nu_{\tau} $&
$2\pi^{-}$,$\pi^{+}$,$\pi^{0}\rightarrow2\gamma$  &  5.18\%  &Mixed had/e.m. \\
\hline
\end{tabular}
\caption{Decay channels for the $\tau$ lepton. Probability and type of
  atmospheric showers are also shown  ("e.m." for electromagnetic;
  "had" for hadronic). The table is reproduced from
  \citet{Fargion:2000iz}.}  
\label{tab:tau}
\end{table}

\paragraph{Tau-neutrino searches with IACTs.} For IACTs, 
the Earth-skimming \nutau observation technique thus requires a suitable orography around the telescopes: The telescopes must be pointed at
the \nutau~ emission point, sufficiently distant, corresponding to the
right amount of crossed matter within the Earth. This is for example
doable if a mountain with right thickness is located in
the telescopes' vicinity, or if the telescopes can be pointed
downward toward a direction in which the amount of crossed matter is
optimal. The MAGIC telescopes are located at the Roque de los Muchachos
Observatory in the Canary Islands. The Roque mountain can be pointed by MAGIC. However, it is only
few hundreds of meters away, and therefore a particle shower would be
initiated through or behind the telescopes. On the other hand, from the MAGIC
location, there is a window of observability toward the Atlantic Ocean
in the Northern direction. If the telescopes are pointed at around 1.5
degrees below horizon, a window of
$5\times80$~deg$^2$ is accessible with an 
extraction point at the Ocean at 165~km, allowing sufficient space for
the shower development. \citet{Gaug:2007fpa} were
among the first to investigate this technique, and performed
analytical computations to find the optimal observation strategy and
the effective area. They also performed the first order-of-magnitude event rate
estimation. However, at that time, a full simulation chain of neutrino
interaction and shower development was not available. A complete Monte
Carlo chain of simulation was provided by \citet{Gora:2016mmy} by adapting the ANIS
code~\citep{Gora:2007,Gazizov:2004va}, as well as with publicly
available orography maps. On this basis, \citet{Gora:2017lsh} computed a precise sensitivity and
angular acceptance for MAGIC.  

\begin{figure}
    \centering
    \includegraphics[width=0.447\linewidth]{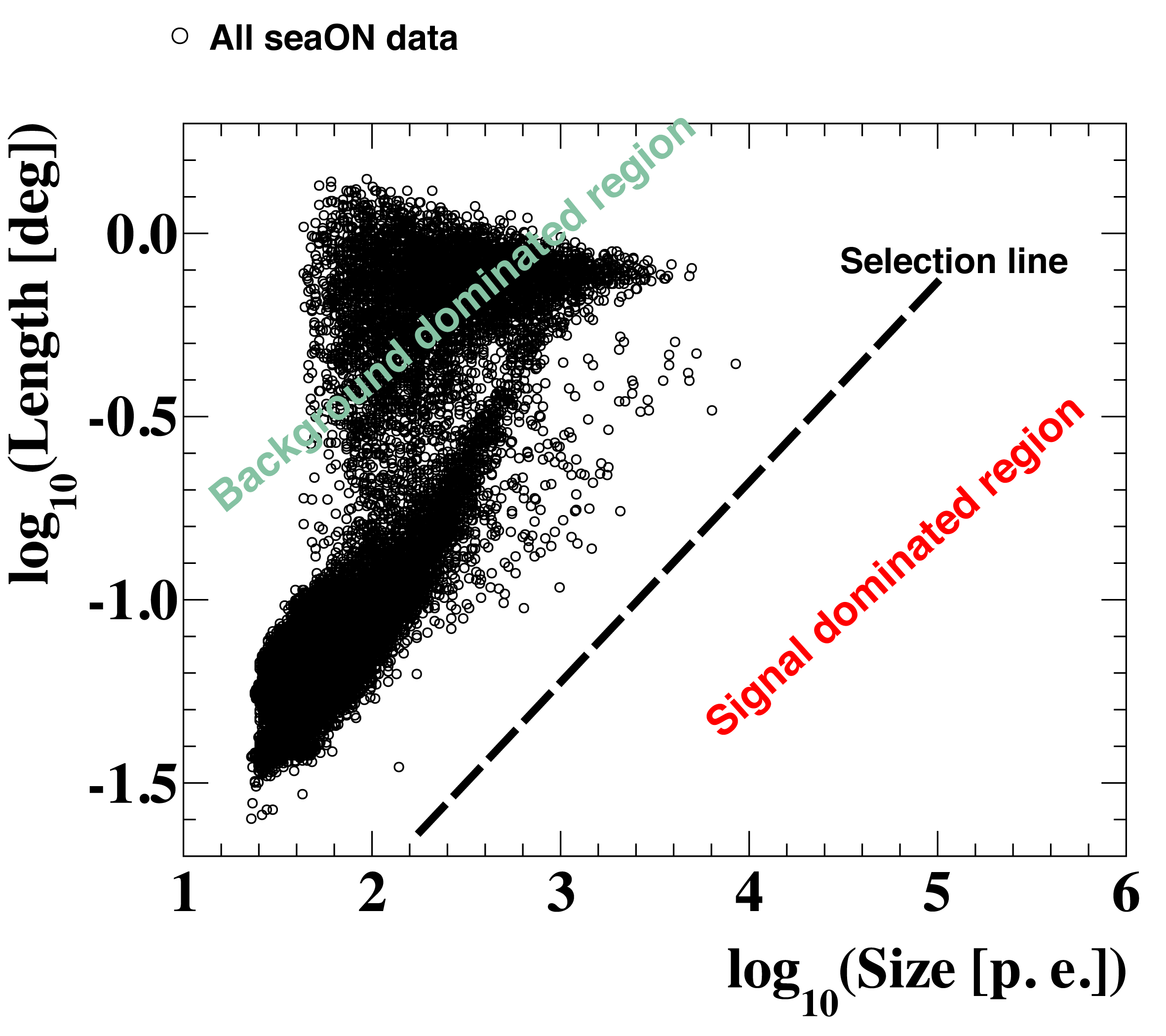}
    \includegraphics[width=0.46\linewidth]{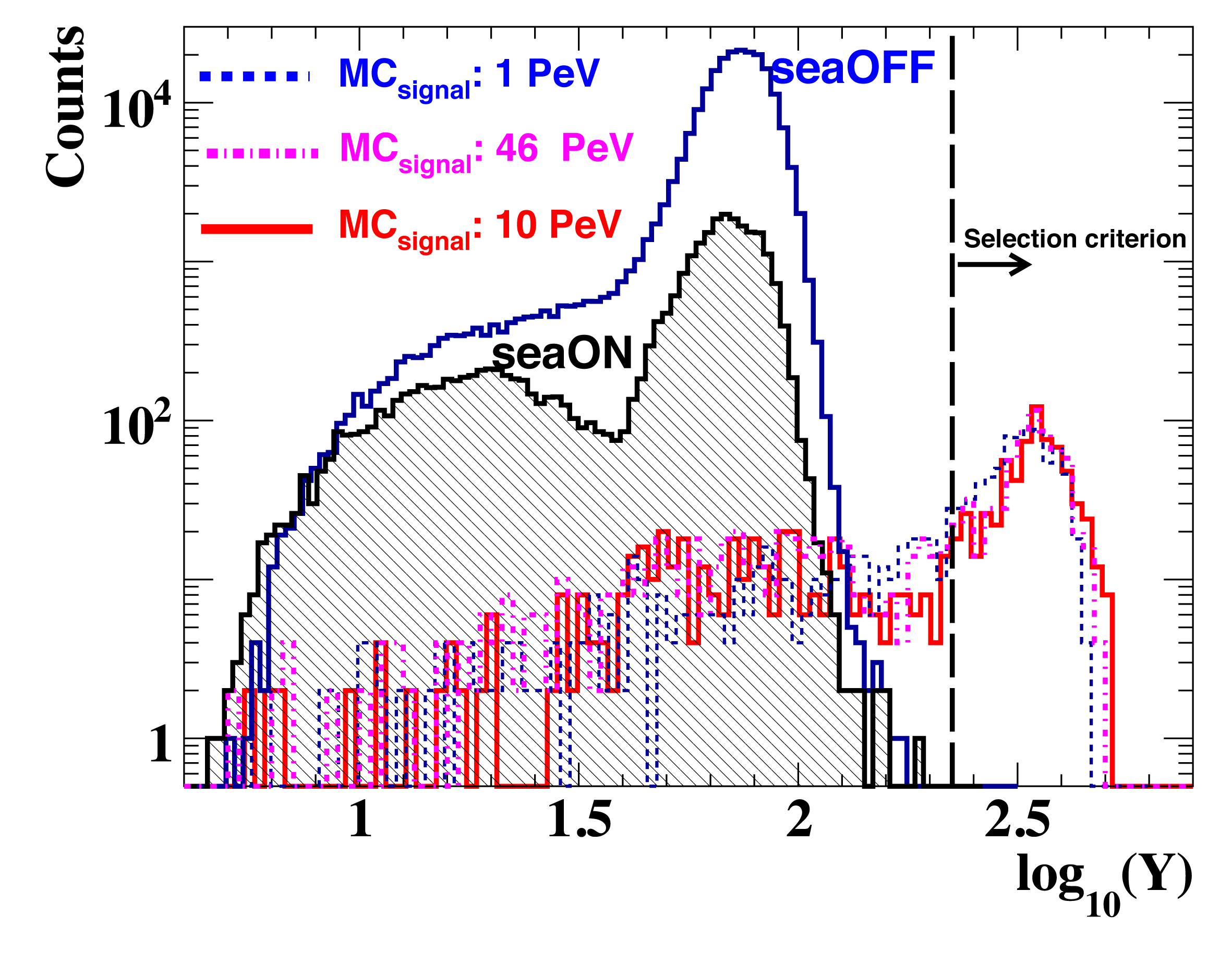}
    \caption{Left: The distribution of events in the space length-size of the image parameters. Background events and signal events are clearly separated by the cut selection $Y$. Right: Distribution of events after the selection cuts based on the same $Y$ parameter (see text) in search for tau-neutrino signatures
      with MAGIC. `seaON', `seaOFF' and signal MC events represent the
      data taken pointing at the signal region (`seaON'), at the
      background control region (`seaOFF') and the MC simulated tau
      neutrinos. In the region $log10(Y)>2.35$, no neutrino candidates
      survive. The selection criterion keeps about 40\% of MC
      signal events. Courtesy from MAGIC Coll.~\citep{Ahnen:2018ocv}.} 
    \label{fig:magic_tau_events}
\end{figure}

\begin{figure}
    \centering
        \includegraphics[width=0.445\linewidth]{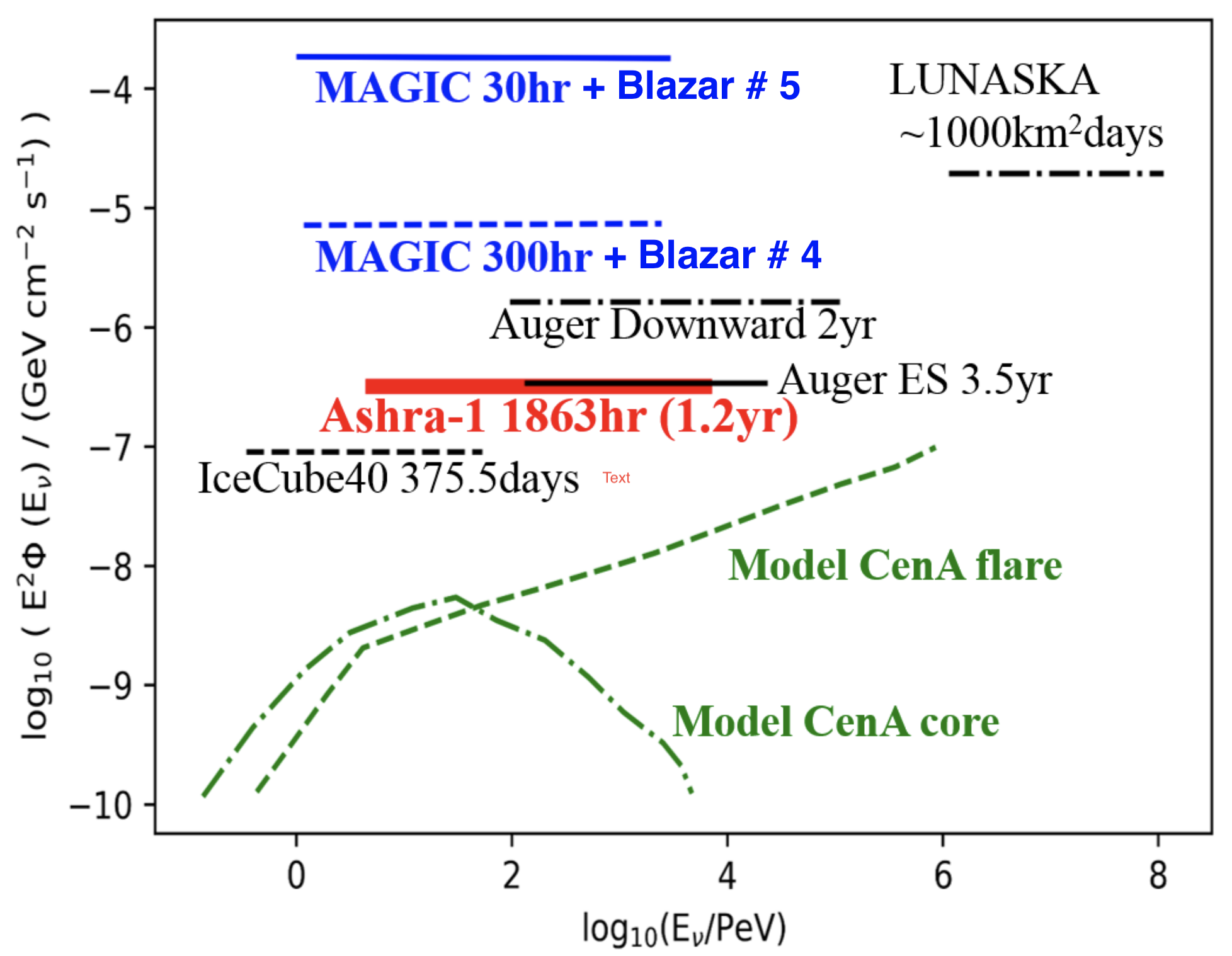}\hspace{0.4cm}
    \includegraphics[width=0.45\linewidth]{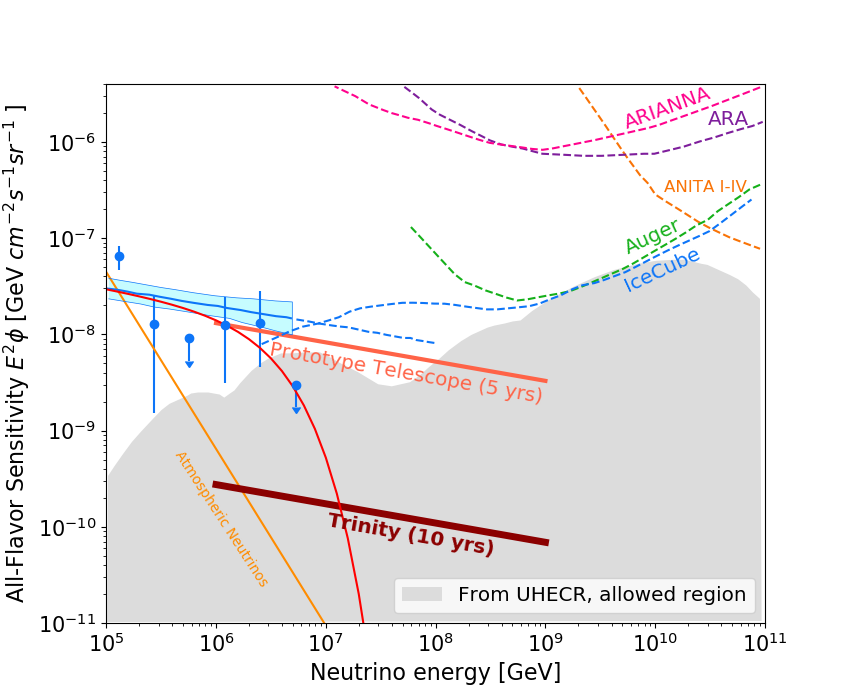}
    \caption{(Left) 90\% C.L. upper limit on the point source tau neutrino flux obtained
      with the MAGIC telescopes with 30 hrs of observation (solid upper horizontal
      line) assuming a specific flux level from flaring blazars, as
      well as extrapolated to 300~h (dashed upper horizontal
      line)~\citep[Fig3 of][]{Ahnen:2018ocv}. (Right) Comparison of sensitivity for VHE and UHE neutrino experiments~\citep[Fig 5 of][]{Brown:2021lef}.} 
    \label{fig:magic_tau}
\end{figure}

\paragraph{Results.}
MAGIC published the results from the observation of 30~h of
data in the direction of the Ocean~\citep{Ahnen:2018ocv}, summarizing the
previous findings by~\citet{Gora:2017lsh,Gora:2017cfs}. The observation
was performed by pointing at an optimized direction toward the Ocean
during several nights, in which high cirrus clouds were present. Such
a condition normally does not allow for standard astronomical
observation, however, it  guarantees good visibility toward
the Ocean. The most complex part of this technique is the
discrimination of the signal parameter space from the
background, as well as the corresponding Monte Carlo simulations. The central analysis steps and results are briefly summarized in the following: First of all, the background was estimated by
pointing the telescope for
5.5 hours slightly above the Ocean, at 2.5~deg above horizon. This allows to gather a significant number of background events as expected in the later below-horizon observation. A combination of
the Hillas size and length image parameter, dubbed
$Y$\footnote{$\log_{10}(Y)=\log_{10}(\textit{Size}[\mathrm{p.e.}])*\cos(\alpha)-\log_{10}(
  \textit{Length}[\mathrm{deg}])*\sin(\alpha)$, where
  $\alpha=63.435^{\circ}$} was found to clearly separate signal and
background, as shown in Fig.\ref{fig:magic_tau_events}. The figure
also shows the $Y$ parameter for \nutau of energy $1,10,46$~PeV. Muon
bundles are the dominant background contribution whereas contributions
from cosmic-ray electrons can be neglected: At horizontal directions
the Cherenkov light from electromagnetic showers is strongly
attenuated, even more than for proton primaries of similar energy, and
below the detection threshold. To build the Monte
Carlo simulations, besides the ANIS modified
version~\citep{Gora:2007,Gazizov:2004va}, also CORSIKA
\citep[][(v6.99)]{Heck:1998vt} and PYTHIA \citep[][(v6.4)]{Sjostrand:2006za})
were used. No event was found, after selection cuts, in 30~h of
observation. The null detection was translated into an upper limit for
\nutau emission after a thorough estimation of the acceptance for
these particles, which is not straightforward considering the geometry
of the observation, the presence of mixed water and rocks in the
\nutau path, the inclusion of atmospheric transparency and the fact
that multiple atmospheric showers can occur. The results are reported
in Fig.~\ref{fig:magic_tau} considering the diffuse neutrino flux as well as different neutrino emission
models for AGNs. The latter were built based on models of real blazar flaring events. In particular, Blazar \#4 and
Blazar \#5 of Fig.~\ref{fig:magic_tau} represent predictions for PKS
2155-304 in high-state and 3C 279
respectively~\citep{Becker:2007sv,Atoyan:2001}. The point-source sensitivity
corresponding to 30~h of observation (Blazar \#4 model are poorly
constraining, of the order of $2\times10^{-4}$~GeV cm$^{-2}$
s$^{-1}$. Only if 300~h are considered, and a more optimistic flux
model is adopted (Blazar \#5), the sensitivity becomes of the order of the one obtained with
downward (all-flavor) neutrinos by Auger~\citep{Aab:2015kma}. This is still far away from the sensitivity obtained with IceCube~\citep{Aartsen:2015exf}. Yet, the sensitive
region is different, being MAGIC indeed sensitive at somewhat higher energies than IceCube but at smaller energies
than PAO, thus allowing for searches for a wider class of astrophysical
emitters. 

\bigskip
Similar results were obtained with the \gls{ashra}-1 Cherenkov telescopes at Mauna Loa~\citep{Ogawa:2020pwy}. This telescope, composed of electrostatic lenses with an optical system to generate convergent beams, observes \nutau skimming through the Mauna Kea mountain. It performed a campaign of 1863~h over 1.2 years that led to limits on the diffuse flux more constraining that those obtained with MAGIC (see Fig.~\ref{fig:magic_tau}), further demonstrating the validity of the technique.

\paragraph{Outlook.} Albeit these results may not look impressive, they demonstrate that IACTs do have a potential for
\nutau searches, and that the data reconstruction is well known. ASHRA-1 will continue gathering hundreds of observational hours each year. Improving on this result is possible with MAGIC considering
that approximately 100~h per year are affected by the presence of
cirrus clouds, and the telescope can be pointed at the Ocean. However, it is currently a non-standard operation and it will require time to integrate more than 300~h. In the future, CTA could be a better suited instrument. CTA has a wider \gls{fov}, especially at high energies where the SSTs operate, thus allowing for an increase in the acceptance which can be roughly estimated at a factor of 4. This will be accompanied with increased sensitivity. The larger number of telescopes may also
allow to investigate less stringent selection cuts too. For the
Northern Hemisphere, CTA will be located where MAGIC telescopes
are. In this sense, CTA telescopes could adopt the same "Ocean
window" used by MAGIC. If this was combined with clever scheduling
(e.g. automatic pointing to the Ocean whenever clouds are present, as
well as an automatic scheduling of flaring objects
that pass behind the Ocean window at a certain time of the night), this could improve the performance by at least one order of magnitude or more with respect to MAGIC, in case bright transients events to happen. In the Southern 
Hemisphere's CTA site, the orography does not allow such kind of observations. 

\bigskip
Recently, an ad-hoc IACT array was proposed to overcome the limits to this technique present for MAGIC and CTA: the rather small \gls{fov} and a non-dedicated observation program. The nane of this project is Trinity~\citep{Otte:2018uxj}. Trinity is composed by 18 units organized in three arrays of six units each. A single telescope covers a $5\times 60$~deg \gls{fov}, so that a single array covers the whole circumference. Each array is placed on mountain tops to cover the whole horizon. Due to this much wider \gls{fov} and dedicated observation mode, up to~1,200 hours/year would be available to this technique. With such performance, sensitivities of the order of 
 $10^{-10}$~GeV cm$^{-2}$ s$^{-1}$ sr$^{-1}$ (see Fig.~\ref{fig:magic_tau}, right panel) could be reached, with tens to hundreds of diffuse neutrinos detectable in 10 years of operation. If approved, such instrument would provide very important results in a gap region between IceCube and radio-neutrino instruments such as \gls{grand}  or \gls{poemma}.

\section*{Verification}
\begin{itemize}
    \item Q1. Why IACTs are sensitive to Earth-skimming \nutaus and not to other neutrino families?
    \item Q2. Why the expected background in IACT observation of Earth-skimming tau neutrinos is expected to be small?
    \item Q3. Why the sensitivity of IACT observation of Earth-skimming tau neutrinos is concentrated in the region of PeV to EeV?
    \item Q4. What are the main limiting factors for the observation of Earth-skimming tau neutrinos for MAGIC and CTA?
\end{itemize}
%\newpage
\section{Magnetic Monopoles}
\label{sec:monopoles}

In the classical Maxwellian formulation of electromagnetism, if 
a magnetic charge $g$ and a magnetic current $\rho_m$ are introduced, the
Maxwell equations would become symmetric if one swaps electric and
magnetic fields, and the electric and magnetic constants:
\begin{eqnarray}
\vv{\nabla}\cdot\epsilon\vv{E}=\rho_e   \qquad&&  \vv{\nabla} \times \vv{E}=-\vv{j_m}-\frac{\partial\vv{B}}{\partial t} \\
\vv{\nabla}\cdot\vv{B}=\rho_m   \qquad  &&\vv{\nabla} \times \vv{B}/\mu=\vv{j_e}+\frac{\partial\epsilon\vv{E}}{\partial t} 
\end{eqnarray}

This symmetry, albeit tantalizing, is valid only in the trivial case
in which neither an electric charge nor an electric current are present. In
any other case, it is incompatible with the concept of vector
potential $\vv{A}$, either because the divergence theorem states that the monopole
field would be null, or because any magnetic field line
is closed.

An elegant solution to this conundrum was found almost a
century ago when \citet{Dirac:1931} introduced the concept of the (now-called)
\emph{Dirac string}, a hypothetical dimensionless entity that can be
visualized as a solenoid, that shifts the
magnetic field phase by multiples of $2\pi$.
%, and as such, allow the null vector potential divergence.
This kind of a phase-shift
is similar to the observed Aharonov--Bohm solenoid
effect~\citep{Aharonov:1959}, a quantum-mechanical effect in
which the electromagnetic field of a charged particle undergoes a phase-shift even in regions of
null electric and magnetic fields such as within a long
solenoid.
%However, the dimensionless condition reflects an unphysical
%condition for the Dirac \gls{mm}.   
The Dirac string theory embodies the  \emph{classical} \gls{mm} and it is
still suggestive because it 
naturally predicts the quantization of the electric charge (so-called \emph{Dirac quantization rule}). This
rule affirms that the product of magnetic $g$ and electric $e$ charge is
quantized as an integer multiple of $2\pi$, as a consequence of the
$2\pi$ phase shift induced by the Dirac string:
\begin{equation}\label{eq:quantization}
    e\,g=2\pi\hbar\,n%=3.3\times 10^{-8} \quad\mbox{cgs}
\end{equation}
with $g=g_D$ defined as the Dirac elementary magnetic charge, or the
unit \emph{Dirac charge}. It naturally motivates the
hitherto unexplained quantization of the electric charge.
%Similarly, the magnetic constant would be
                          %$\alpha_m=g^2/\hbar\,c=34.3$, in analogy to
                          %the fine-structure constant.  
By assuming a Dirac string size such as that of the electron, it is possible to
estimate that the Dirac string would have a mass of 2.4~GeV. The
signatures of the Dirac string would have been seen at
accelerators and the missing detection provides strong motivation for rejecting this
 elegant solution~\citep{Patrizii:2015uea}. Dirac himself 
said of \gls{mm} \textit{``One would be surprised if Nature had made no use of it.''}

A change of paradigm happened few decades later, when
\citet{Hooft:1974} and \citet{Polyakov:1974} realised that \gls{mm}s
were not only a feature arising from the asymmetry in the Maxwell
equations, but also naturally predicted in \gls{guts}. Unlike the
Dirac \gls{mm}, \gls{guts} \gls{mm}s would be extremely heavy,
because directly related to the mass of the mediator of the
unified interaction, in the range
$10^{16}\div10^{17}$~GeV~\citep{Preskill:1984}, or even above, in
\gls{guts} including gravity~\citep{Giacomelli:2001ag}. Considering
they arise from a complete particle physics theory, \gls{guts}
\gls{mm}s have the advantage of allowing a more complete theoretical
mapping, with predictable properties such as internal structure and
interactions with other fields. They are not structure-less as the
Dirac \gls{mm}, but on the contrary a huge conglomerate of particle fields
arranged in layers, formed by a core of gauge bosons of the unified
interaction, surrounded by shells of electroweak bosons and confined
by a fermion-antifermion condensate region~\citep{Patrizii:2015uea}.  

At distances larger than few fm, such a \gls{mm} would behave like a pointlike magnetic charge generating a radial magnetic field. These objects are also called a
\emph{topological soliton} and they could even show excited states in
which they acquire a net electrical charge, going under the more
general name of dyon~\citep{Schwinger:1969}. These \gls{mm} would be
stable relics of events happened in the early Universe, such as the
\emph{freeze-out} of some conditions or a symmetry breaking. Their
decay would be thereafter forbidden due to the conservation of the
magnetic charge.  
A further scenario is that of intermediate mass \gls{mm}, generated at
later cosmic times, during other phase
transitions~\citep{Polyakov:1974}, allowing for masses of the order of
$10^{7}\div10^{13}$~GeV~\citep{Patrizii:2015uea}. Depending on their
mass and number density, \gls{mm}s could actually constitute all or a
 fraction of the total \gls{dm} of the
Universe~\citep{Khoze:2014woa}. For further readings see,
e.g.,~\citet{Shnir:2005}. 

The Universe expansion must have slowed down the initial velocity of
\gls{mm}s down to the non-relativistic speed. However, dynamo effects in magnetic
fields of galaxy clusters, active galactic nuclei or even possible
magnetars could have re-accelerated a fraction of \gls{mm}s: such
acceleration process could be in fact highly efficient. A \gls{mm}
traversing a magnetic field $B$ with coherence length $L$ gains
$K=g\,B\,L$ in kinetic energy \citep{Wick:2003}. For values of
$B=3~\mu{\rm G}$ and $L=1$~kpc, like those typical of galaxy clusters, the
gain corresponds to $1.8\times 10^{11}~{\rm GeV}$. This mechanism could
even allow the thesis that \gls{mm}s could be associated to
ultra-high-energy (EeV) cosmic rays~\citep{Kephart:1995bi}. The
acceleration of \gls{mm} happens at the expense of local magnetic
fields. Out of this, \citet{Parker:1970} obtained a conservative
limit on the \gls{mm} density requiring that this 'draw' of magnetic
energy was sufficiently slow not to disrupt galactic magnetic fields
completely. This is the so-called Parker bound. Galactic objects such
as the Sun could accumulate \gls{mm}s in their core. The enhanced overdensity
could ignite  \gls{mm} catalized nucleon decay. Because of the presence of
\gls{guts} gauge bosons in their core, interactions such as
$p+\mbox{MM}\rightarrow e^+ +\mbox{MM}+\pi_0$, may have
ample cross sections if the interaction is independent from the grand unification gause boson mass \citep[Callan-Rubakov process;][]{Callan:1982,Rubakov:1982}, generating huge pion fluxes
and providing strong photons or neutrino bursts after the pion
decays~\citep{PRD2018}. Not long ago,  Super Kamiokande 
provided very strong constraints on
\gls{mm}s~\citep{Ueno:2012md}. The Callan-Rubakov process is
efficient only for \gls{guts} \gls{mm}s of moderate
speed $\beta<0.3$.  

\par
In the passage through matter, a \gls{mm} with $\beta>0.1$ would behave like an equivalent huge electric charge, with strong exciting and ionizing power. Dyons, being electrically charged, would also directly generate Cherenkov light. The number of photons $N$ per unit length $x$ and wavelength $\lambda$ for charged particles is computed by \citet{Tamm:1937} in air:
\begin{equation}\label{eq:tamm-frank}
 \frac{d^2N}{dx\,d\lambda}(\text{MM})\simeq \frac{n^2}{4\,\alpha^2}\;\frac{d^2N}{dx\,d\lambda}(e,\mu),   
\end{equation}
and computed for \gls{mm} using Eq.~\ref{eq:quantization}. In Eq.~\ref{eq:tamm-frank}, $n$ is the air refraction index and $\alpha$ is the
electromagnetic fine structure constant. In air,
$n^2/(4\,\alpha^2)\simeq4,700$ and therefore the number of photons
produced by a \gls{mm} is significantly larger than that by an
electron or a muon, and it
corresponds to an energy loss of about $8$ GeV g$^{-1}$ cm$^{-2}$
compared to $2$ MeV g$^{-1}$ cm$^{-2}$ for the
muons~\citep{Tompkins:1965,Cecchini:2016vrw}. 
Such a strong interaction with matter opens up interesting
possibilities for the detection of \gls{mm}s, for example by generating local damages on high-Z material slabs when traversed by
\gls{mm}s that are inspected off-line. For masses above $10^3$~GeV, the
MACRO \citep{Ambrosio:2002qq} and MoEDAL \citep{Patrizii:2015uea} instruments  are successful examples of highly sensitive nuclear track detectors.

\gls{mm}s could be also identified by their passage through the Earth's
atmosphere. In this occurrence, they will strongly radiate Cherenkov
light, either directly, if electrically charged, or through the
secondary charged particle generated through ionization, as well as
through luminescence~\citep{ObertackePollmann:2016uvi}. Even a
globally neutral \gls{mm}, due to its charged outer layer, would
excite atmospheric nuclei throughout the full length of the
atmosphere. Such an effect does not happen for cosmic rays, that are
absorbed in the high atmosphere. The search for such a signature was
investigated with the \gls{pao}. \citet{Aab:2016poe} reported strong
constraints searching for fluorescence light induced by \gls{mm}s. 
%, in the range $10^{-19}\div10^{-21}$ cm$^{-2}$ s$^{-1}$ sr$^{-1}$ for speed of $\beta\sim1$. 
The IceCube Collaboration
\citep{Abbasi:2012eda,Aartsen:2014awd,Aartsen:2015exf}, improving on
the results obtained with AMANDA~\citep{Achterberg:2010}, performed
extensive searches for non-relativistic and relativistic
\gls{mm}s. \gls{guts} \gls{mm}s could catalyze proto decays in the
surrounding Antarctic ice via the Callan-Rubakov effect. This would
generate relativistic charged particles emitting Cherenkov light with
a characteristic hit pattern in the detector.  
No such pattern was observed, allowing to limit the flux of non-relativistic GUT \gls{mm}s to %$<10^{-17}\div10^{-18}$ cm$^{-2}$s$^{-1}$sr$^{-1}$~\citep{Aartsen:2014awd}, 
three orders of magnitude below the Parker bound. 
%For relativistc\gls{mm}s, the limit was set to similar values of $<1.55\times10^{-18}$ cm$^{-2}$s$^{-1}$sr$^{-1}$~\citep{Aartsen:2015exf}. 
A collection of the limits discussed above on the \gls{mm} flux is shown in Fig.\ref{fig:monopole_limits}.

\paragraph{Magnetic Monopoles searches with IACTs.}
The search for \gls{mm} signatures was so far only investigated with a
simplified approach by \citet{Spengler:2009} and 
\citet{Spengler:2011}. \citet{Spengler:2009} provided preliminary
discussion of the sensitivity to \gls{mm}s by studying dedicated \gls{mc}
simulations, optimization of the selection cuts, and real data selection.
%,
%and from this computation of upper limits in flux by computing the
%proper acceptance.
%Throughout the work, several simplifications were
%utilized, mostly because of conciness of the work, and sometimes
%because lack of proper knowledge of all aspects of the \gls{mm}
%interactions within the Earth atmosphere. It was previosly  mentioned
%in fact, that when a \gls{mm} crosses the Earth atmosphere, it will
%lose energy through direct and induced Cherenkov emission. The
%relative contribution of these effects depends on the \gls{mm} actual
%mass, speed and magnetic charge. 

\begin{figure}
    \centering
    \includegraphics[height=6cm]{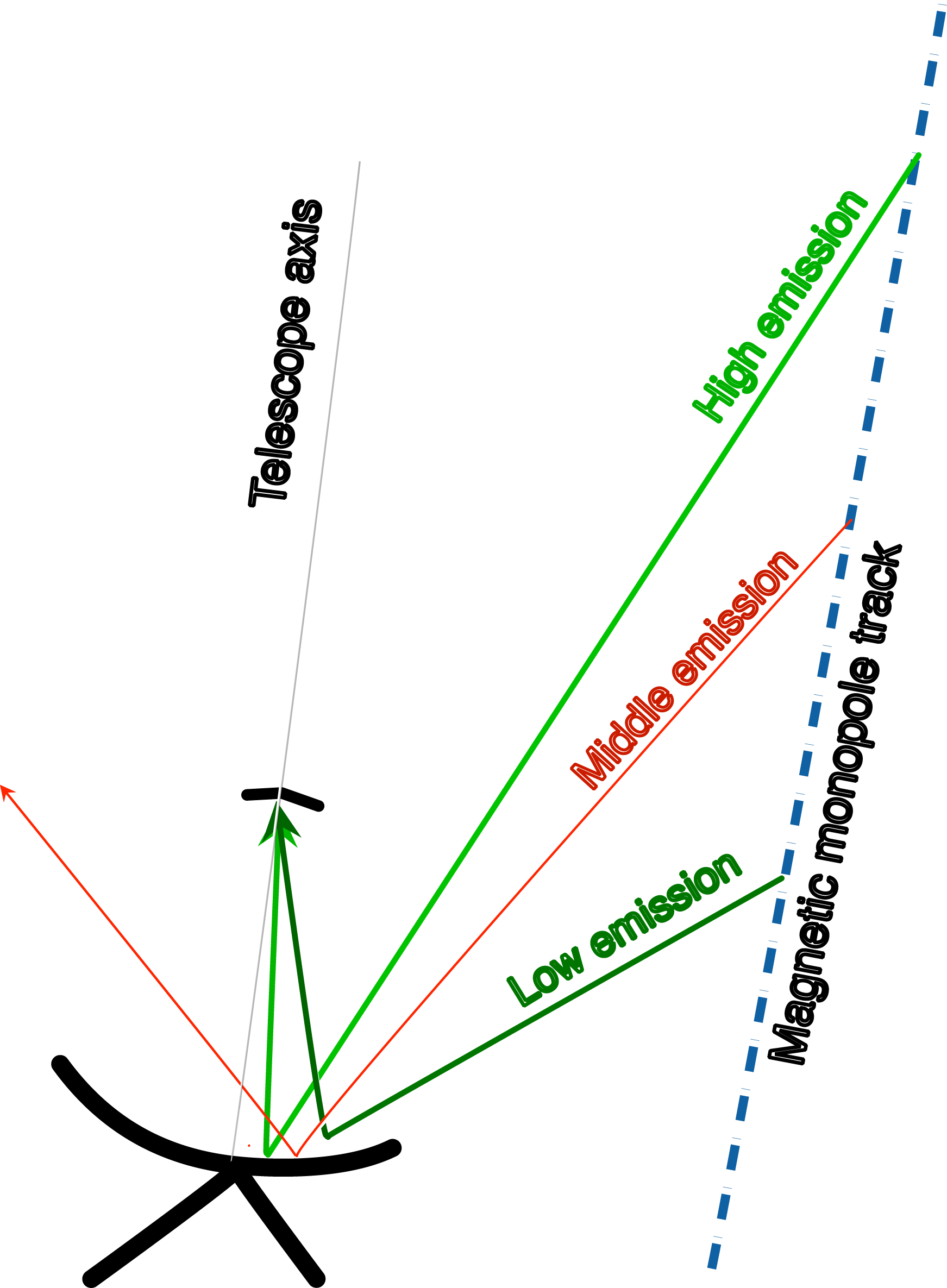}
    \hspace{2cm}
    \includegraphics[width=6cm,height=6cm]{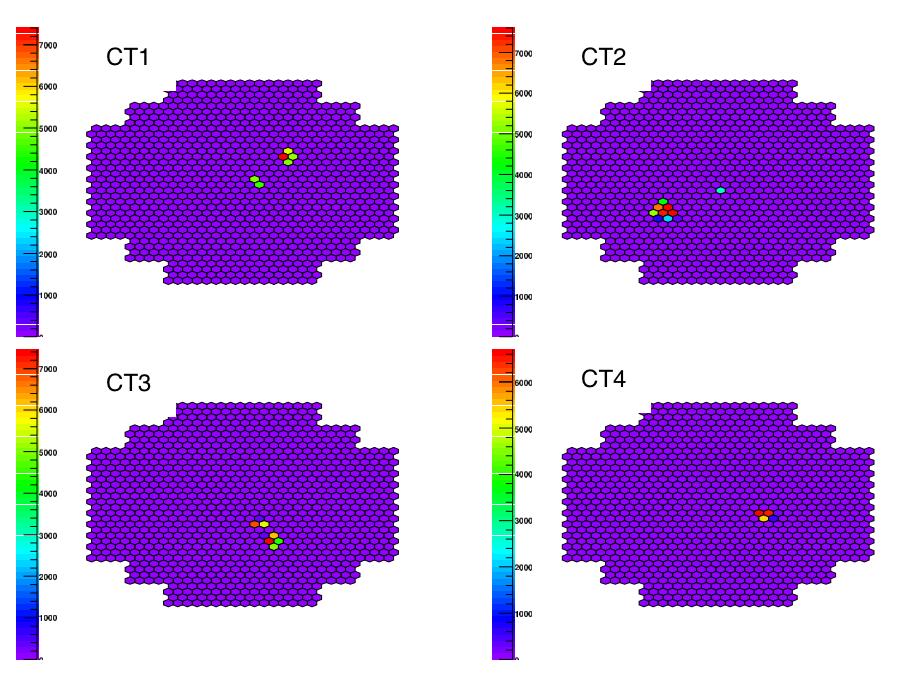}
    \caption{Left: Emission scheme from an ultrarelativistic \gls{mm}
      emitting Cherenkov radiation throughout the full length of the
      atmosphere. Right: A simulated \gls{mm} event on
      \gls{hess} cameras. Courtesy of
      \citep{Spengler:2009}.} 
    \label{fig:monopole}
\end{figure}

\begin{figure}[h!t]
    \centering
    \includegraphics[width=0.9\linewidth]{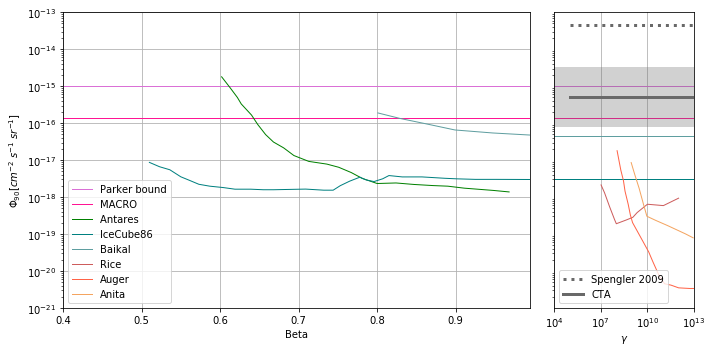}
    \caption{\label{fig:monopole_limits}Current imits on the flux of
      \gls{mm} as a function of the speed of \gls{mm} at the
      detector. Upper limits obtained with the \gls{hess} instrument
      are shown as a grey dashed line in the inset to the right
      \citep{Spengler:2009}. Preliminary predictions for CTA are also shown
      as a thick gray line \citep{Doro:2017vjf}.} 
\end{figure}

A \gls{mm} entering the Earth's atmosphere will start generating
Cherenkov light in the high atmosphere, at 80~km altitude or
above. This emission will start within a narrow Cherenkov cone angle
of 0.1~deg, increasing to angles of 1.2~deg when the \gls{mm} will
hit the ground. Differently than cosmic rays, the Cherenkov emission will
last throughout the full length of the \gls{mm} track. The Cherenkov
emission depends on the \gls{mm} Lorentz factor $\gamma$, which is related to the
air local refraction index $n$ by the 
relation:
\begin{equation}
\gamma > \frac{1}{\sqrt{2\,(n-1)}}
\end{equation}
For the atmosphere close to the ground $n-1=10^{-4}$ and therefore a
minimum boost of $\gamma>70$ is required to have Cherenkov emission
throughout the full length of the track. This condition in turn allows
for a significant effectiveness in the separation from the
background. This is illustrated in Fig.~\ref{fig:monopole}: the
\gls{mm} would be observed as a double-spot system in the IACT
cameras: a first due to the very high altitude emission, and a second
due to the low-altitude emission, while the emission from the central
part of the track would not be geometrically focused.
A second simplification was applied in order to guarantee minimal
energy loss of the \gls{mm} during the track. \citet{Spengler:2009} computed a total
energy loss of 22.5~TeV throughout the atmosphere. This required
therefore that $\gamma>22.5~\rm{TeV}/m_{MM}c^2\sim10^5$ or, in other
words, the analysis is limited to ultra-relativistic \gls{mm}.
We will discuss the effect of these limitations further below.

The images formed by \gls{mm}s in the IACTs cameras will be mostly
composed of small clusters of very illuminated pixels, for the
high-altitude signal, or very bright fraction or rings, similarly to
the muon case, for low-altitude originating emission. An example of a
simulated event is shown in the right panel of Fig.~\ref{fig:monopole}. 
Considering the large Cherenkov photon yield, thousands of times larger
than that of a muon, all illuminated pixels will show extremely high
signals, often saturating the reach of the camera \gls{pmt}s. This
further characteristic was used by \citet{Spengler:2009} as main
selection cut: an event was to show a large fraction of saturated
pixels in at least two out of four telescopes. Further cuts, also
optimised through \gls{mc}, were applied to exclude all the remaining
background. For example, less than 200 pixels were to trigger in each
camera, in order to remove high energy hadronic showers surviving
previous cuts. The selection cuts were optimized on a sample of  about
$10^6$ \gls{mc} events from 0 to 60~deg zenith angle, and applied to 5
years of (selected) \gls{hess} data, for a total of $2\,400$~h. No
\gls{mm} candidate event was found, and from this, limits of
$<4.5\times10^{-14}$ cm$^{-2}$s$^{-1}$sr$^{-1}$ were computed and shown
in Fig.~\ref{fig:monopole_limits}. These values are extremely poorly
constraining, less than the conservative Parker bound, and orders of
magnitude less constraining than Auger or IceCube limits. This is due
to the larger aperture and duty cycle of these instruments. In any case, these
limits should be taken as conservative limits in a very simplified
scenario, and further exploration is required.

\paragraph{Outlook.}
Improving on this work would require to reasses the \gls{mc}
simulations, by allowing lower speed connected effect such as more
significant energy loss during the \gls{mm} tracking, appeareance of
secondary hadronic sub-showers which would alter the signatures in the
IACTs camera and therefore the simplified picture of background/signal
separation. Furthermore, \citet{Spengler:2009} did not use the time of
arrival of photons to separate candidate events from background, which
requires a isochronous paraboloid such as that of the CTA \glspl{lst}.
CTA also features a wider \gls{fov} that allows to increase the
acceptance to \gls{mm} events, allowing those with farther impact
parameters. Finally, one should mention that by pointing the telescope
to very high zenith angles (low altitude above the horizon) one could
in principle modify the geometry of the event and increase largely the
effective area. \gls{mm}s investigation with CTA would also profit by the
10-fold increase in effective area, and the improvement in energy
resolution, angular resolution and off-axis performance. A back of the
envelope estimation for CTA performance was discussed in
\citet{Doro:2012xx}, however, a full quantification of the performance
is yet to be done. CTA also features a more flexible software inter-trigger
system, which could allow dedicated search campaigns. 
Still, \gls{mm} events could survive trigger
criteria and be hidden in current IACT data archive. In MAGIC, the
number of background events stored on disk is roughly $1.2\times10^9$ 
events/year for two telescopes \citep{Doro:2017vjf},\footnote{for CTA this
number could be a factor of 50 larger in principle} thus providing a
rich search sample. Nonetheless, specific on-line or off-line fast
selection criteria need to be envisaged to avoid rejecting possible
triggers from \glspl{mm} and thus increasing the chances of detection or the
constraining power of any consequent limit.  If detected, \glspl{mm}
would provide possibly clear signature in an IACT. Despite
the smaller strengths of IACTs compared to other techniques, it is
important to notice that this search would be achievable at no cost of
dedicated telescope pointings nor specifically allocated observation
time, but making use of archival data and tailored analyses. 

\section*{Verification}
\begin{itemize}
    \item Q1. Why does a \gls{mm} leave a different imprint in the cameras of IACTs with respect to a primary $\gamma$ shower?
    \item Q2. What is the time imprint of a shower initiated by a \gls{mm}?
    \item Q3. What is the main reason why IACTs are mostly sensitive to ultra-relativistic \gls{mm}s?
    \item Q4. If you were to optimize your search analysis on \gls{mm}s, on what would you base your classification algorithm?
\end{itemize}

\section{Summary and Conclusions}
\label{sec:conclusion}
Fundamental physics is probably a slippery wording, because every piece of evidence found in scientific pursues is fundamental in the increase of knowledge. Yet, there is no single word that can properly encompass the searches for massive particle \gls{dm}, axion-like particles, primordial black holes, tau neutrinos, and magnetic monopoles that have been summarized in this chapter. In most cases, we do not possess a full theoretical mapping. Therefore we often cannot sharpen our instruments to aim for detection, but merely to set yet-another constraints. Nevertheless, we believe that placing many independent and complementary constraints to exotic phenomena -  and, in some cases, \glspl{iact} provide the strongest ones -- is how we build trust in successful  theories, and how we rule out alternative explanations or extended  models. It is a crucial part of physical observation to continuously challenge the current worldview and to test predictions of novel phenomena. This is the reason for us to have devoted thousands of hours of \gls{iact} observations and invested tenfold hours of data analysis and investigation in the search for the \emph{unknown}. 
%
%Michele's original text: is how we build trust in successful  theories, or how we distrust others. This is the reason for us to have devoted thousands of hours of \gls{iact} observations and invested tenfold hours of data analysis and investigation in the search for signatures of these \emph{unknowns}.  Nevertheless, we believe that placing independent and complementary constraints (and, in some cases, the strongest ones) is how we build trust on theories, or how we distrust others. This is the reason for us to have devoted thousands of hours of \gls{iact} observations and invested tenfold hours of data analysis and investigation in the search for signatures of these \emph{unknowns}.
%
All in all, astrophysical gamma rays are wonderful probes: they do not only trace back to the emitting source, but reveal to us nuclear and sub-nuclear interactions that have taken place billions of billions of kilometers away, that is, tens or hundreds of millions of years back in time, and at energy scales hardly accessible in the laboratory. For \gls{dm}, if the mass of its constituent, elusive particle is above several hundreds GeV, gamma rays may represent one of the few probes  -- perhaps the only one -- to unveil its true nature beyond gravitational pull. We may be hunting far from the right spot, yet constraints set by \glspl{iact} on the nature of \gls{dm} are the strongest in the TeV regime as of today, and they will likely stay so for many years to come.  Current limits from the Galactic centre halo by \gls{hess} have reached the thermal cross section value of $3\times10^{-26}\,$cm$^3$s$^{-1}$, a reference value for WIMP \gls{dm} models. 
% very good comments around this by Graciala Gelmini at Kashiwa Symposium 2019: https://indico.icrr.u-tokyo.ac.jp/event/259/contributions/1659/attachments/1196/1425/Crossroads-WIMP.pdf
However, these limits could suffer from significant uncertainties regarding the precise amount and distribution of \gls{dm} in the Galactic centre region. This kind of uncertainty is less severe in the case of using Milky Way dwarf satellite galaxies. Therefore, not in vain \gls{hess}, \gls{magic} and \gls{veritas} have also extensively searched for \gls{dm} in the best candidates among them, although as well without detecting a signal so far.  

While on searches for \gls{wimp} \gls{dm} \glspl{iact} are bannermen, their role in the search for axion-like particles, primordial black holes, and magnetic monopoles  is more that of pawns compared to other instruments. Nevertheless, the potential of these searches is large, and it is important to recall that these are made on archival data. There is no need of new specific observations but rather of careful and complex analyses. A similar situation occurs for tau neutrinos. Even if the sensitivity of \glspl{iact} is not competitive to instruments entirely designed for the detection of astrophysical neutrinos, it has been shown that \gls{iact} may play a significant role, specially if specific conditions are met at the site. 

The upcoming \gls{cta} will provide a performance boost in all of these fields. We might finally unequivocally exclude -- or confirm -- the hypothesis of massive particle \gls{dm}, and also search for axion-like particles, primordial black holes, magnetic monopoles, and tau neutrinos with in all cases a several factors higher sensitivity. Yet, in order to adequately profit of this boost, some prerequisites have to be guaranteed: first, that the theoretical and analysis experience gained from the pioneering studies here presented should not be lost, and we hope to have contributed to that. Second, and more importantly: the \gls{iact} community should be aware of the relevance of these pursues for fundamental physics, and take as an obligation to devote time and resources to them, a fact that often clashes with the need of longer observations for other 
%\mh{\st{more ``classical''}} 
astrophysical endeavours. Last, but by no means least: we should guarantee that \gls{cta} -- or any future \gls{iact} experiment -- and its corresponding data processing pipelines are put in a position to potentially catch fundamental physics events. As examples, along these pages we have shown how peculiar magnetic monopoles signatures are expected to be, or how serendipitous primordial black hole events may show up, lasting just a few seconds anywhere in the field of view. We should therefore make an extra effort to guarantee that at no place we simplify our systems so to 
%be sensitive only to ``conventional'' gamma-ray signals and not 
drop information to catch fundamental physics events. For example, by selecting only the brightest pixels in the camera and throwing away all others from the data acquisition and analysis, we may lose all means to identify a rare event occurring in the camera~\citep{Doro:2017vjf}. It is important to revive in the community the discussion on the best strategy to search for fundamental physics with \gls{cta}; one that will also be in a good balance with the rest of physics cases. Let nature provide the rest in our hope to ever catch signatures of new physics with \gls{iact} data.

\bigskip
\paragraph{Acknowledgements.} The authors would like to warmly thank for useful discussions to E. Bernardini, A. De Angelis, M. Gaug, D. Gora, D. Krennrich, M.~Meyer, N.~Otte, L. Patrizii,  A.~Raccanelli, J.~Rico, U.~Schwanke, and G. Spengler.

\bibliographystyle{abbrvnat}
\bibliography{biblio}

\end{document}